\documentclass[traditabstract]{aa}
\usepackage{natbib}
 \bibpunct{(}{)}{;}{a}{}{,}    
\usepackage{txfonts}
\usepackage{times}
\usepackage{amsmath}
\usepackage{amssymb}
\usepackage{graphicx}

\begin{document}

\title{Low-redshift compact star-forming galaxies as analogues
of high-redshift star-forming galaxies}
\author{Y. I. \ Izotov \inst{1,2}    
\and N. G. \ Guseva \inst{1,2}
\and K. J. \ Fricke \inst{2,3}
\and C. \ Henkel \inst{2,4}
\and D. \ Schaerer \inst{5.6}
\and T. X. \ Thuan \inst{7}
}
\offprints{Y. I. Izotov, yizotov@bitp.kiev.ua}
\institute{          
                     Bogolyubov Institute for Theoretical Physics,
                     National Academy of Sciences of Ukraine, 
                     14-b Metrolohichna str., Kyiv, 03143, Ukraine
\and
                     Max-Planck-Institut f\"ur Radioastronomie, Auf dem H\"ugel 
                     69, 53121 Bonn, Germany
\and 
                     Institut f\"ur Astrophysik, G\"ottingen Universit\"at, 
                     Friedrich-Hund-Platz 1, 37077 G\"ottingen, Germany 
\and
                     Astronomy Department, King Abdulaziz University, 
                     P.O.Box 80203, Jeddah 21589, Saudi Arabia
\and
                     Observatoire de Gen\`eve, Universit\'e de Gen\`eve, 
                     51 Ch. des Maillettes, 1290, Versoix, Switzerland
\and
                     IRAP/CNRS, 14, Av. E. Belin, 31400 Toulouse, France
\and
                     Astronomy Department, University of Virginia, P.O. 
                     Box 400325, Charlottesville, VA 22904-4325, USA
}
\date{Received \hskip 2cm; Accepted}

\abstract{
We compare the relations among various integrated characteristics 
of $\sim$ 25,000 low-redshift ($z$ $\protect\la$ 1.0)
compact star-forming galaxies (CSFGs) from Data Release 16 (DR16) of the 
Sloan Digital Sky Survey (SDSS) and of high-redshift ($z$ $\protect\ga$ 1.5) star-forming 
galaxies (SFGs) with respect to oxygen 
abundances, stellar masses $M_{\star}$, far-UV absolute magnitudes
$M_{\rm FUV}$, star-formation rates SFR and specific star-formation rates sSFR,
Lyman-continuum photon production efficiencies ($\xi_{\rm ion}$), UV continuum 
slopes $\beta$, [O~{\sc iii}]~$\lambda$5007/[O~{\sc ii}]~$\lambda$3727 and
[Ne~{\sc iii}]~$\lambda$3868/[O~{\sc ii}]~$\lambda$3727 ratios, and emission-line equivalent 
widths EW([O~{\sc ii}] $\lambda$3727), EW([O~{\sc iii}] $\lambda$5007), and 
EW(H$\alpha$). We find that the relations for low-$z$ CSFGs with high equivalent
widths of the H$\beta$ emission line, EW(H$\beta$) $\geq$ 100\AA, and 
high-$z$ SFGs are very similar, implying close physical properties in these two 
categories of galaxies. Thus, CSFGs are likely
excellent proxies for the SFGs in the high-$z$ Universe. They also extend to
galaxies with lower stellar masses, down to $\sim$ 10$^6$ M$_\odot$, and to 
absolute FUV magnitudes as faint as $-$14 mag. Thanks to their proximity, CSFGs can 
be studied in much greater detail than distant SFGs. Therefore, the 
relations between the integrated characteristics of the large sample of CSFGs studied here can prove 
very useful for our understanding of high-$z$ dwarf galaxies in future 
observations with large ground-based and space telescopes.
}
\keywords{galaxies: abundances --- galaxies: irregular --- 
galaxies: evolution --- galaxies: formation
--- galaxies: ISM --- H {\sc ii} regions --- ISM: abundances}
\titlerunning{SDSS compact star-forming galaxies}
\authorrunning{Y. I. Izotov et al.}
\maketitle

\section {Introduction}\label{intro}

Considerable progress has been made over the past decade in extending our
understanding of the physical properties of high-redshift star-forming galaxies
(SFGs) and their evolution in the redshift range extending from $z$~$>$~2  to
$z$ $\sim$ 8 -- 11 \citep[e.g. ][]{Oe16,Wi16,Sa18,Is18,St20,St20b,La19b,DB19,Bo19}.
Large samples of galaxies at $z$ $>$ 2 have become available thanks to the release of extensive
spectroscopic surveys \citep[e.g. ][]{St14,Kr15,ML18,Ba19}.

The deep spectra of high-$z$ galaxies have revealed that their properties are different 
from those of typical $z$~$\sim$~0 SFGs. 
In the Baldwin-Phillips-Terlevich (BPT) 
diagram \citep{BPT81}, these galaxies are offset to higher 
[O~{\sc iii}] $\lambda$5007/H$\beta$ ratios at a fixed 
[N~{\sc ii}] $\lambda$6584/H$\alpha$ ratio \citep{St14,Sh15,Sa20,To20a,To20b}, as compared 
to $z$ $\sim$ 0 main-sequence galaxies \citep{Ke13}. \citet{St17}, \citet{Sa20}, 
and \citet{To20a} attributed this difference to the  harder ionising spectrum at a 
fixed nebular metallicity in high-$z$ SFGs, whereas \citet{Ke13} and \citet{St14}
suggested that in addition to a harder ionising radiation, a higher 
ionisation parameter and a higher number density or a higher N/O abundance ratio
can also shift the relation for high-$z$ SFGs to higher 
[O~{\sc iii}] $\lambda$5007/H$\beta$ and [N~{\sc ii}] $\lambda$6584/H$\alpha$ 
ratios.

A common feature of high-$z$ SFGs is the presence of
hydrogen Ly$\alpha$, H$\beta,$ and H$\alpha$ emission lines, as well as the 
presence of some forbidden emission lines -- most often of the
[O~{\sc ii}] $\lambda$3727, [O~{\sc iii}] $\lambda$4959, 5007,
[N~{\sc ii}] $\lambda$6584 emission lines (in the rest-frame optical spectra), 
or C~{\sc iii}] $\lambda$1906, 1909 emission lines (in the rest-frame UV spectra).
This allows us to derive oxygen abundances 12 + log O/H, most commonly with the use
of the strong-line calibrations \citep[e.g. ][]{PP04}, which link strong-line 
fluxes and 12 + log O/H. The results of these determinations have been presented
by \citet{Cu14}, \citet{Tr14}, \citet{On16}, \citet{San15}, \citet{Va20},
\citet{Sa20b}, and others. In the rarer cases
when the weak [O~{\sc iii}] $\lambda$4363 emission-line is detected
\citep{A16,Ko17,Ko20}, the direct $T_{\rm e}$ method may be used. 
Various studies have established that $z$ $\geq$ 2 mass-metallicity
relations (MZR) based on oxygen are offset by $\sim$ 0.3 -- 0.7 dex to lower 
metallicities relative to typical main-sequence SFGs at $z$ $\sim$ 0. 

\citet{Ma10} introduced an additional parameter, namely, the
star formation rate (SFR), which is higher in SFGs at larger $z$.
They proposed the use of a more general relation, that is, the fundamental mass-metallicity
relation (FMR) and replacing $M_\star$ with SFR$^{-\alpha}$$M_\star$. \citet{Ma10} 
found that the (12+logO/H) -- SFR$^{-\alpha}$$M_\star$ relation is unique for SFGs 
at any redshift if $\alpha$ = 0.32. On the other hand, \citet{Cu20} derived $\alpha$ = 0.55. 

The $M_\star$ -- SFR and $M_\star$ -- specific SFR 
(defined as sSFR = SFR/$M_\star$) relations are most commonly derived for high-$z$ 
SFGs \citep[e.g. ][]{Wh14,Sa15,Shi15,On16,MQ16,Sm16,Ka17,Sa17,Iy18,Da18,Kh20}. 
Stellar masses are usually obtained from the spectral energy distribution (SED) 
in the rest-frame UV range, which is derived adopting a 
certain star formation history (SFH). The SFR and sSFR are derived either from 
the UV luminosity, from the UV SED, or from the luminosities of hydrogen
emission lines (Ly$\alpha$, H$\beta$, H$\alpha$). Some of these above-cited works found that
the slope of the $M_\star$ SFR relation is steeper at low stellar masses and 
that the SFR increases with $z$ at a fixed $M_\star$ \citep[e.g. ][]{Wh14}. 
However, \citet{Iy18} found that log SFR -- log $M_\star$ at $z$ = 6 remains 
linear down to log~($M_\star$/M$_\odot$)~=~6. Similarly, the 
average sSFR systematically increases with increasing $z$ \citep{MQ16,Da18},
whereas \citet{Sa15} found an unevolving correlation between stellar 
mass and SFR between $z$~=~6 and $z$~=~4. Furthermore, for instance, \citet{Sa17} have 
shown that the increase in the sSFR with redshift is milder than predicted by 
theoretical models and that the sSFR is nearly constant at $z$~$>$~2 with an
average value of $\sim$ 3 Gyr$^{-1}$. For comparison, \citet{St17} found 
sSFR $\sim$ 3 -- 40 Gyr$^{-1}$ in the redshift range of $z$ $\sim$ 2 -- 3.

It was found that high-$z$ SFGs have small sizes in both the FUV continuum and
Ly$\alpha$ emission line \citep[e.g. ][]{Ha16,Bo17,PA17,PA18,Ma18,Ma19}. 
In particular, \citet{Ha16} and \citet{PA17} noted that the brightness 
distribution in $z$ $\sim$ 2
SFGs are characterised by exponential scales $r_{\rm e}$ = 0.5 -- 3.0 kpc.
\citet{Bo17} found extremely small sizes of faint $z$~$\sim$~2~--~8 
galaxies  with half-light radii $\sim$ 100 -- 200 pc for $M_{\rm FUV}$~$\sim$~--15
mag, whereas \citet{Ma18} derived $r_{50,{\rm Ly}\alpha}$ $\sim$ 0.3 kpc for the
luminous ($M_{\rm FUV}$ = --21.6) SFG COLA1 at $z$ = 6.593.
It was also found that galaxies with brighter Ly$\alpha$ emission lines
tend to be more compact both in the UV and in Ly$\alpha$ \citep{PA18,Ma19}.
For comparison, the confirmed low-$z$ LyC leakers and CSFGs with
extremely high O$_{32}$ = [O~{\sc iii}]$\lambda$5007/[O~{\sc ii}]$\lambda$3727
ratios of $>$ 20 show, in the {\sl HST} NUV images, a
compact morphology with $r_{\rm e}$ $\sim$ 0.1 -- 1.4 kpc
\citep{I16a,I16b,I18a,I18b,I20}.
 
The slope $\beta$ determined from the relation 
$F_\lambda$ $\propto$ $\lambda^\beta$ of the UV spectra 
in high-$z$ SFGs is steep. \citet{Bo12,Bo14} and \citet{Ya19}
found that $\beta$ in SFGs at $z$ $\sim$ 4 -- 8 is in the range of --1.5 - --2.4,
whereas \citet{Wi16} derived $\beta$~$\sim$~--1.9~-~--2.3 for $z$~$\sim$~10 
galaxies, with the UV slope becoming steeper for the higher redshift SFGs.
Additionally, \citet{Bo12,Bo14} found a steeper $\beta$ for SFGs with 
lower $M_\star$ and fainter UV luminosities. \citet{Pe18} derived a similar range
of $\beta$ for $z$ $\sim$ 6. \citet{San20} determined steep $\beta$~$\sim$~--2.0
and $M_\star$ $\sim$ 2$\times$10$^9$ M$_\odot$ for $\sim$ 4000 Lyman-alpha 
emitters (LAEs) at 
$z$ = 2 -- 6. They concluded that typical star-forming galaxies at high redshift
effectively become LAEs. 

The steep slopes in the UV range and the strong emission lines of high-$z$ SFGs indicate 
the presence of massive stellar populations which produce copious amounts of 
ionising photons. It has thus been suggested that SFGs are the main source of the
reionisation of the Universe given that the product 
$\xi_{\rm ion}$$\times$$f_{\rm esc}$(LyC) is sufficiently high, where $\xi_{\rm ion}$
is the Lyman-continuum production efficiency and $f_{\rm esc}$(LyC) is the
fraction of the LyC radiation escaping galaxies. Currently, direct 
observations of the LyC have resulted in the discovery of approximately
three dozens of LyC leakers at high redshifts, with reliably high LyC escape fractions 
\citep{Va15,DB16,Sh16,Bi17,Va18,RT19,St18,Fl19}.
The parameter $\xi_{\rm ion}$ is derived from synthesis models and it
depends on metallicity, star formation history, age, and assumptions on  
stellar evolution. Values of 
log [$\xi_{\rm ion}$/Hz erg$^{-1}$] $\sim$ 25.2 -- 25.3 
are typically assumed in studies of the reionisation of the Universe.

Several analyses have indicated that $\xi_{\rm ion}$ in high-$z$ SFGs is similar to
or above the canonical value. \citet{Bo15,Bo16} found 
log [$\xi_{\rm ion}$/Hz erg$^{-1}$]  
$\sim$ 25.3 for $z$ = 4 -- 5 galaxies and 25.5-25.8 for galaxies 
with the largest $\beta$. \citet{Na18,Na20} and \citet{Fa19} found an 
increase of $\xi_{\rm ion}$ for fainter objects at $z$ $\sim$ 3 -- 6.
\citet{Fi19} explored scenarios for reionising the intergalactic medium with 
low galactic ionising photon escape fractions. They found that the ionising
emissivity from the faintest galaxies ($M_{\rm FUV}$ $>$ --15 mag) is expected to be
dominant. The ionising emissivity from this model is consistent with 
observations at $z$ = 4 -- 5.

An extreme [O~{\sc iii}] $\lambda$5007 line emission associated with a high 
O$_{32}$ ratio are common features of the early-lifetime phase for SFGs 
at $z$ $>$ 2.5 \citep{Co18,Ho16,Ta19,Re18}. A high fraction of SFGs at $z$ $>$ 2
with extreme O$_{32}$ are LAEs \citep{Er16}. It was found that 
[O~{\sc iii}]5007/H$\beta$ emission-line ratios increase with redshift
out to $z$ $\sim$ 6 \citep{Cu16,Fa16}, attaining values similar to the ratios in
local SFGs with high EW(H$\alpha$). This increase results in high 
EW([O~{\sc iii}] 5007) $>$ 1000$\AA$\ for bright $z$~$>$~7 galaxies \citep{RB16}.

The properties of the high-$z$ SFGs discussed above contrast with those 
of the $z$ $\sim$ 0 main-sequence SFGs. The latter exhibit, on 
average, a much lower level of star formation activity and, thus, they are characterised by much 
lower SFRs, sSFRs, and EWs of emission lines at a fixed stellar mass. However, there
is a small fraction of low-$z$ SFGs with compact structure and other properties 
that are similar to those of high-$z$ galaxies. Subsets of these compact 
star-forming galaxies (CSFGs) are variously called 
blue compact dwarf (BCD) galaxies, `green pea' (GP) galaxies, and luminous 
compact galaxies (LCGs) in the literature.

BCDs are nearby dwarf galaxies at redshift $z$ typically 
$\leq$ 0.01 -- 0.02, with strong bursts of star formation and low oxygen
abundances of 12 + logO/H $\la$ 7.9, extending down to 12 + logO/H $\la$ 7.0
\citep[e.g. ][]{TM81,ITL94,I18c}. Typically, their stellar masses are low, 
$M_\star$ $\la$ 10$^8$M$_\odot$. A subset of the lowest-mass BCDs with
$M_\star$ $\sim$ 10$^6$ -- 10$^7$ M$_\odot$ and very high O$_{32}$ $\ga$ 10
has been named `blueberry' galaxies by \citet{Ya17} because of their compact 
structure and intense blue colour on the SDSS composite images.

\citet{Ca09} considered `green pea' (GP) galaxies
in the redshift range of $z$ = 0.112 -- 0.360, with unusually strong 
[O~{\sc iii}]$\lambda$5007 emission lines, similar to high-$z$ UV-luminous 
galaxies, such as Lyman-break galaxies and LAEs, with sSFRs up to 
$\sim$ 10 Gyr$^{-1}$. The green colour of these 
galaxies on composite SDSS images is due to strong 
[O~{\sc iii}]$\lambda$5007 line emission (with EWs up to 1000$\AA$),  
redshifted into the SDSS $r$-band.

\citet{I11} studied a much larger  sample of 803 star-forming luminous 
compact galaxies (LCGs) in a larger redshift range $z$ = 0.02 -- 0.63, with 
global properties similar to those of the GPs. The sSFRs in LCGs are extremely 
high and vary in the range $\sim$ 1 -- 100 Gyr$^{-1}$, comparable to those 
derived in high-$z$ SFGs. Similar high sSFRs were also found for low-$z$ LyC leaking
galaxies \citep{I16a,I16b,I18a,I18b,Sc16} and some other low-$z$ SFGs
\citep[e.g. ][]{Se17,Yu19}. \citet{Iz16c} found sSFRs of up to 
1000 Gyr$^{-1}$ in a sample of $\sim$ 13000 CSFGs selected from the SDSS.
Compared to BCDs, both GPs and LCGs are characterised by higher stellar masses 
$\sim$ 10$^9$ M$_\odot$ and higher oxygen abundances $\sim$ 7.8 -- 8.3.

\begin{figure*}[t]
\hbox{
\hspace{0.0cm}\includegraphics[angle=-90,width=0.49\linewidth]{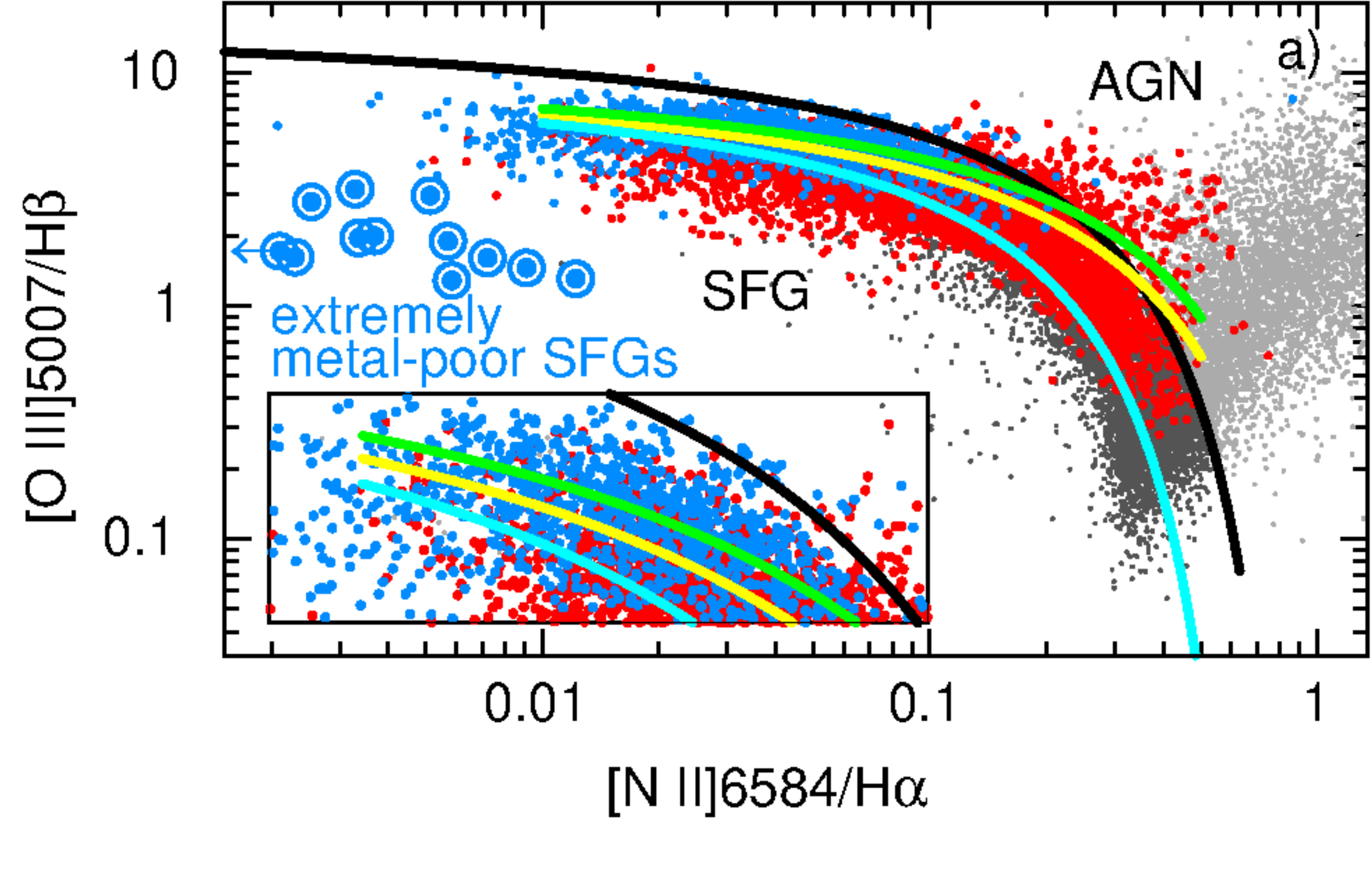}
\hspace{0.0cm}\includegraphics[angle=-90,width=0.49\linewidth]{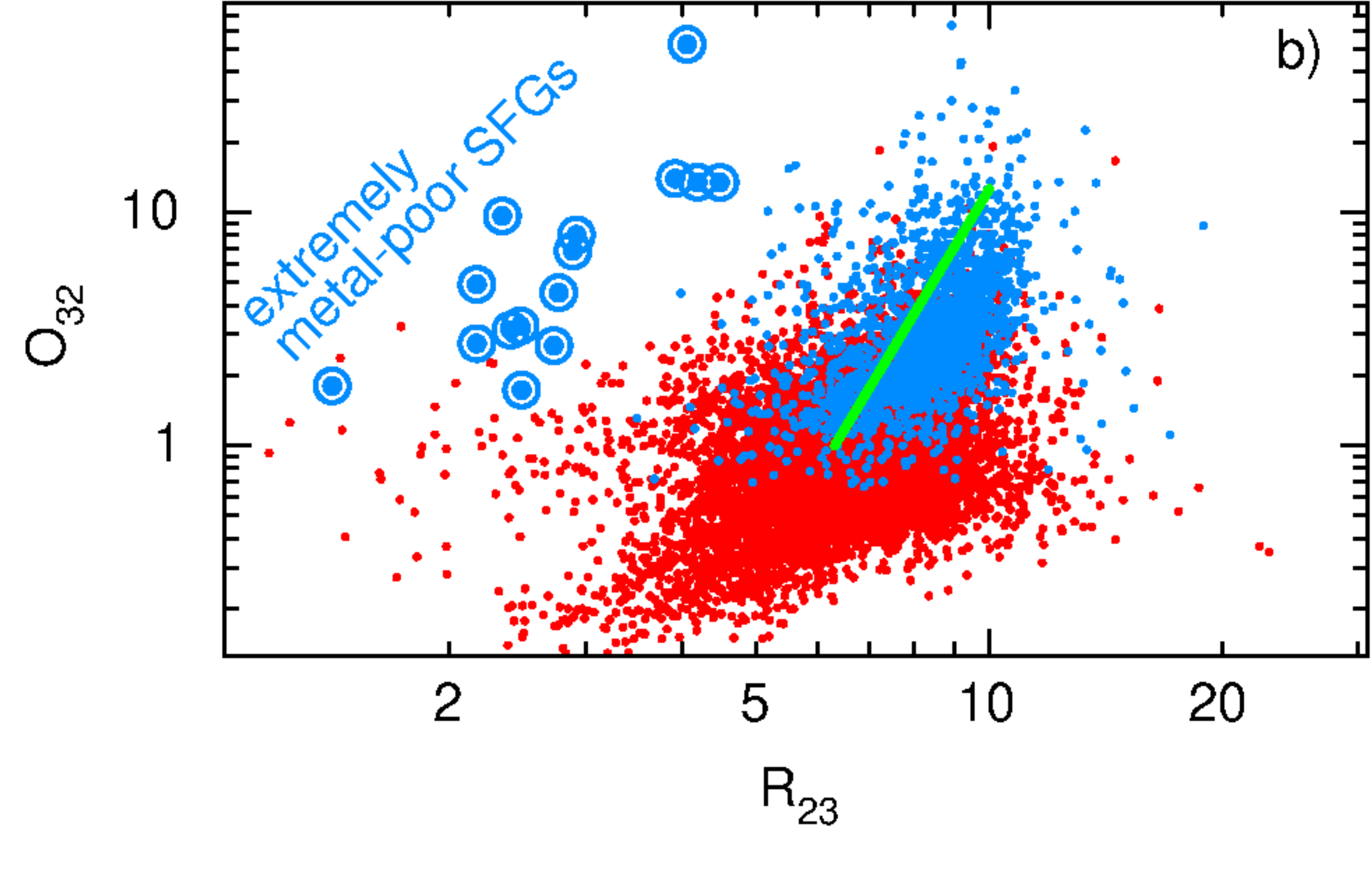}
}
\hbox{
\hspace{0.0cm}\includegraphics[angle=-90,width=0.49\linewidth]{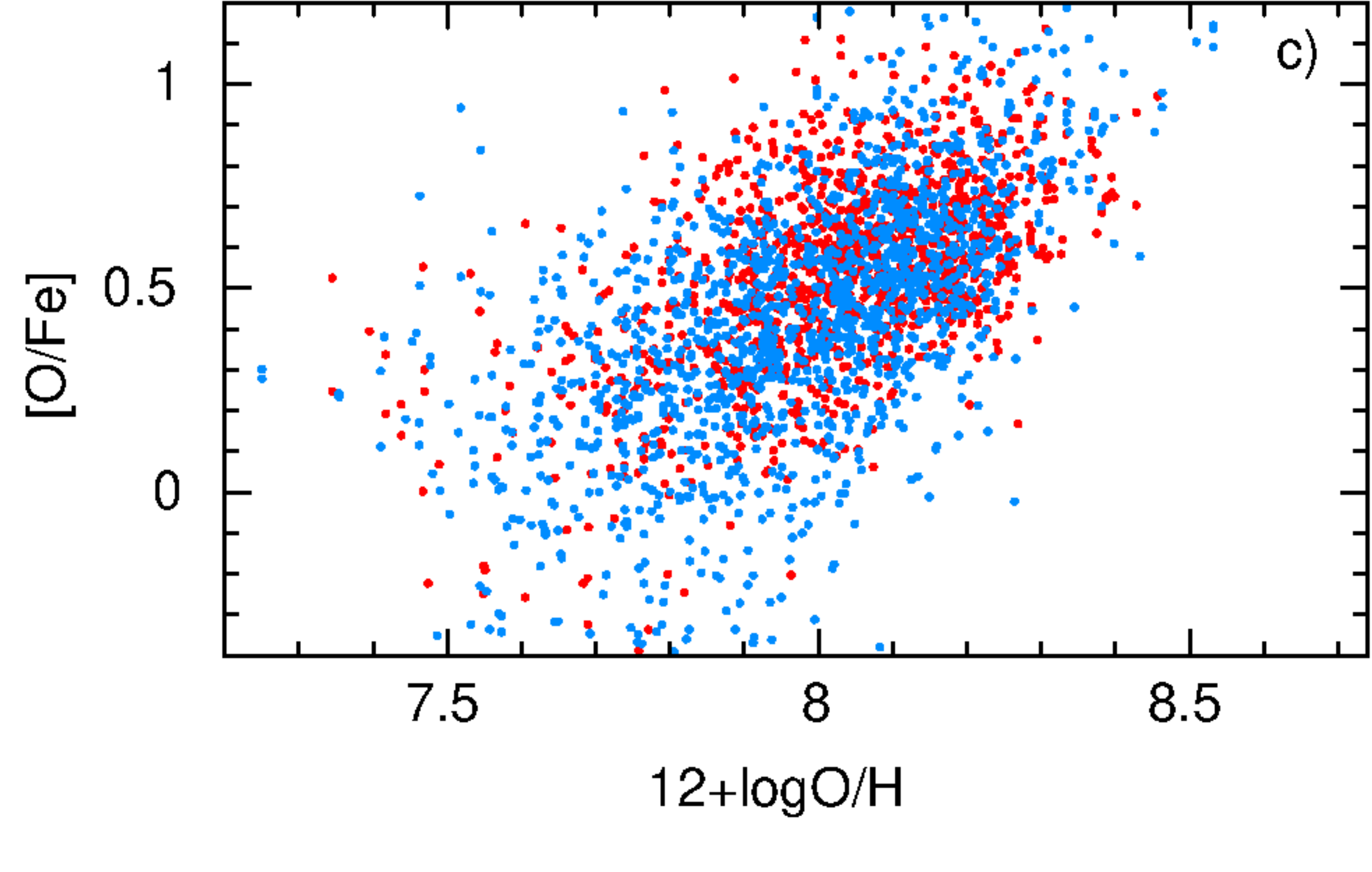}
\hspace{0.0cm}\includegraphics[angle=-90,width=0.49\linewidth]{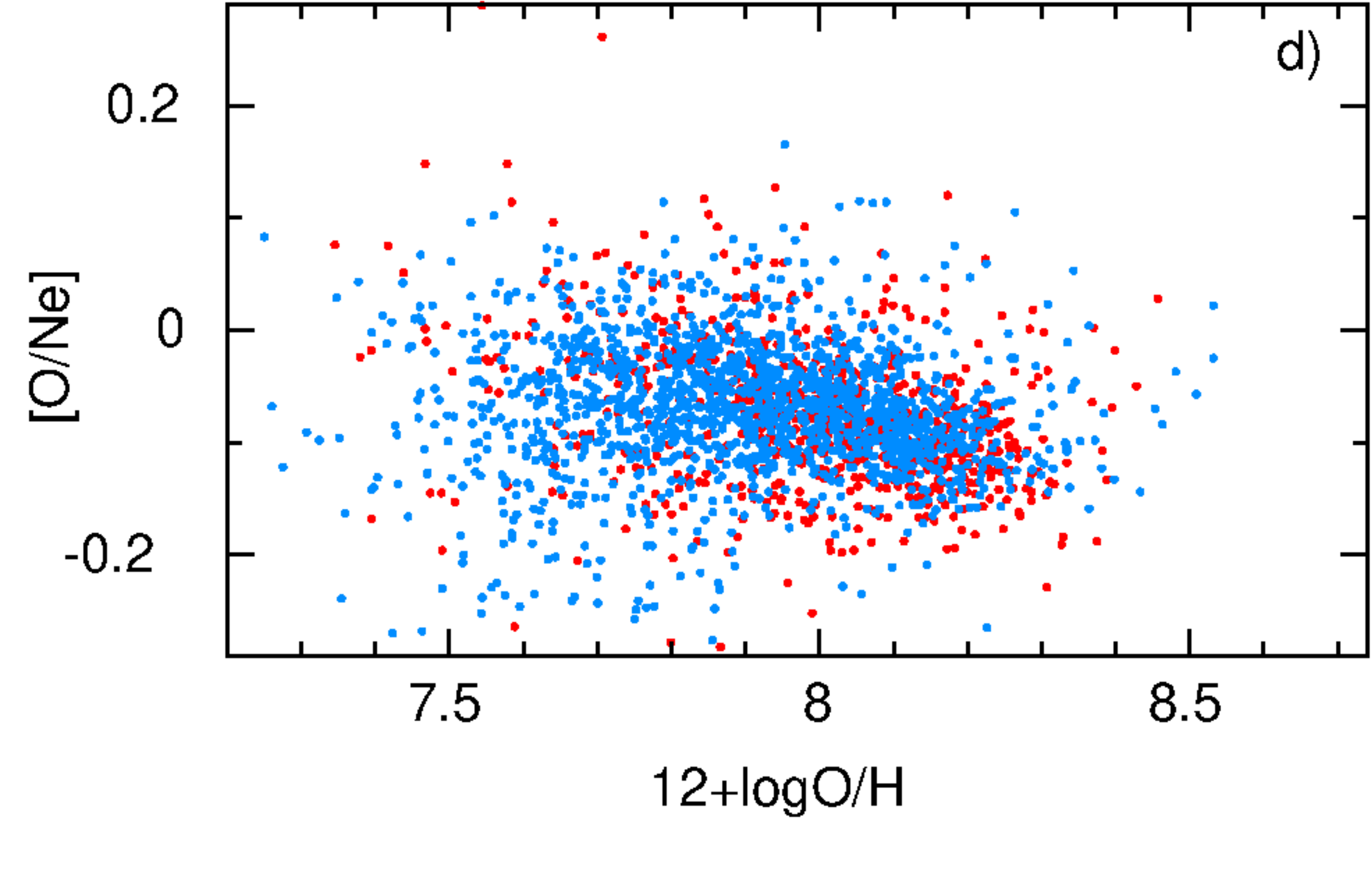}
}
\caption{{\bf (a)} [O~{\sc iii}]~$\lambda$5007/H$\beta$ vs.
[N~{\sc ii}]~$\lambda$6584/H$\alpha$ BPT diagram \citep{BPT81} for SDSS DR16
CSFGs with EW(H$\beta$) $\ge$ 100$\AA$\ (blue dots) and with 
EW(H$\beta$) $<$ 100$\AA$\ (red dots). The SDSS DR7 galaxies
including both SFGs (without constraints used for selection of CSFGs) and
AGN are represented by dark grey dots and light grey dots, respectively. 
The black line \citep{K03}
separates SFGs from AGN, whereas the cyan line by \citet{Ke13} represents the 
best fit relation for the total sample of $z$ $\sim$ 0 SFGs from the SDSS DR7.
Yellow and green lines indicate 
relations for $z$~$\sim$~2~--~3 SFGs by \citet{Sh15} and \citet{St17},
respectively. An expanded version of the upper part of the
diagram in the ranges [N~{\sc ii}]$\lambda$6584/H$\alpha$ of 0.006 - 0.16
and [O~{\sc iii}]~$\lambda$5007/H$\beta$ of 4 - 8 is shown in the inset. 
{\bf (b)} Relation O$_{32}$ -- R$_{23}$,
where O$_{32}$ = [O~{\sc iii}]~$\lambda$5007/[O~{\sc ii}]~$\lambda$3727,
R$_{23}$ = ([O~{\sc iii}]~$\lambda$4959~+~[O~{\sc iii}]~$\lambda$5007~+~[O~{\sc ii}]~$\lambda$3727)/H$\beta$. The green line represents the relation for 
$z$ $\sim$ 2 -- 3 SFGs by \citet{St17}.
The most metal-poor nearby galaxies with 12 + log O/H $\sim$ 6.9 -- 7.25 
\citep[][Izotov et al., in preparation]{I18c,Ko20} are shown 
in {\bf (a)} and {\bf (b)} by encircled blue-filled circles.  
{\bf (c)} Dependence of the oxygen overabundance 
[O/Fe] $\equiv$ log(O/Fe) -- log(O/Fe)$_\odot$ on the oxygen abundance 
12 + logO/H. {\bf (d)} Dependence of [O/Ne] $\equiv$ 
log(O/Ne) -- log(O/Ne)$_\odot$ on the oxygen abundance 12 + logO/H. 
Abundances are derived using the $T_{\rm e}$ method. 
About $\sim$~2300 galaxies with detected [O~{\sc iii}]$\lambda$4363 
emission and an error less than 25\% of the line flux are shown in
{\bf (c)} and {\bf (d)}. The meanings of the
symbols for SDSS galaxies in {\bf (b)} -- {\bf (d)} are the same as in 
{\bf (a)}.}
\label{fig1}
\end{figure*}

\citet{Iz15} selected $\sim$ 5200 low-redshift (0~$<$~$z$~$<$ 1) CSFGs and studied the relations between the global characteristics of this sample, including 
absolute optical magnitudes, SFRs, stellar masses, and oxygen abundances.
They found that for all relations, low-$z$ and high-$z$ SFGs are closely related, 
indicating a very weak dependence of metallicity on stellar mass, redshift, 
and star formation rate. This finding favours the assumption of a universal 
character for the global relations with regard to CSFGs with high-excitation H~{\sc ii}
regions over the redshift range 0 $<$ $z$ $<$ 3. 

Furthermore, \citet{I16b,I18a,I18b,I20}, in studying high-resolution UV images of
LyC leakers and $z$ $<$ 0.1 CSFGs, found that these objects show a 
disc-like structure with exponential scale lengths in the range 
$\sim$ 0.1 -- 1.4 kpc, similarly to high-$z$ galaxies. 

Thus,   evidence exists to support the claim that CSFGs are likely good proxies of high-$z$ SFGs.
Their proximity, compared to the more distant high-$z$ galaxies, allows us to 
study them in considerably more detail across a wide range of wavelengths, from the UV to the radio, and to compare their 
global properties with those of the galaxies in the early Universe. The sample
of CSFGs extends to much lower galactic stellar masses, 
$\sim$ 10$^5$ -- 10$^6$ M$_\odot$, a range which is not yet accessible at high 
$z$. Their study is 
important for comparison with future observations of low-mass 
high-$z$ SFGs with the James Webb Space Telescope (JWST). In this work,
we extend the previous analysis of CSFG properties by using a larger sample and by 
including many more global parameters that can be compared with parameters of
the rapidly growing observations of high-$z$ SFGs.

The selection criteria and the sample are described in Section~\ref{selection}. 
Methods for the determination of various global parameters are presented 
in Section~\ref{global}. 
For SFR and $M_\star$ determinations, we assume a Salpeter initial mass function (IMF) from 0.1 to 100 
M$_\odot$, and all results from the literature have been rescaled to this IMF 
for comparison.
Relations between global parameters and their comparison with those for 
high-$z$ galaxies are discussed in Section~\ref{results}.
Our main conclusions are summarised in Section~\ref{summary}.

\begin{figure*}[t]
\centering
\hspace{0.0cm}\includegraphics[angle=-90,width=1.02\linewidth]{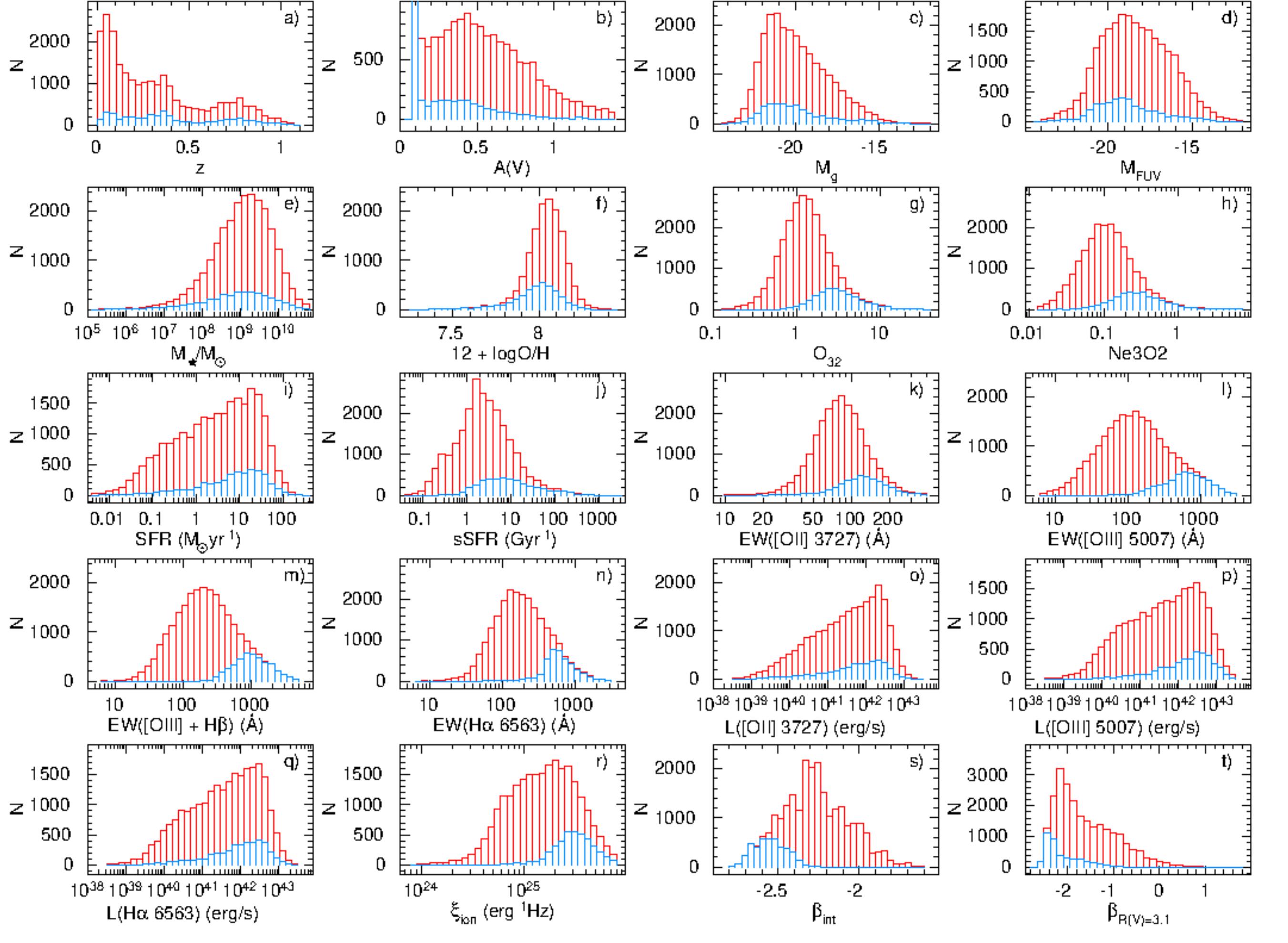}
\caption{Histograms of the {\bf (a)} redshift $z$, {\bf (b)} extinction $A(V)$
in the $V$ band derived from the hydrogen Balmer decrement,
{\bf (c)} absolute $g$-band magnitude $M_g$ corrected 
for the Milky Way extinction, {\bf (d)} `observed' absolute FUV magnitude $M_{\rm FUV}$, i.e. the absolute magnitude  
derived from the extrapolation of the extinction-corrected SDSS spectrum and attenuated adopting extinction $A(V)$ obtained from 
the Balmer hydrogen decrement, {\bf (e)} stellar mass derived from the
SED fitting of the extinction-corrected SDSS spectrum, {\bf (f)} oxygen
abundance 12 + logO/H, 
{\bf (g)} [O~{\sc iii}]~$\lambda$5007/[O~{\sc ii}]~$\lambda$3727 ratio denoted as
O$_{32}$, {\bf (h)} [Ne~{\sc iii}]~$\lambda$3868/[O~{\sc ii}]~$\lambda$3727 ratio 
denoted as Ne3O2, {\bf (i)} star-formation rate SFR derived from the 
extinction-corrected H$\beta$ luminosity, {\bf (j)} specific star-formation rate 
sSFR = SFR/($M_\star$/M$_\odot$),
{\bf (k) - (n)} equivalent widths of the [O~{\sc ii}]~$\lambda$3727,
[O~{\sc iii}]~$\lambda$5007, H$\beta$ + [O~{\sc iii}]~$\lambda$4959 + 
[O~{\sc iii}]~$\lambda$5007 and H$\alpha$ emission lines, respectively,
{\bf (o) - (q)} extinction-corrected luminosities of the 
[O~{\sc ii}]~$\lambda$3727, [O~{\sc iii}]~$\lambda$5007 and 
H$\alpha$~$\lambda$6563 emission lines, respectively,
{\bf (r)} ionising photon production efficiency derived
from the extinction-corrected H$\beta$ luminosity and extinction-corrected
monochromatic luminosity at the rest-frame wavelength of 1500$\AA$, 
{\bf (s)} extinction-corrected UV slope derived from the modelled rest-frame 
SED, {\bf (t)} UV slope derived 
from the obscured SEDs with the extinction coefficient derived from the observed
Balmer decrement adopting the \citet{C89} reddening law with $R(V)$ = 3.1.
In all panels, histograms represented 
by the red and blue lines are for the total SDSS sample 
of CSFGs and for CSFGs with EW(H$\beta$) $\ge$ 100$\AA$, respectively.}
\label{fig2}
\end{figure*}

\begin{figure*}[t]
\centering
\hspace{0.0cm}\includegraphics[angle=-90,width=0.49\linewidth]{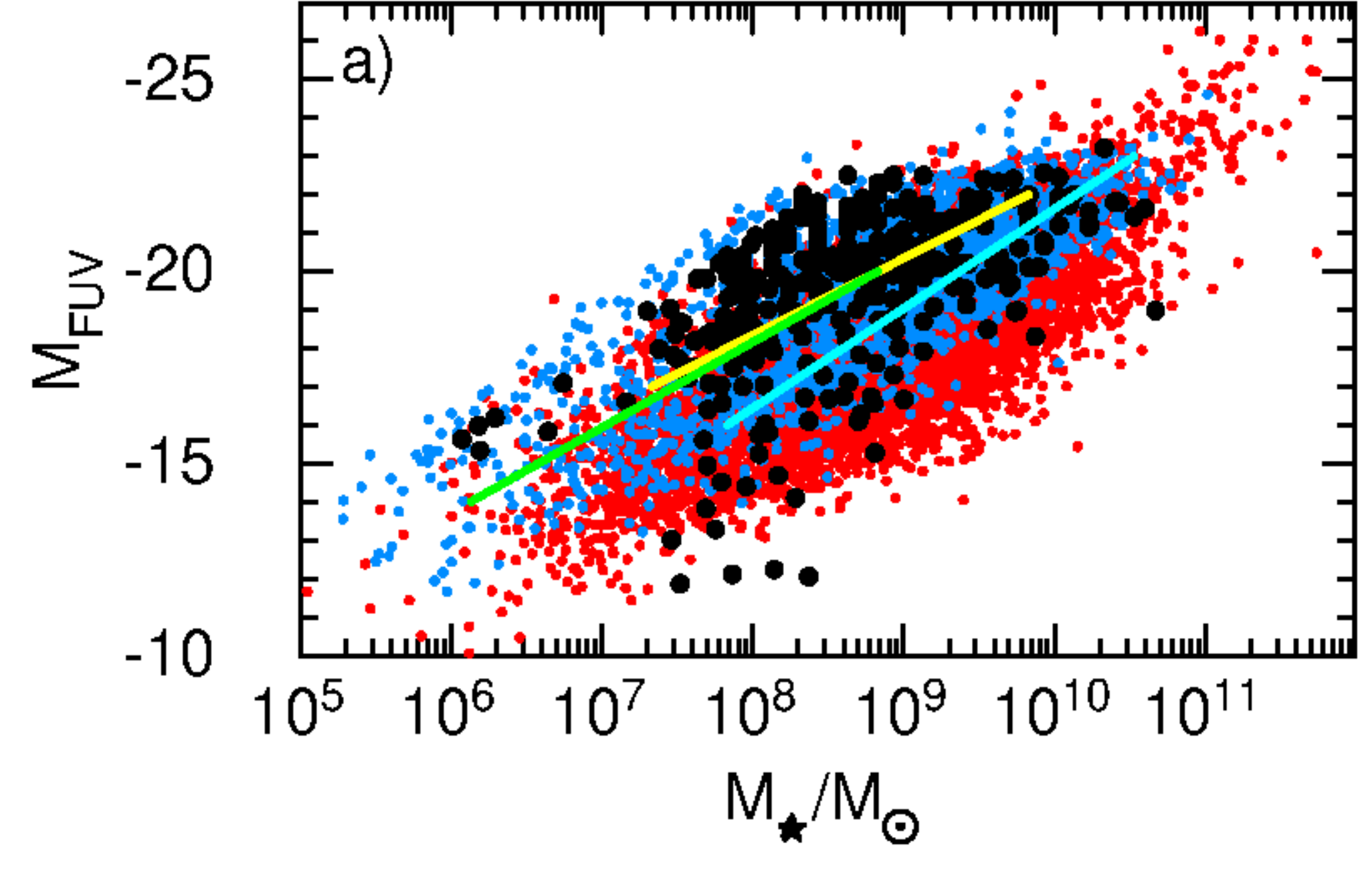}
\hspace{0.0cm}\includegraphics[angle=-90,width=0.49\linewidth]{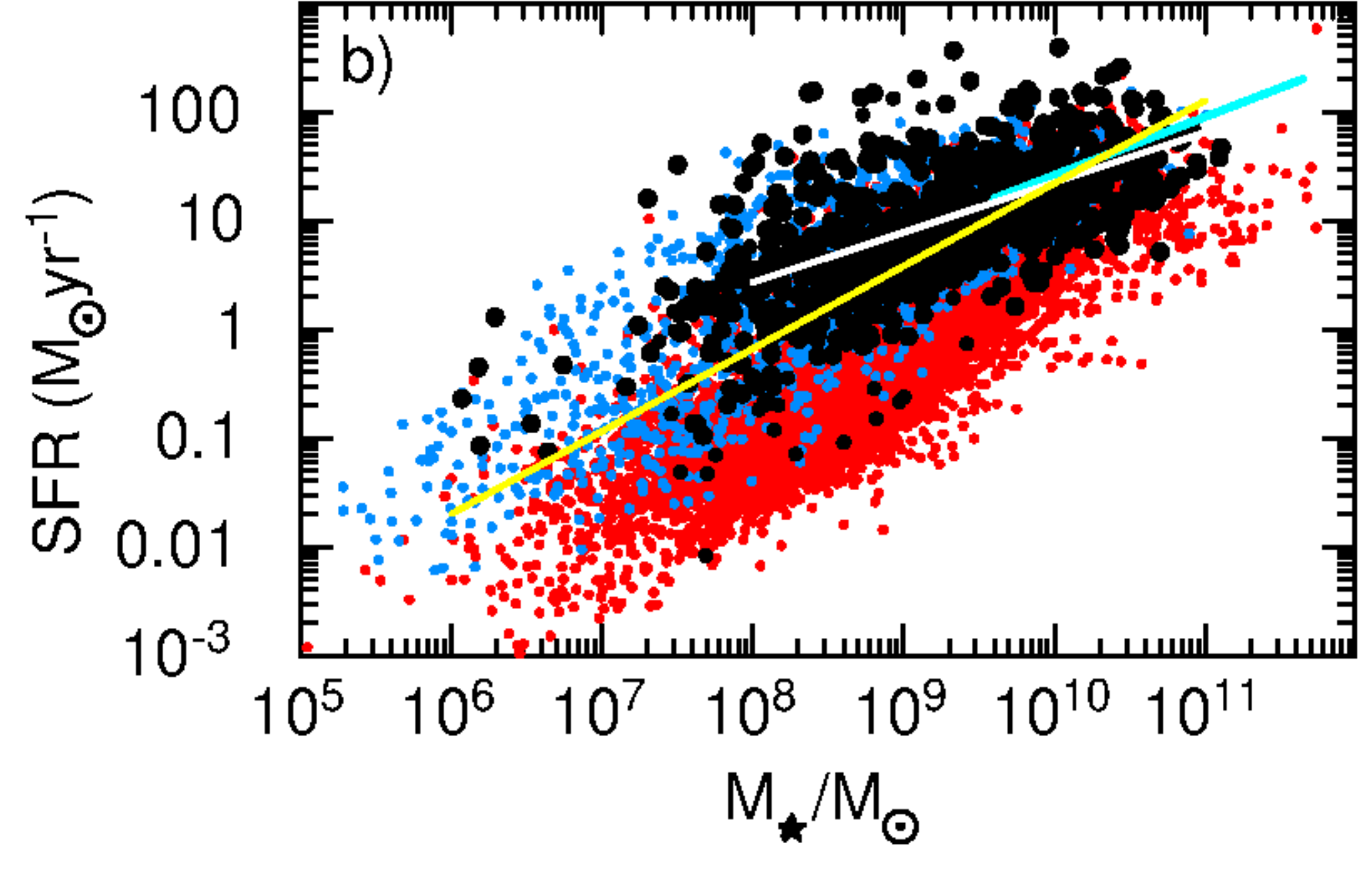}
\hspace{0.0cm}\includegraphics[angle=-90,width=0.49\linewidth]{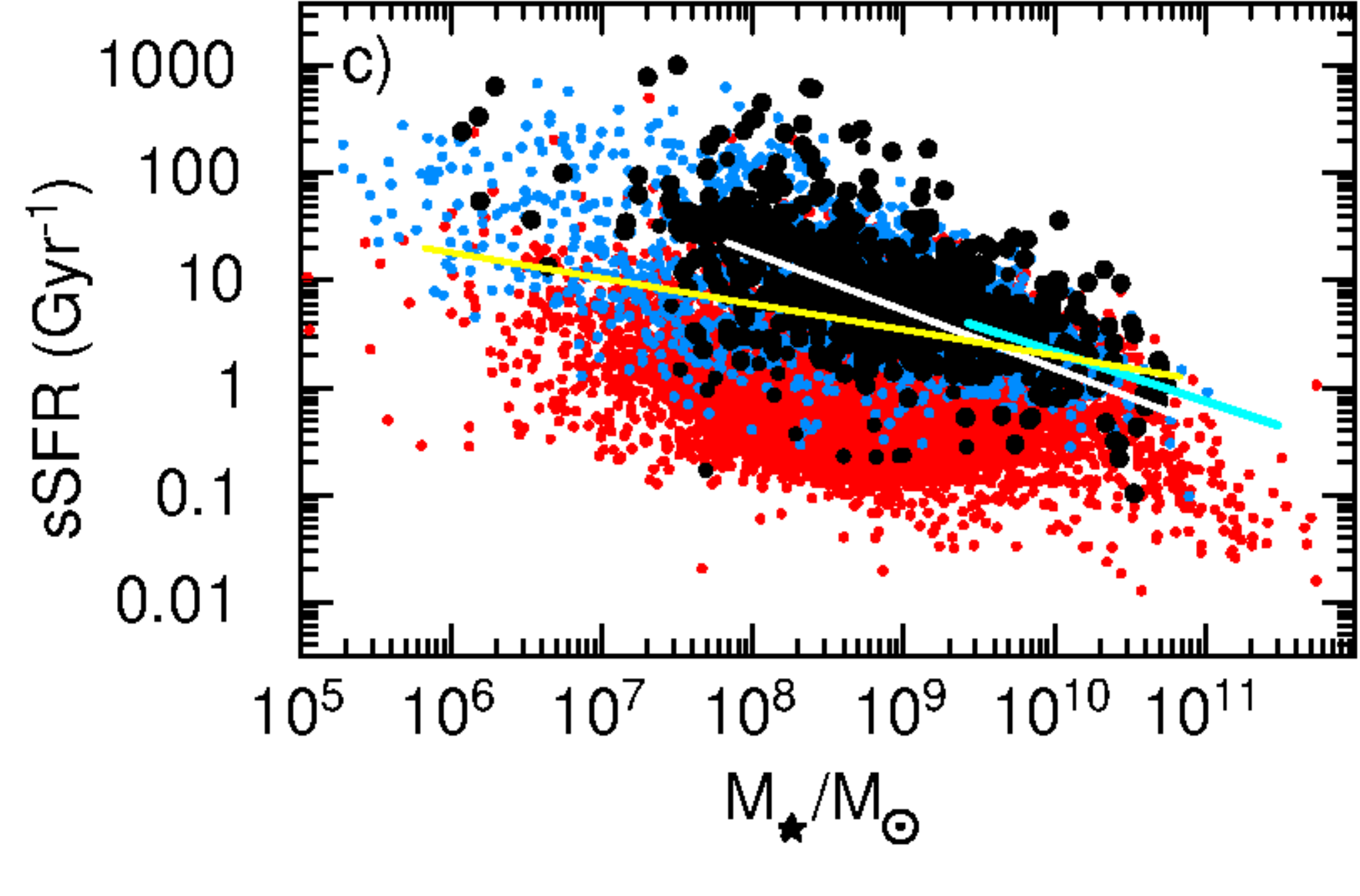}
\hspace{0.0cm}\includegraphics[angle=-90,width=0.49\linewidth]{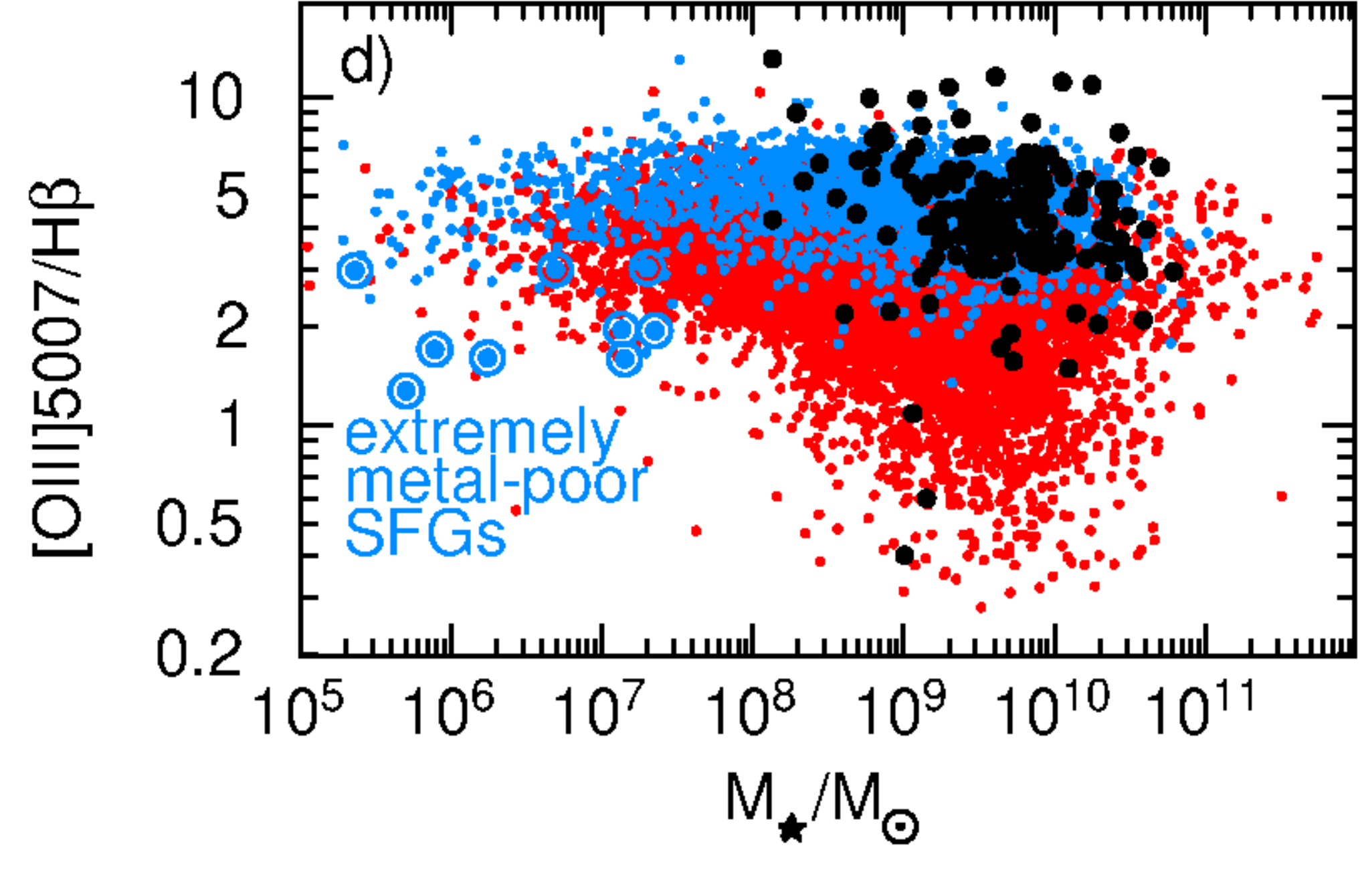}
\caption{{\bf (a)} Dependence of the FUV absolute magnitude on the stellar
mass $M_\star$. Black-filled circles are high-$z$
galaxies \citep{Br19,En21,Ka17,Ra16,Kh20,Ta20,St20b,San20}. Relations for
high-$z$ galaxies by \citet{So16}, \citet{Gr15} and \citet{Yun19} are 
represented by yellow, cyan and green lines, respectively.
{\bf (b),}  {\bf (c)} SFR versus the stellar mass and 
the sSFR versus the stellar mass, respectively.
Black-filled circles in both panels are high-$z$ galaxies 
\citep{Ka17,A16,Sa20,Er16,Ho16,Kh20,Ra16,San20,Tr14,St20,Re06,Ha16,On16,Sa17,Cu14,Sax20,En21,En20,Ta20,St20b,Jo20}. White solid lines, yellow solid lines and cyan solid lines in
{\bf (b)} and {\bf (c)} represent relations for high-$z$ galaxies by \citet{A20}, \citet{Iy18} and \citet{Fl20}, respectively.
 {\bf (d)} The dependence of the [O~{\sc iii}]~$\lambda$5007/H$\beta$ flux ratio
on the stellar mass $M_\star$. Black filled citcles represent high-$z$ galaxies 
\citep{Ri11,Sc13,St14,Tr14,Ho16}.
The most metal-poor nearby galaxies with 12 + log O/H $\sim$ 6.9 -- 7.25 
\citep[][Izotov et al., in preparation]{I18c,Ko20} are shown by encircled blue-filled circles. The meanings of other symbols in all panels 
are the same as in Fig.~\ref{fig1}.}
\label{fig3}
\end{figure*}

\section {Selection criteria and the sample} \label{selection}

Our sample of CSFGs has been extracted from the spectroscopic database of the
SDSS DR16 \citep{Ah20}. The selection criteria are described, for example, in 
\citet{Iz16c}.  
These criteria are as follows: (1) the angular galaxy radius on the SDSS images is
$R_{50}$ $\leq$ 3 arcsec, where $R_{50}$ is the galaxy's Petrosian radius
within which 50 per cent of the galaxy's flux in the SDSS $r$-band is contained.
With this criterion, the fraction of the light outside the 3 arcsec diameter SDSS 
spectroscopic aperture is small. This allows us to compare SDSS spectroscopic 
data with photometric data in the UV, optical and infrared ranges without 
large aperture corrections;
(2) the equivalent width of the H$\beta$ emission line is $\ga$~10$\AA$.
This criterion selects galaxies with the most extreme starbursts;
(3) galaxies with AGN activity (the presence of broad 
Mg~{\sc ii}~$\lambda$2800 and hydrogen emission lines, 
a strong high-ionisation emission line He~{\sc ii} $\lambda$4686 with an intensity of 
$\ga$ 10\% that of H$\beta$ and the presence of high-ionisation 
[Ne~{\sc v}] $\lambda$3426 emission line) 
were excluded. Applying these 
criteria, $\sim$ 25000 CSFGs at redshifts $z$ $\la$ 1 were selected.

  The emission-line fluxes, equivalent widths and errors were measured for 
each SDSS-selected spectrum, using the 
{\sc iraf}\footnote{{\sc iraf} is the Image 
Reduction and Analysis Facility distributed by the National Optical Astronomy 
Observatory, which is operated by the Association of Universities for Research 
in Astronomy (AURA) under cooperative agreement with the National Science 
Foundation (NSF).} {\it splot} routine. They were corrected for extinction from 
the observed decrement of all Balmer hydrogen emission lines
in the SDSS spectra usable for measurements. The range of derived extinctions 
in our sample galaxies is discussed in Section~\ref{histo}.
The procedure consists of two steps. Firstly, the emission line fluxes are
corrected for the Milky Way (MW) reddening, adopting the extinction $A(V)$ from the 
NASA Extragalactic Database (NED) and the reddening law of \citet{C89} with 
$R(V)$ = $A(V)$ / $E(B-V)$ = 3.1.
This correction is applied adopting the observed wavelengths of the emission lines.
Secondly, the internal extinction coefficients $C$(H$\beta$)
 were iteratively derived,  following the prescriptions of \citet{ITL94},
 from the Balmer decrement 
of hydrogen emission lines, together with the equivalent widths of stellar hydrogen
absorption lines, underlying the emission-line spectra. The rest-frame 
wavelengths of hydrogen lines and \citet{C89} reddening law were also used.
For clarity, we adopted $R(V)$ = 3.1, although \citet{Iz15} found that
$R(V)$ = 2.7 is more likely in the far-UV range for CSFGs.

In general, the internal extinction coefficients $C$(H$\beta$) 
for our CSFGs are small.
Therefore, the choice of the $R(V)$ value, defining the 
slope of the reddening law, is not important for the extinction
correction in the optical range, giving small differences in the continuum and
emission line fluxes for different $R(V)$s. However, it is important in the UV range.
The same extinction-corrected SDSS spectra were used for the
determination of some integrated characteristics, including absolute magnitudes,
SFRs, $M_\star$s and some other characteristics (see Sect.~\ref{histo}).

\section{Global characteristics of CSFGs} \label{global}

Here, we consider various photometric and spectroscopic charactersistics of CSFGs. 
As mentioned before, they are global quantities as the aperture correction is small.
The characteristics directly derived from the SDSS photometric and 
spectroscopic data are: redshifts, luminosities, and equivalent widths of emission
lines, reddening, element abundances, absolute SDSS $g$ magnitudes, and star 
formation rates from hydrogen emission-line luminosities.

On the other hand, fitting the spectral energy distribution with specified
star-formation histories is needed for the determination of some other 
characteristics, such as stellar mass, specific star-formation rates, 
monochromatic absolute magnitudes and luminosities in the UV range, the spectral
slopes in the UV range, and production efficiencies of ionising radiation
$\xi_{\rm ion}$. A description of the SED modelling is given in Sect.~\ref{histo}.

\subsection{Diagnostic diagrams} \label{diag}

The BPT diagnostic diagram 
[O~{\sc iii}]$\lambda$5007/H$\beta$ -- [N~{\sc ii}]$\lambda$6584/H$\alpha$
 \citep{BPT81} for $\sim$16500 CSFGs with a measured 
[N~{\sc ii}]$\lambda$6584
emission line is shown in Fig.~\ref{fig1}a. For the remaining $\sim$ 8500 
CSFGs,
that emission line is either not detected or has a redshifted wavelength outside
the SDSS spectral range. It is shown in Fig.~\ref{fig1}a that
most of the CSFGs are located in the region of star-forming galaxies, below
the black solid line of \citet{K03}, which approximately separates SFGs and AGN.

A number of authors have found \citep[e.g. ][]{St14,Sh15,Cu16,St17,Bi20} 
that the distribution of high-$z$ SFGs in the BPT
diagram is offset from the locus of $z$ $\sim$ 0 SDSS
SFGs. This difference can be caused by a combination of a harder stellar
ionising radiation field, a higher ionisation parameter, and a higher N/O at a 
given O/H, compared to the 'plateau' value found for local low-metallicity galaxies.

Our sample of CSFGs is shown in the BPT diagram  (Fig.~\ref{fig1}a) by red dots 
(galaxies with EW(H$\beta$) $<$ 100$\AA$) and blue dots (galaxies with 
EW(H$\beta$) $\ge$ 
100$\AA$). Although this division is somewhat arbitrary, it allows us to 
study the properties of galaxies with different excitation conditions in their
H~{\sc ii} regions. Furthermore, high-redshift SFGs tend to have high
EW(H$\beta$) $\ga$ 100$\AA$. The galaxies from our sample with high 
EW(H$\beta$) occupy the upper part of the sequence for 
SFGs. We can see that high-EW(H$\beta$) CSFGs are offset to higher 
[O~{\sc iii}]$\lambda$5007/H$\beta$ and lower 
[N~{\sc ii}]$\lambda$6584/H$\alpha$, as compared to the total sample
of SDSS SFGs (grey dots). The distribution of the latter galaxies can be
approximated by the cyan line in Fig.~\ref{fig1}a \citep{Ke13}. 
We also show the relations derived by
\citet{Sh15} (yellow line) and by \citet{St17} (green line) for 
$z$ $\sim$ 2 -- 3 SFGs, which also fit quite well the distribution of the CSFGs
in our sample with the strongest emission lines (CSFGs with EW(H$\beta$) 
$\ge$ 100 $\AA$, blue dots). The agreement between the locations of 
high-EW(H$\beta$) CSFGs and $z$ $\sim$ 2 -- 3 SFGs is best seen in the inset
of Fig.~\ref{fig1}a. On the other hand, CSFGs with lower EW(H$\beta$) $<$ 
100$\AA$\ (red dots) are located somewhat below the relations for 
$z$ $\sim$ 2 -- 3 SFGs (yellow and green lines), but they are still offset from
the relation for the main population of $z$ $\sim$ 0 SFGs (cyan line). 
We conclude that the locus of CSFGs with EW(H$\beta$) $\geq$ 100$\AA$\ 
(blue dots) and of a considerable fraction of CSFGs with 
EW(H$\beta$) $<$ 100$\AA$\ (red dots) is similar to that of 
$z$ $\sim$ 2 -- 3 SFGs, implying a common cause for the observed
offset. One of the likely causes is the higher ionisation parameter of
CSFGs and of $z$~$\sim$~2~--~3 SFGs, compared to that of galaxies in the main 
sequence $z$~$\sim$~0 sample.

We note, however, that the location of SFGs in the BPT diagram should also
depend on metallicity. To illustrate this effect, we show in Fig.~\ref{fig1}a, using
 encircled blue-filled circles, the sample of the most metal-deficient nearby galaxies known,
with 12 + log O/H $\sim$ 6.9 -- 7.25, from \citet{I18c} and \citet{Ko20}. No
galaxy with such low metallicity has been reported thus far at high redshifts. These galaxies are mostly
compact and are characterised by EW(H$\beta$) $\ga$ 100$\AA$ and, thus, by
high ionisation parameters, which increase from the right to the left 
of the BPT diagram. The most metal-deficient galaxies strongly deviate from the CSFG sequence
and high-$z$ galaxies with typical 12 + log O/H $\sim$ 8.0, towards lower 
[O~{\sc iii}]$\lambda$5007/H$\beta$ ratios. It is clear that they follow a 
sequence that is different from those of CSFGs and $z$ $\sim$ 2 SFGs. 
It is likely that this sequence corresponds to galaxies at high redshifts that are least 
enriched with heavy elements. Therefore, the dependence on metallicity (O/H) 
should be taken into account in the analysis of the BPT diagram, even if N/O 
remains constant.

The distribution of $\sim$ 25000 CSFGs in the diagram O$_{32}$ = 
[O~{\sc iii}]$\lambda$5007 / [O~{\sc ii}]$\lambda$3727 versus
R$_{23}$ = ([O~{\sc iii}]$\lambda$4959 + [O~{\sc iii}]$\lambda$5007 + [O~{\sc ii}]$\lambda$3727) / H$\beta$ 
is shown in Fig.~\ref{fig1}b. 
The H~{\sc ii} regions in the galaxies from our sample are characterised by a wide
range of ionisation parameter as evidenced by a large variation of O$_{32}$.
Most of the low-EW(H$\beta$) CSFGs have low O$_{32}$ $\la$ 1, whereas CSFGs
with EW(H$\beta$) $\geq$ 100~$\AA,$\ have O$_{32}$ $\ga$ 1. There is a small 
number of CSFGs with O$_{32}$ $\ga$ 10, reaching in some cases extremely high 
values up to 60. For comparison, the relation for $z$ $\sim$ 2 - 3 SFGs by 
\citet{St17} is also shown with a green line. It shows good 
agreement with the distribution of high-EW(H$\beta$) CSFGs (blue dots) and with 
that of a large fraction of low-EW(H$\beta$) CSFGs (red dots).
This agreement implies similar properties of CSFGs and high-$z$ SFGs.

On the other hand, the most metal-deficient nearby galaxies with 
12 + logO/H $\sim$ 6.9 -- 7.25 and high O$_{32}$ $>$ 1 from 
\citet{I18c}, \citet{Ko20} and Izotov et al. (in preparation) 
(encircled blue-filled circles in Fig.~\ref{fig1}b) show a strong deviation 
from the sequences of most CSFGs and high-$z$ SFGs, indicating again a strong 
metallicity dependence.

\citet{Sa20} and \citet{To20b} considered constraints on the properties of 
massive stars and ionised gas for a sample of SFGs at $z$ $\sim$ 2.3.
Oxygen abundances in the \citet{Sa20} sample were derived by the direct
$T_{\rm e}$ method and range from 12 + log O/H $\sim$ 7.5 to 8.2,
whereas \citet{To20b} derived 12 + log O/H $\sim$ 8.1 -- 8.6 using the 
strong-line method. They concluded that high-$z$
SFGs differ from the local main-sequence galaxies by considerably harder
ionising spectra at fixed oxygen abundance, resulting in the offset in the BPT
diagram to higher [O~{\sc iii}]$\lambda$5007/H$\beta$ values. They also found
that stellar models with super-solar O/Fe ratios and binary evolution
of massive stars are required to reproduce the observed strong-line ratios
in high-$z$ SFGs. 

\subsection{Dependence of [O/Fe] on oxygen abundance}

In Fig.~\ref{fig1}c, we show the dependence of [O/Fe] on 12 + log O/H for the CSFGs 
in the SDSS. We note that iron abundances are derived directly from SDSS 
spectra, using the [Fe~{\sc iii}]$\lambda$4658 and 4986 emission lines. This is 
different from the \citet{Sa20} and \citet{To20b} data, where iron abundances 
are derived indirectly from stellar models. The fact that our oxygen and iron 
abundances are derived directly from observations also means that our 
[O/Fe] values may be subject to the depletion of iron and oxygen onto dust grains.

We confirm the steady increase of [O/Fe] with increasing 12~+~log~O/H
found previously, for instance, by \citet{I06} for a smaller sample of nearby SFGs.
Our values of [O/Fe] are in the range $\sim$ 0.4 -- 1.0 for a range of
oxygen abundances 12~+~log~O/H $\sim$ 8.1 -- 8.6, in full agreement with
the values of [O/Fe] for $z$~$\sim$~2.3 SFGs in the same range of oxygen
abundances. We also note that there is no difference in distributions of CSFGs
with low and high EW(H$\beta$). 

\citet{Sa20} found a similarly high, 
3$\sigma,$ lower limit for [O/Fe] $\sim$ 0.5, but in $z$ $\sim$ 2 SFGs with lower 
metallicities, in apparent disagreement with our CSFGs, which have
[O/Fe]~$<$~0.5 at the same lower metallicities. 
However, their conclusion is based on a very small
sample consisting of only four galaxies. Two of these galaxies show oxygen 
overabundance with respect to the iron abundance, 
while the other two do not show this effect. Furthermore, the [O/Fe]
values derived by \citet{Sa20} and \citet{To20b} depend on assumptions related to the 
star formation history and initial mass function for stars, and sets of stellar
evolution models, whereas both the O and Fe abundances of our CSFGs are derived 
directly from observations.

The increase of [O/Fe] with 12 + log O/H likely contradicts the
idea that the high [O/Fe] $\ga$ 0.5 at the highest metallicities 
(Fig.~\ref{fig1}c) are due to harder radiation. In fact, we expect the ionising radiation to become harder at lower, not at higher 
metallicities. However, no high
overabundance of oxygen relative to iron ([O/Fe] $\ga$ 0.3) is observed in 
CSFGs with 12 + log O/H $\la$ 7.8. 

\citet{I06} have proposed an alternative explanation. They have suggested that
the trend seen in Fig.~\ref{fig1}c is not due to an oxygen overabundance, but
to iron and oxygen depletion onto dust grains. That depletion is more
severe at higher metallicities due to the larger content of dust. The effect is
stronger for iron because it is $\sim$ 20 times less abundant than oxygen. 
The dust depletion hypothesis is in line with a 
slight increase of the noble gas neon to oxygen abundance ratio or a decrease of
[O/Ne] with increasing oxygen abundance, which has been found by \citet{I06}
and is shown for the CSFG sample in Fig.~\ref{fig1}d. Therefore, it is important
to increase the sample of $z$ $\sim$ 2.3 SFGs by \citet{To20b} and to extend it
to lower oxygen abundances to check both hypotheses. 

Independently of explaining the high [O/Fe] values,
 we emphasise again that the distributions of
high-redshift SFGs and local CSFGs in Fig.~\ref{fig1}a - \ref{fig1}b are quite similar, 
implying similar physical properties for the galaxies in these samples.

\subsection{Physical properties of CSFGs} \label{histo}

To represent the distributions of physical properties of the CSFGs, we show in 
Fig.~\ref{fig2} histograms of various physical parameters. 
The histograms in red and blue are for the entire sample and 
for CSFGs with EW(H$\beta$)~$\ge$~100$\AA$, respectively. 
The redshift distribution of the galaxies is shown in Fig.~\ref{fig2}a
with three maxima at $z$ = 0.05, 0.35 and 0.8. Most CSFGs are at 
$z$ $\la$ 0.5, but $\sim$ 20\% 
are at higher redshifts. Most of our CSFGs (57\%) are characterised by low 
extinction $A(V)$ $\la$ 0.1 mag. The observed H$\alpha$/H$\beta$ ratio
in a considerable number of galaxies is lower that the theoretical recombination
value, mainly in objects at low redshift with high intensities of emission 
lines. The most likely reason for that is the clipping of emission lines, which is
not rare in SDSS spectra. This effect is larger for the stronger H$\alpha$ and
[O~{\sc iii}] $\lambda$5007 emission lines. In those cases, we adopted 
$A(V)$ = 0 and the intensity of [O~{\sc iii}] $\lambda$5007 emission line to be equal to
three times the intensity of the [O~{\sc iii}] $\lambda$4959
 emission line. The remaining galaxies (43\%) are distributed 
over a wide range of $A(V)$ with the maximum at $\sim$ 0.45 mag 
(Fig.~\ref{fig2}b).

\subsubsection{Absolute magnitudes and H$\beta$ luminosity}

The extinction-corrected absolute $g$-band magnitudes $M_g$ and 
extinction-corrected H$\beta$ luminosities $L$(H$\beta$) were obtained 
respectively from the apparent SDSS $g$-band model magnitude for the entire 
galaxy, corrected for Milky Way extinction, and the H$\beta$ 
emission-line flux measured in the SDSS spectrum and corrected for both 
Milky Way and internal galaxy extinction.

In principle, UV characteristics such as luminosities can be 
derived from the FUV and NUV magnitudes of the {\sl GALEX} data base. However, CSFGs
are located at very different redshifts, in the range of $z$ $\sim$ 0 -- 1.
Therefore, a correction for redshift is needed to derive UV magnitudes and 
monochromatic luminosities at fixed rest-frame wavelengths. 

Instead, we decided to derive
UV characteristics from the UV SEDs extrapolated from the modelled SED to the 
rest-frame SDSS spectrum. For the sake of comparison with high-$z$ galaxies, 
these SEDs are attenuated by adopting the \citet{C89} reddening law with 
$R(V)$ = 3.1 and an internal extinction derived from the hydrogen Balmer 
decrement. Then the attenuated rest-frame SEDs are convolved with the {\sl GALEX} 
FUV transmission curve. The details and validity
of this approach have been described and justified previously, for instance, by \citet{Iz17}, who 
showed that the extrapolation of the optical SDSS SED to the UV and adopting 
$R(V)$ = 2.7 -- 3.1 reliably reproduces the 
observed {\sl GALEX} FUV and NUV apparent magnitudes of CSFGs. 
The average accuracy is better 
than 0.7 mag, translating to a $\sim$ 0.28 dex uncertainty in the UV luminosity
and the ionising photon production efficiency $\xi_{\rm ion}$. We also note that
the typical uncertainties of FUV and NUV magnitudes of CSFGs in the {\sl GALEX} 
database are $\sim$ 0.2 -- 0.4 mag. Furthermore, \citet{I16a,I16b,I18a,I18b}
have shown that our procedure for extrapolating the optical SED to the UV range 
reproduces  very well the observed {\sl HST} UV spectra of $z$ $\sim$ 0.3 -- 0.4 
LyC leakers, which constitute a subset of our CSFG sample.

CSFGs are generally bright in both the visible and the far-UV ranges,
with the absolute magnitude distributions peaking at about $-$21 mag in both 
wavelength ranges (Figs.~\ref{fig2}c,d), similarly to $M_{\rm FUV}$s of the 
$z$~$\sim$~2~--~8 galaxies \citep{Br19,En21,Ra16,Kh20,San20,So16,Gr15}.
However, the CSFG sample shows a tail extending to 
very faint galaxies with $M_g$, $M_{\rm FUV}$~$\sim$~$-$13 to $-$14 mag. Similar low-luminosity galaxies are also present in samples of
$z$ $\sim$ 3 -- 6 low-luminosity galaxies \citep{Ka17,So16,Gr15}.
We also note that the distributions of $M_g$ and $M_{\rm FUV}$ for the entire CSFG sample
(red lines in Fig.~\ref{fig2}c,d) and for galaxies with 
EW(H$\beta$)~$\ge$~100$\AA$\ (blue lines in Fig.~\ref{fig2}c,d) are similar and 
peak at similar magnitudes.

\subsubsection{Spectral energy distribution (SED) and stellar masses $M_\star$} 
\label{sed}

The stellar mass is one of the most important global galaxy characteristics. 
For its determination, we follow the prescriptions described by \citet{G06,G07}
and \citet{I11,Iz14}. SED fits were performed for each extinction-corrected
rest-frame SDSS spectrum. We take into account both stellar and ionised gas emission. 
The ionised gas continuum is strong in galaxies with high 
EW(H$\beta$)~$>$~50~$\AA$. Its contribution should be subtracted from the total
SED in the determination of the stellar mass.
Otherwise, stellar masses of galaxies with high 
EW(H$\beta$)’s in their spectra would be overestimated by a factor of $\sim$~3 
or more \citep{I11}. 

Monte Carlo simulations were carried out to reproduce the stellar and
nebular SEDs of each galaxy in our sample. 
The stellar SEDs were calculated with the PEGASE.2 package 
\citep{Fi97} and used to derive the stellar SED of the galaxy.
The star-formation history 
in each galaxy has been approximated by a recent short burst with age 
$t_b$~$<$~10~Myr and a prior continuous star formation with a constant SFR
between ages $t_{c1}$ and $t_{c2}$ for the older stars \citep{I11,Iz14}. 
Furthermore, a stellar initial mass 
function with a Salpeter slope, an upper mass limit of 100 M$_\odot$, and a lower
mass limit of 0.1 M$_\odot$ was adopted.
The SED of the gaseous continuum includes hydrogen and helium free-bound, 
free-free, and two-photon emission \citep{A84}. The fraction of the gaseous
continuum in the total SED is defined by the ratio between 
EW(H$\beta$)$_{\rm rec}$ for pure gaseous emission and the observed 
EW(H$\beta$)$_{\rm obs}$ value. The EW(H$\beta$)$_{\rm rec}$ varies in the range of 
$\sim$ 900 -- 1100$\AA$\ and also depends on the electron temperatures
$T_{\rm e}$ = 10000 -- 20000K, where $T_{\rm e}$ is derived from the observed 
spectrum.

The best modelled SED was 
found from $\chi^2$ minimisation of the deviation between the modelled and the 
observed continuum, varying $t_b$, $t_{c1}$, $t_{c2}$ and the ratio of masses of 
young to old stellar populations. Additionally, the best model should also
reproduce the observed equivalent widths of the H$\beta$ and H$\alpha$
emission lines. Typical stellar mass uncertainties 
for our sample galaxies are $\sim$ 0.1 -- 0.2 dex.

The distributions of stellar masses $M_\star$ for the entire CSFG sample and
for the subsample of CSFGs with EW(H$\beta$)~$\ge$~100~$\AA$, both with their maxima 
at $\sim$~10$^9$~M$_\odot$, are shown in Fig.~\ref{fig2}e. The mass range covered
by our sample overlaps with the stellar masses currently observed in 
many high-redshifts studies \citep[e.g. ][]{Ho16,Re18,Ta20,En21}. We note
that the number of galaxies with stellar masses $>$~10$^9$~M$_\odot$ 
decreases, likely due to the adopted selection criteria. Many massive 
galaxies do not satisfy the compactness criterion.

\subsubsection{Oxygen abundances 12 + log O/H} \label{abund}

One of the most reliable methods for oxygen abundance determination is the
$T_{\rm e}$-method, based on the electron temperature derived from the 
([O~{\sc iii}]~$\lambda$4959~+~$\lambda$5007)~/~[O~{\sc iii}]~$\lambda$4363 flux ratio. 
However, it requires the measurement of the weak [O~{\sc iii}]~$\lambda$4363 
flux with good accuracy. In this study, we could apply the $T_{\rm e}$-method 
to $\sim$ 2300 galaxies,
with [O~{\sc iii}]~$\lambda$4363 emission-line fluxes in their SDSS
spectra measured with an accuracy better than 4$\sigma$. We use equations
from \citet{I06} for the determination of electron temperatures, electron
number densities, ionic and total oxygen, and other element abundances. 

However, the SDSS spectra of most CSFGs  are noisy, preventing us from 
applying the $T_{\rm e}$-method to derive oxygen abundances in these galaxies. 
Thus, for these galaxies, we resort to strong emission line (SEL) methods, with the use of strong 
[O~{\sc ii}]~$\lambda$3727, [O~{\sc iii}]~$\lambda$4959, 
[O~{\sc iii}]~$\lambda$5007 and [N~{\sc ii}]~$\lambda$6584 emission lines 
to derive oxygen abundances. 

For the determination of the oxygen abundance in galaxies where the
$T_{\rm e}$-method cannot be applied, we use the relation given by \citet{Iz15}:
\begin{equation}
12 + \log\frac{\rm O}{\rm H} = 8.59\pm0.02 - (0.28\pm0.01)\times{\rm O3N2}, \label{sel2}
\end{equation}
where O3N2 = log ([O~{\sc iii}] 5007/H$\beta$) -- log ([N~{\sc ii}] 6584/H$\alpha$).
This calibration was obtained from the relation between O3N2s and oxygen
abundances derived by the direct $T_{\rm e}$-method for a large sample of SDSS
low-metallicity CSFGs, in which the 
[O~{\sc iii}]$\lambda$4363$\AA$\ flux is derived with good accuracy. This 
calibration is very similar 
to that obtained by \citet{PP04} who used the same relation between O3N2s and
oxygen abundances derived by the direct $T_{\rm e}$-method, but for a different
sample. 

However, the [N~{\sc ii}] $\lambda$6584 emission line in faint SDSS CSFGs 
is weak and cannot be measured in many cases. Furthermore, this 
line is outside the spectral range for galaxies with $z$ $\ga$ 0.4 in earlier 
SDSS releases (DR1 -- DR9), and for galaxies with $z$ $\ga$ 0.55 in later SDSS
releases (DR10 -- DR16). For these galaxies we use another calibration by
\citet{Iz15}:
\begin{equation}
12 + \log\frac{\rm O}{\rm H} = 8.18\pm0.01 - (0.43\pm0.03)\times\log{\rm O}_{32}. \label{sel3}
\end{equation}
Equations~\ref{sel2} and \ref{sel3} are compatible because both calibrations were 
based on samples with oxygen abundances derived by the 
$T_{\rm e}$-method.

The distributions of oxygen abundances 12 + logO/H are shown in 
Fig.~\ref{fig2}f.
We note that these distributions are relatively narrow, with maxima at
12 + logO/H $\sim$ 8.0 and with similar FWHMs of $\sim$ 0.3 dex for both the entire 
and high-EW(H$\beta$) galaxy samples.

\subsubsection{O$_{32}$ and Ne3O2}

The parameters O$_{32}$ and Ne3O2 defined as the 
[O~{\sc iii}]$\lambda$5007/[O~{\sc ii}]$\lambda$3727 and
[Ne~{\sc iii}]$\lambda$3868/[O~{\sc ii}]$\lambda$3727 ratios, respectively, 
trace the ionisation parameter \citep[e.g. ][]{St15}. Alternatively, these 
ratios are expected to be higher in density-bounded H~{\sc ii} regions. 
The distributions of O$_{32}$ and Ne3O2 are shown in 
Figs.~\ref{fig2}g -- \ref{fig2}h.  They have 
maxima at O$_{32}$ $\sim$ 1 and Ne3O2 $\sim$ 0.1 for the entire sample. 
However, for galaxies with EW(H$\beta$) $\ge$ 100$\AA$, these 
maxima are at considerably higher values, $\sim$ 3 and $\sim$ 0.3, respectively.
These differences can be due to 
the younger starburst age of high-EW(H$\beta$) galaxies and thus to higher 
luminosities of ionising radiation.
Furthermore, H~{\sc ii} regions powered by younger bursts are likely more 
compact, again implying a higher ionisation parameter.
The O$_{32}$ ratios for CSFGs are similar to those found in high-$z$ galaxies by \citet{Tr14}, 
\citet{Cu14}, \citet{Er16}, and \citet{On16}.

  \begin{table*}
  \caption{Relations between global parameters for CSFGs
\label{tab1}}
\centering
\begin{tabular}{llll} \hline
\multicolumn{1}{c}{$x$}&\multicolumn{1}{c}{$y$}&\multicolumn{1}{c}{Expression}&\multicolumn{1}{c}{Note} \\ \hline
log (SFR$^{-0.5}$$M_\star$/M$_\odot$) &12+logO/H            &$y$=(0.13$\pm$0.03)$x$+(6.87$\pm$0.12)&Fig.~\ref{fig4}b \\
log (SFR$^{-0.9}$$M_\star$/M$_\odot$) &log O$_{32}$          &$y$=$-$(0.42$\pm$0.03)$x$+(3.78$\pm$0.24)&Fig.~\ref{fig5}b \\
log EW([O~{\sc ii}] 3727)         &log ($M_\star$/M$_\odot$SFR$^{-0.9}$) &$y$=$-$(2.02$\pm$0.23)$x$+(12.60$\pm$0.64)&Fig.~\ref{fig6}b \\
log EW([O~{\sc iii}] 5007)        &log ($M_\star$/M$_\odot$SFR$^{-0.9}$) &$y$=$-$(1.05$\pm$0.06)$x$+(10.98$\pm$0.09)&Fig.~\ref{fig6}d \\
log EW(H$\alpha$ 6563)            &log ($M_\star$/M$_\odot$SFR$^{-0.9}$) &$y$=$-$(1.10$\pm$0.09)$x$+(11.26$\pm$0.18)&Fig.~\ref{fig6}f \\
log EW([O~{\sc iii}] 5007)        &log O$_{32}$          &$y$=(0.56$\pm$0.06)$x$$-$(1.08$\pm$0.09)&Fig.~\ref{fig7}a \\
log EW([O~{\sc iii}] 5007)        &log Ne3O2            &$y$=(0.71$\pm$0.07)$x$$-$(2.52$\pm$0.14)&Fig.~\ref{fig7}b \\
log O$_{32}$                       &log Ne3O2            &$y$=(1.22$\pm$0.03)$x$$-$(1.14$\pm$0.07)&Fig.~\ref{fig8}b \\
log EW([O~{\sc iii}] 5007)        &log [(10$^{-9}$$M_\star$/M$_\odot$)$^{-0.9}$SFR]&$y$=(0.95$\pm$0.09)$x$$-$(1.76$\pm$0.10)&Fig.~\ref{fig9}b \\
log EW(H$\alpha$ 6563)            &log [(10$^{-9}$$M_\star$/M$_\odot$)$^{-0.9}$SFR]&$y$=(1.40$\pm$0.09)$x$$-$(2.90$\pm$0.18)&Fig.~\ref{fig9}d \\
log EW([O~{\sc iii}] 5007)        &log sSFR                                    &$y$=(1.06$\pm$0.10)$x$$-$(1.96$\pm$0.17)&Fig.~\ref{fig10}a \\
log EW(H$\alpha$ 6563)            &log sSFR                                    &$y$=(1.30$\pm$0.14)$x$$-$(2.68$\pm$0.24)&Fig.~\ref{fig10}c \\
log EW([O~{\sc iii}] 5007)        &log [(10$^{-9}$$M_\star$/M$_\odot$)$^{0.1}$sSFR]&$y$=(1.01$\pm$0.09)$x$$-$(1.84$\pm$0.15)&Fig.~\ref{fig10}b \\
log EW(H$\alpha$ 6563)            &log [(10$^{-9}$$M_\star$/M$_\odot$)$^{0.1}$sSFR]&$y$=(1.20$\pm$0.12)$x$$-$(2.53$\pm$0.21)&Fig.~\ref{fig10}d \\
log (SFR$^{-0.9}$$M_\star$/M$_\odot$) &log $\xi_{\rm ion}$    &$y$=$-$(0.70$\pm$0.09)$x$+(31.16$\pm$2.10)&Fig.~\ref{fig11}b \\
log EW([O~{\sc iii}] 5007)        &log $\xi_{\rm ion}$ &$y$=(0.50$\pm$0.12)$x$+(23.95$\pm$0.96)&Fig.~\ref{fig12}a \\
log EW(H$\alpha$ 6563)            &log $\xi_{\rm ion}$ &$y$=(0.80$\pm$0.09)$x$+(23.28$\pm$0.75)&Fig.~\ref{fig12}b \\
\hline
\end{tabular}
  \end{table*}

\subsubsection{The SFR and sSFR}
\label{sfr}

 The star formation rate is obtained from the extinction-corrected H$\beta$ 
luminosity $L$(H$\beta$) using the relation of \citet{K98} between the SFR and 
the H$\alpha$ luminosity and adopting an H$\alpha$/H$\beta$ ratio of 2.8:
\begin{equation}
{\rm SFR} = 2.21 \times 10^{-41} L({\rm H}\beta), \label{SFR}
\end{equation}
where $L$(H$\beta$) is in erg~s$^{-1}$ and SFR is in M$_\odot$~yr$^{-1}$.
Correspondingly, the specific star formation rate 
is determined as sSFR = SFR/$M_\star$.

The distribution of SFRs ranging from $\sim$~0.01 M$_\odot$ yr$^{-1}$ to 
$\sim$~200 M$_\odot$ yr$^{-1}$, which is similar to values for high-$z$ SFGs, is 
represented in Fig.~\ref{fig2}i with a maximum at $\sim$ 20 M$_\odot$ yr$^{-1}$ 
for both the entire and high-EW(H$\beta$) samples.
These SFRs offer a sufficient comparison with those  of $\sim$ 0.1 -- 100 M$_\odot$ yr$^{-1}$ found
for high-$z$ SFGs by \citet{Ho16}, \citet{Re18}, \citet{Ta20}, and \citet{En21}.

The distributions of sSFGs are shown in Fig.~\ref{fig2}j with maxima at
$\sim$ 2 Gyr$^{-1}$ and $\sim$ 5 Gyr$^{-1}$ for the entire and high-EW(H$\beta$)
samples, respectively. Again, these values are typical of high-$z$ SFGs 
\citep[e.g. ][]{Ra16,Re18,Co18}.

\subsubsection{Equivalent widths and luminosities of emission lines}

The distributions of EW([O~{\sc ii}]$\lambda$3727), EW([O~{\sc iii}]$\lambda$5007),
EW([O~{\sc iii}]$\lambda$4959+5007 + H$\beta$), and 
EW(H$\alpha$) are shown in Figs.~\ref{fig2}k -- \ref{fig2}n.
We display these distributions because the equivalent widths of strong emission 
lines are commonly reported for high-$z$ SFGs, and among them 
EW([O~{\sc iii}]$\lambda$5007) and EW(H$\alpha$) are most often used.
Sometimes the EW([O~{\sc iii}]$\lambda$4959+5007 + H$\beta$)s are measured in
low-resolution spectra, for instance, in {\sl HST} prism spectra, where the 
[O~{\sc iii}] and H$\beta$ lines are not resolved. 

The maxima for the high-EW(H$\beta$) sample occur at
considerably higher EW values, compared to the entire sample, as expected. In
particular, the values of EW([O~{\sc iii}]$\lambda$5007) and 
EW(H$\alpha$) at the maxima for CSFGs with EW(H$\beta$) $\ge$ 100$\AA$\ are
$\sim$ 700$\AA$\ and $\sim$ 600$\AA$, respectively, with a considerable number
of galaxies with EWs $>$ 1000$\AA$. On the other hand, the EWs at the maxima of
the distributions for the entire sample are around four to five times lower. The smallest
difference between the two samples is found for the distributions of 
EW([O~{\sc ii}]$\lambda$3727) with an EW maximum for high-EW(H$\beta$) 
galaxies that is only $\sim$ 30\% higher than that for the entire sample. 
This is because the [O~{\sc ii}]$\lambda$3727/H$\beta$
flux ratio, in contrast to the [O~{\sc iii}]$\lambda$5007/H$\beta$ ratio, 
decreases with increasing EW(H$\beta$) and thus 
EW([O~{\sc ii}]$\lambda$3727) increases with EW(H$\beta$) more slowly
than EW([O~{\sc iii}]$\lambda$5007). The distributions of extinction-corrected luminosities  
$L$([O~{\sc ii}]$\lambda$3727), $L$([O~{\sc iii}]$\lambda$5007),
and $L$(H$\alpha$) are shown in Figs.~\ref{fig2}o -- \ref{fig2}q.

\subsubsection{Ionising photon production efficiencies $\xi_{\rm ion}$} \label{ksi}

The CSFGs in our sample are characterised by high H$\beta$ and H$\alpha$ 
luminosities (e.g. Fig.~\ref{fig2}q) and thus they
produce copious amounts of ionising photons, which can be estimated by
the production rate $N$(LyC) of the LyC radiation according to \citet{SH95}:
\begin{equation}
N({\rm LyC}) = 2.1\times 10^{12}L({\rm H}\beta), \label{NLyC}
\end{equation}
where $N$(LyC) and the extinction-corrected H$\beta$ luminosity 
$L$(H$\beta$) are in units of photons s$^{-1}$ and erg s$^{-1}$, 
respectively.
Another parameter characterising ionising radiation is the ionising photon 
production efficiency, $\xi_{\rm ion}$, determined as
\begin{equation}
\xi_{\rm ion} = \frac{N({\rm LyC})}{L_\nu}, \label{xi}
\end{equation}
where $L_\nu$ in erg s$^{-1}$ Hz$^{-1}$ is the intrinsic monochromatic luminosity 
at the rest-frame wavelength of 1500$\AA,$\ including the stellar and nebular 
emission, derived from the SED fitting. The production efficiency $\xi_{\rm ion}$ 
depends on metallicity, star-formation history, age of the stellar population, 
and also on assumptions on stellar evolution. It is higher for galaxies with 
higher
EW(H$\beta$), that is, for younger bursts of star formation.

The distribution of $\xi_{\rm ion}$ is shown in Fig.~\ref{fig2}r, with values varying 
over two orders of magnitude for the entire sample. The range of
$\xi_{\rm ion}$ for CSFGs with EW(H$\beta$) $\ge$ 100$\AA$\ is much narrower, and 
almost all galaxies from this smaller sample are above the canonical value 
log ($\xi_{\rm ion}$/[Hz erg$^{-1}$]) $\ga$ 25.2 -- 25.3 required for the
reionisation of the Universe, assuming a
typical LyC escape fraction of 10 -- 20\% \citep{Ro13,Bo15}.

\subsubsection{Slope $\beta$ of the UV continuum} \label{beta}

We define the slopes $\beta_{\rm int}$ and $\beta$ of the modelled 
intrinsic and obscured SEDs as
\begin{equation}
\beta_{\rm int} = \frac{\log I(\lambda_1) - \log I(\lambda_2)}{\log \lambda_1 - \log \lambda_2}, \label{betint}
\end{equation}
\begin{equation}
\beta = \beta_{\rm int} - 0.4\times \frac{A(\lambda_1) - A(\lambda_2)}{\log \lambda_1 - \log \lambda_2}, \label{bet}
\end{equation}
respectively, where $\lambda_1$ =1300$\AA$\ and $\lambda_2$ = 1800$\AA$\ are 
the rest-frame wavelengths, $I(\lambda)$ is the intrinsic flux, which includes 
both the stellar and nebular emission, and $A(\lambda)$ is the extinction in 
mags.

The distribution of the intrinsic UV slope $\beta_{\rm int}$ is shown in 
Fig.~\ref{fig2}s. As expected, the slopes of the younger bursts in CSFGs 
with EW(H$\beta$)~$\ge$~100$\AA$\ are steeper, with a maximum of the distribution at
$\sim$~$-2.55$ and covering the narrow range between $-2.75$ and $-2.3$. The
maximum of the distribution for the entire sample is at $\sim$~$-2.3$ and the 
range of $\beta_{\rm int}$ variations is much larger, between $-2.75$ and $-1.7$.

For the sake of comparison with the observed UV slopes for high-$z$ SFGs, we show
in Fig.~\ref{fig2}t the distributions of obscured UV
slopes of CSFGs derived from the intrinsic UV slopes. The extinction is derived from the hydrogen Balmer decrement in the SDSS spectra, assuming the reddening law by \citet{C89} with $R(V)$ = 3.1. The $\beta$
distribution of CSFGs in the entire sample, in the range of 
$\sim$~--2.5~-~0.0, is similar to that for high-$z$ SFGs (see 
Sect.~\ref{beta1}).

\begin{figure*}
\centering
\hspace{0.0cm}\includegraphics[angle=-90,width=1.00\linewidth]{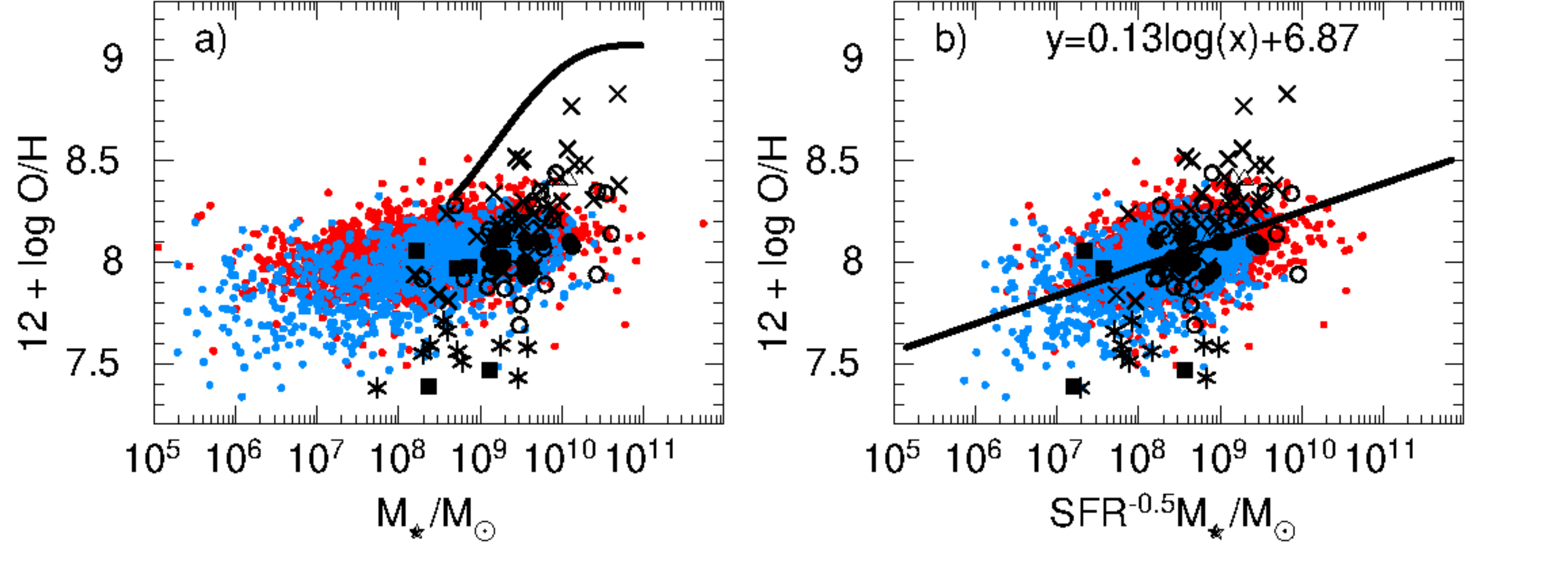}
\caption{Dependences of oxygen abundances
12~+~logO/H on {\bf (a)} stellar masses $M_\star$ and on
{\bf (b)} SFR$^{-0.5}$$M_\star$ for the samples of CSFGs and high-$z$ SFGs.
The maximum likelihood relation to SDSS data in {\bf (b)} is
shown by the straight solid line. The black line in {\bf (a)} is the
relation for $z$ = 0 SDSS SFGs by \citet{Ma10}.
Large symbols represent high-$z$ SFGs by \citet{A16} 
($z$ = 2.4--3.5, asterisks), \citet{Er16} ($z$ $\sim$ 2.3, filled circles),
\citet{Tr14} ($z$ $\sim$ 3--5, open circles), \citet{On16} ($z$ $\sim$ 3 -- 3.7,
crosses), \citet{Cu14} ($z$ $>$ 2, open triangles) and \citet{Jo20} 
($z$ = 7.1 -- 9.1, filled squares). The meanings of symbols for 
SDSS CSFGs in both panels are the same as in Fig.~\ref{fig1}.}
\label{fig4}
\end{figure*}

\begin{figure*}
\centering
\hspace{0.0cm}\includegraphics[angle=-90,width=1.00\linewidth]{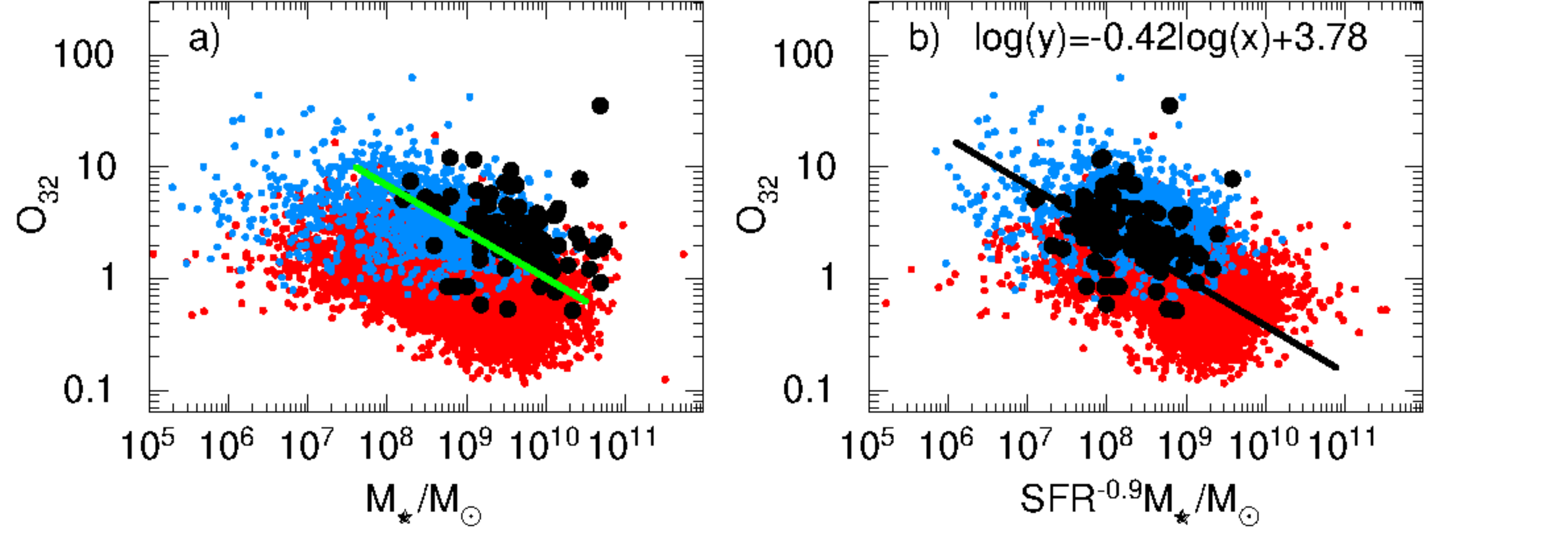}
\caption{Dependences of O$_{32}$ =
[O~{\sc iii}]~$\lambda$5007/[O~{\sc ii}]~$\lambda$3727 emission line ratios 
on {\bf (a)} stellar masses $M_\star$ and on
{\bf (b)} SFR$^{-0.9}$$M_\star$ for samples of CSFGs and high-$z$ SFGs.
The black line in {\bf (b)} is the maximum likelihood relation
whereas the green line in {\bf (a)} is the relation for 
$z$ $\sim$ 1.7 -- 3.6 galaxies \citep{Sa20}. Black-filled circles
represent high-$z$ SFGs by \citet{Er16} ($z$ $\sim$ 2.3), \citet{On16} 
($z$ $\sim$ 3 -- 3.7), \citet{Tr14} ($z$ $\sim$ 3 -- 5), and \citet{Cu14} 
(stacks of $z$ $>$ 2 galaxies).
The meaning of symbols for our SDSS are the same as in Fig.~\ref{fig1}.}
\label{fig5}
\end{figure*}

\begin{figure*}
\centering
\hspace{0.0cm}\includegraphics[angle=-90,width=0.90\linewidth]{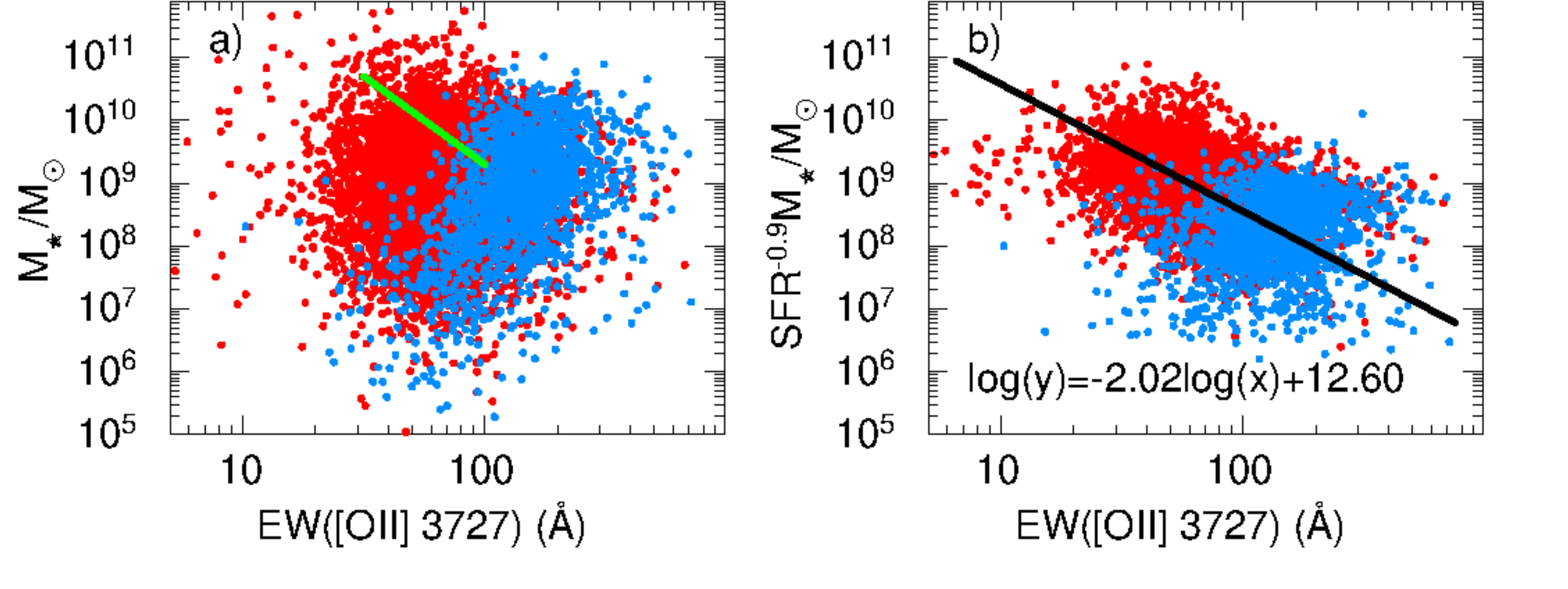}
\hspace{0.0cm}\includegraphics[angle=-90,width=0.90\linewidth]{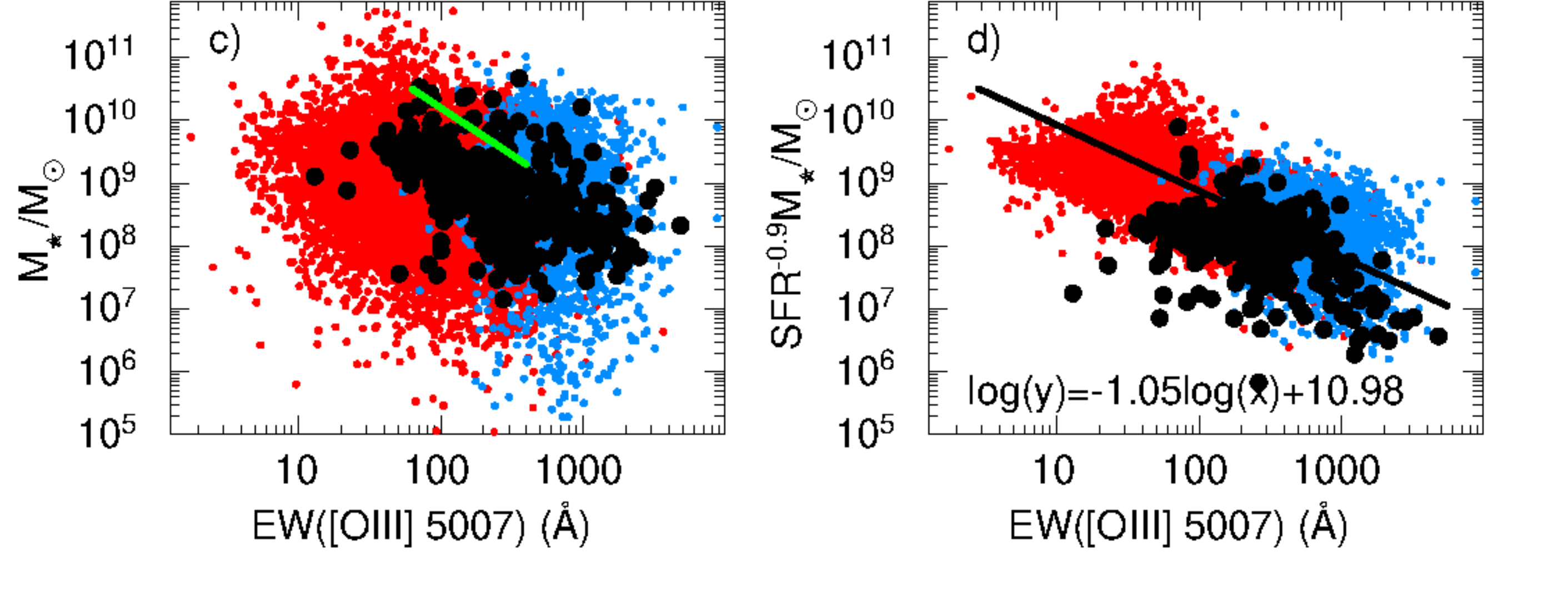}
\hspace{0.0cm}\includegraphics[angle=-90,width=0.90\linewidth]{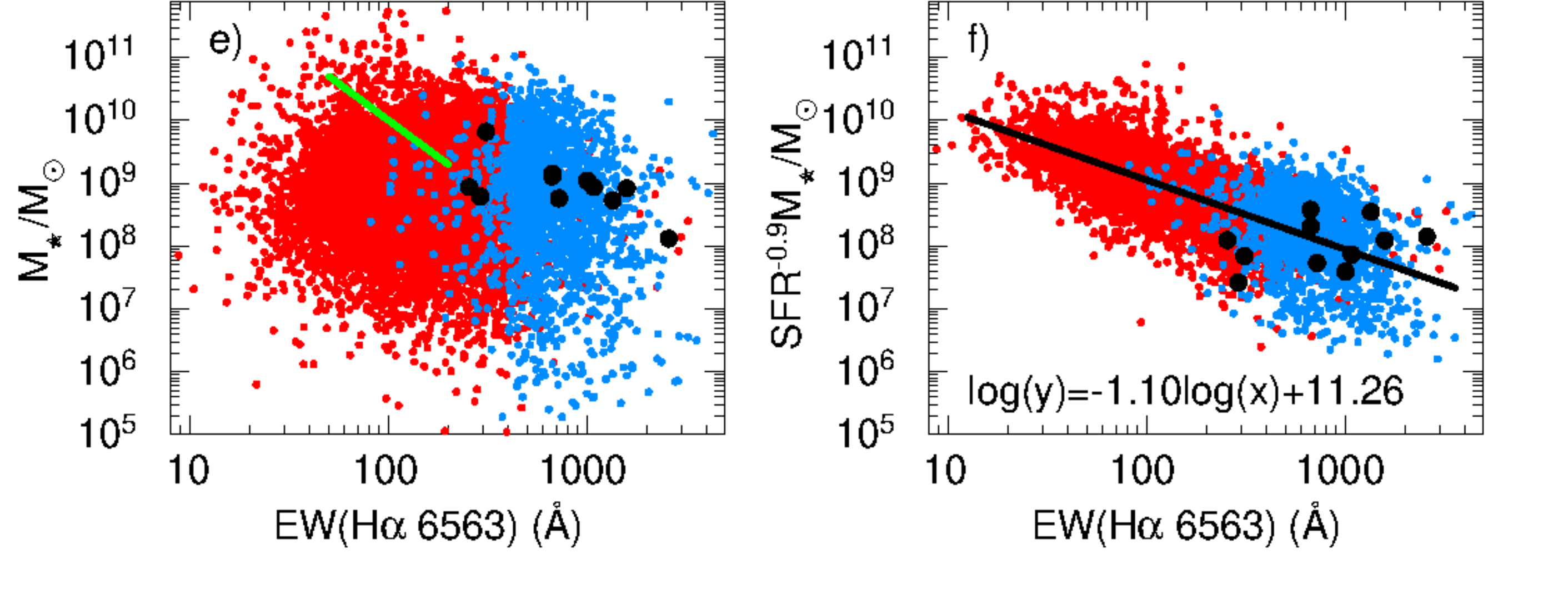}
\caption{Dependences of stellar masses $M_\star$ ({\bf (a)}, {\bf (c)}, {\bf (e)})
and of SFR$^{-0.9}$$M_\star$ ({\bf (b)}, {\bf (d)}, {\bf (f)}) on the equivalent 
widths of the
[O~{\sc ii}]~$\lambda$3727, [O~{\sc iii}]~$\lambda$5007 and 
H$\alpha$~$\lambda$6563 emission lines, respectively, for samples of CSFGs and 
high-$z$ SFGs.
The solid lines in {\bf (b)}, {\bf (d)}, {\bf (f)} are the maximum likelihood 
relations, whereas green lines in {\bf (a)}, {\bf (c)}, {\bf (e)}
are relations for $z$ $\sim$ 1.4 -- 3.8 SFGs \citep{Re18}.
 Galaxies denoted in {\bf (c)} and {\bf (d)}
by black-filled circles are $z$ $\sim$ 3.5 LBGs 
\citep{Ho16}, $z$ $\sim$ 2 SFGs \citep{Ha16}, $z$ $\sim$ 1.3 -- 3.7 
galaxies with high EW([O~{\sc iii}]~+~H$\beta$) \citep{Ta20}, 
$z$ $\sim$ 7 SFGs with high EW([O~{\sc iii}]~+~H$\beta$) 
\citep{En21}, 
$z$ $\sim$ 6.6 SFGs with Ly$\alpha$ emission 
\citep{En20} and $z$ $\sim$ 8 galaxies 
\citep{DB19}. Galaxies represented in {\bf (e)} and {\bf (f)} by black-filled
circles are $z$ $\sim$ 5 SFGs by \citet{Ra16}.
The meanings of symbols for our SDSS CSFGs are the same as in Fig.~\ref{fig1}.}
\label{fig6}
\end{figure*}

\begin{figure*}
\hbox{
\hspace{0.0cm}\includegraphics[angle=-90,width=0.55\linewidth]{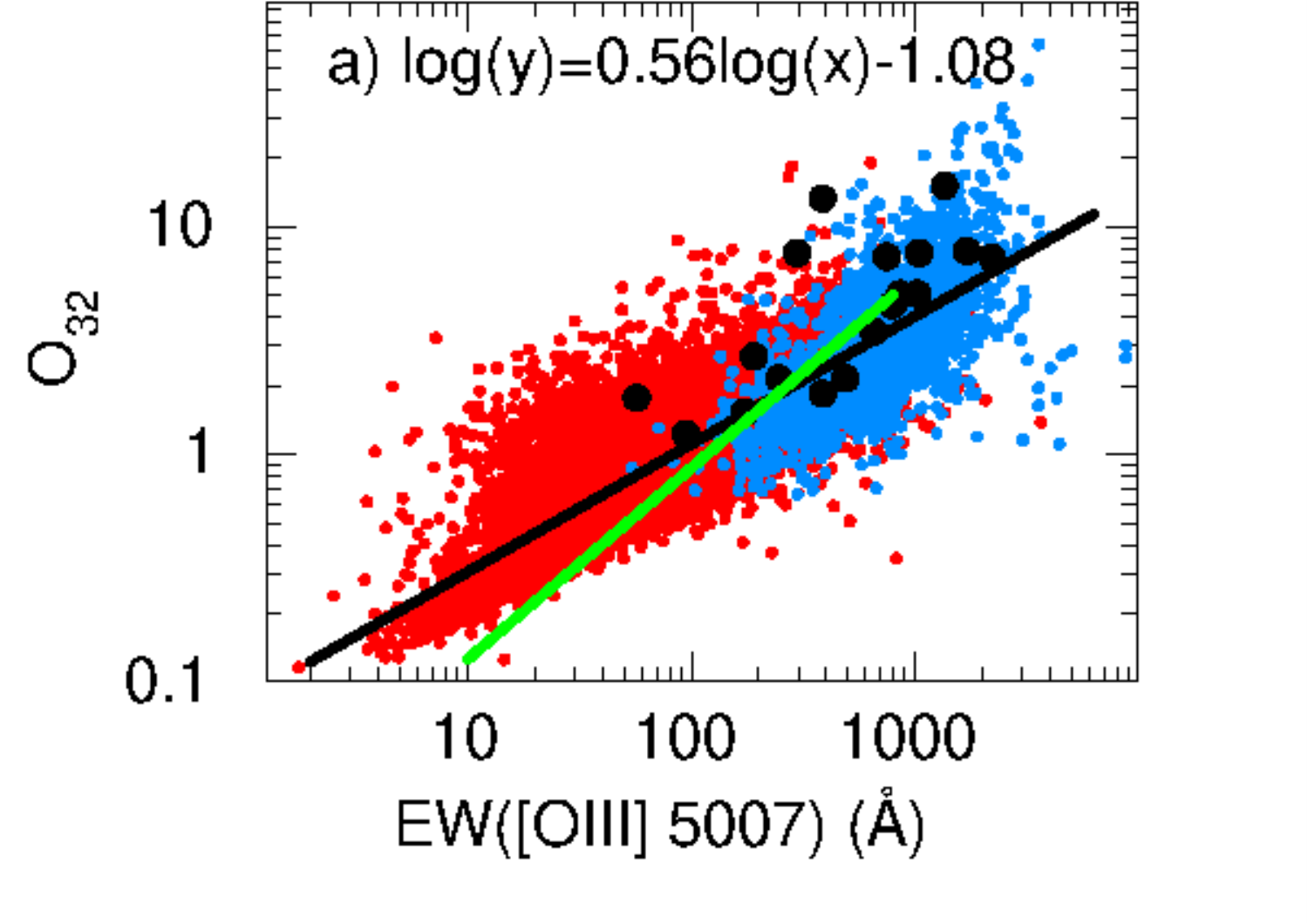}
\hspace{-1.3cm}\includegraphics[angle=-90,width=0.55\linewidth]{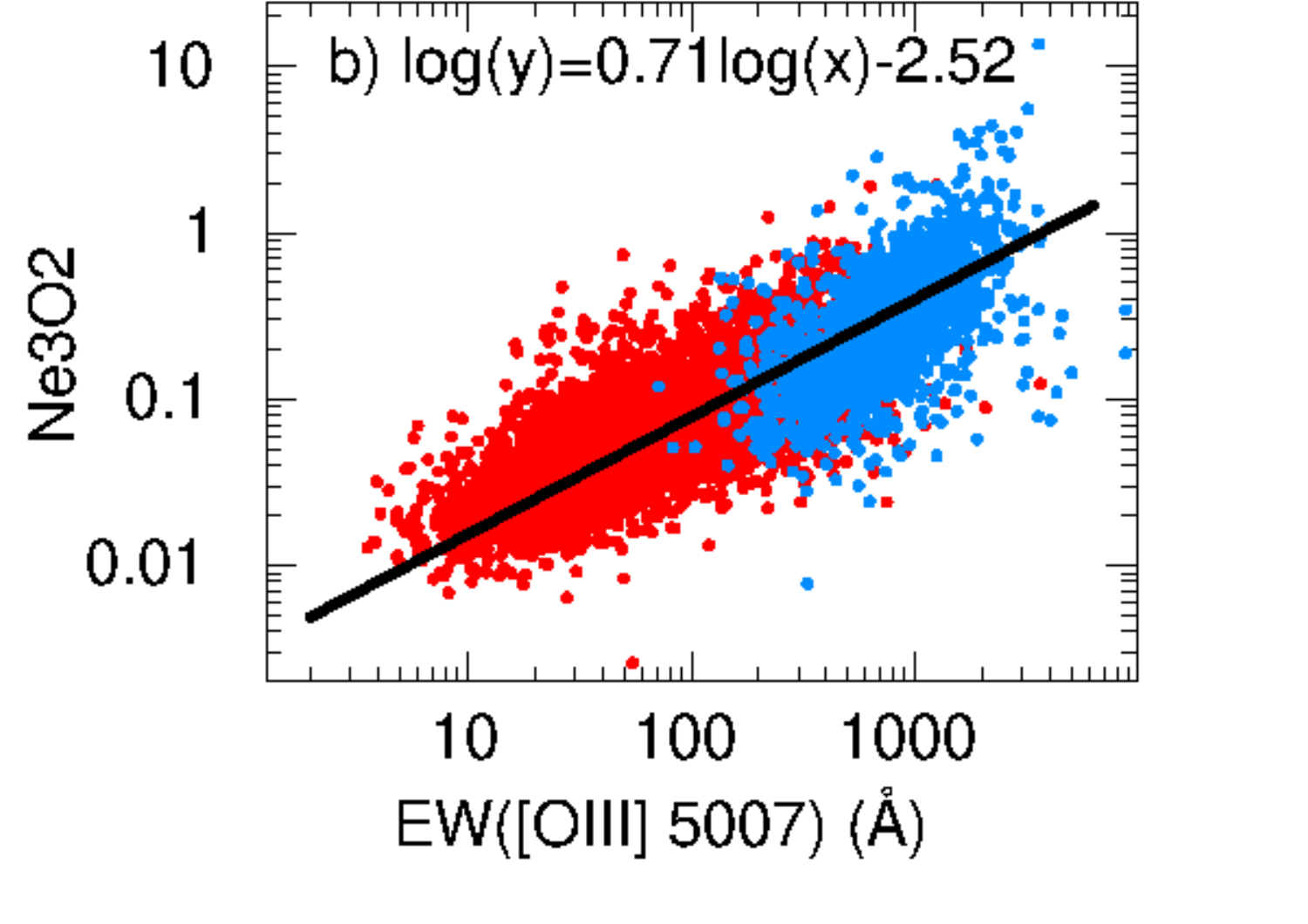}
}
\caption{Relations between equivalent widths of 
the [O~{\sc iii}]~$\lambda$5007 emission line and extinction-corrected 
{\bf (a)} O$_{32}$ = [O~{\sc iii}]~$\lambda$5007/[O~{\sc ii}]~$\lambda$3727 and 
{\bf (b)} Ne3O2 = [Ne~{\sc iii}]~$\lambda$3868/[O~{\sc ii}]~$\lambda$3727 for 
samples of CSFGs and high-$z$ SFGs. The solid lines are 
the maximum likelihood relations, whereas the green line in {\bf (a)} 
is the relation for $z$ $\sim$ 1.4 -- 3.8 galaxies \citep{Re18}. LAEs at 
$z$ $\sim$ 3 by \citet{Na20} are shown in {\bf (a)} by filled black circles. 
The meanings of symbols for SDSS CSFGs are 
the same as in Fig.~\ref{fig1}.}
\label{fig7}
\end{figure*}



\begin{figure*}
\centering
\hspace{0.0cm}\includegraphics[angle=-90,width=0.49\linewidth]{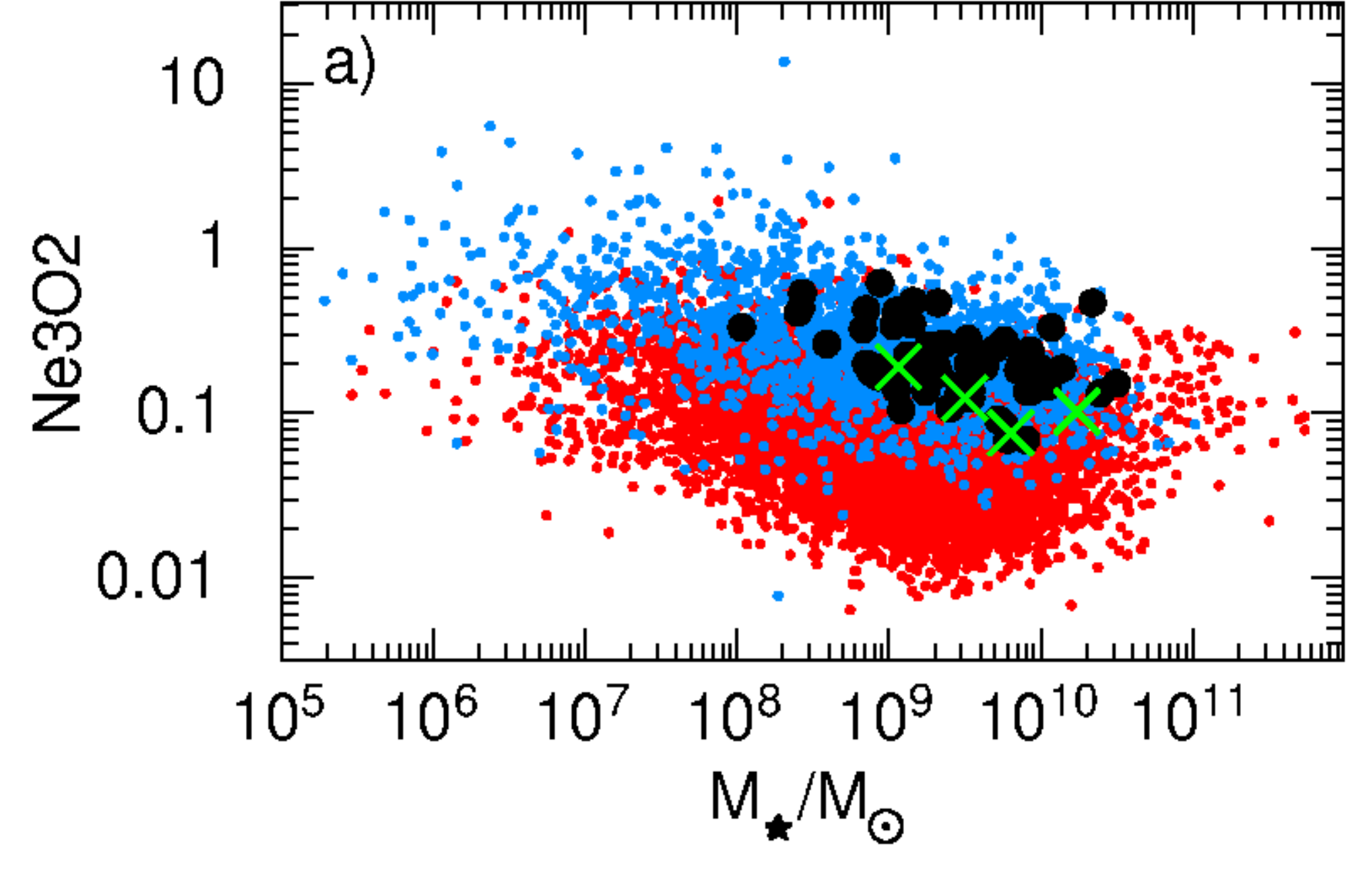}
\hspace{0.0cm}\includegraphics[angle=-90,width=0.49\linewidth]{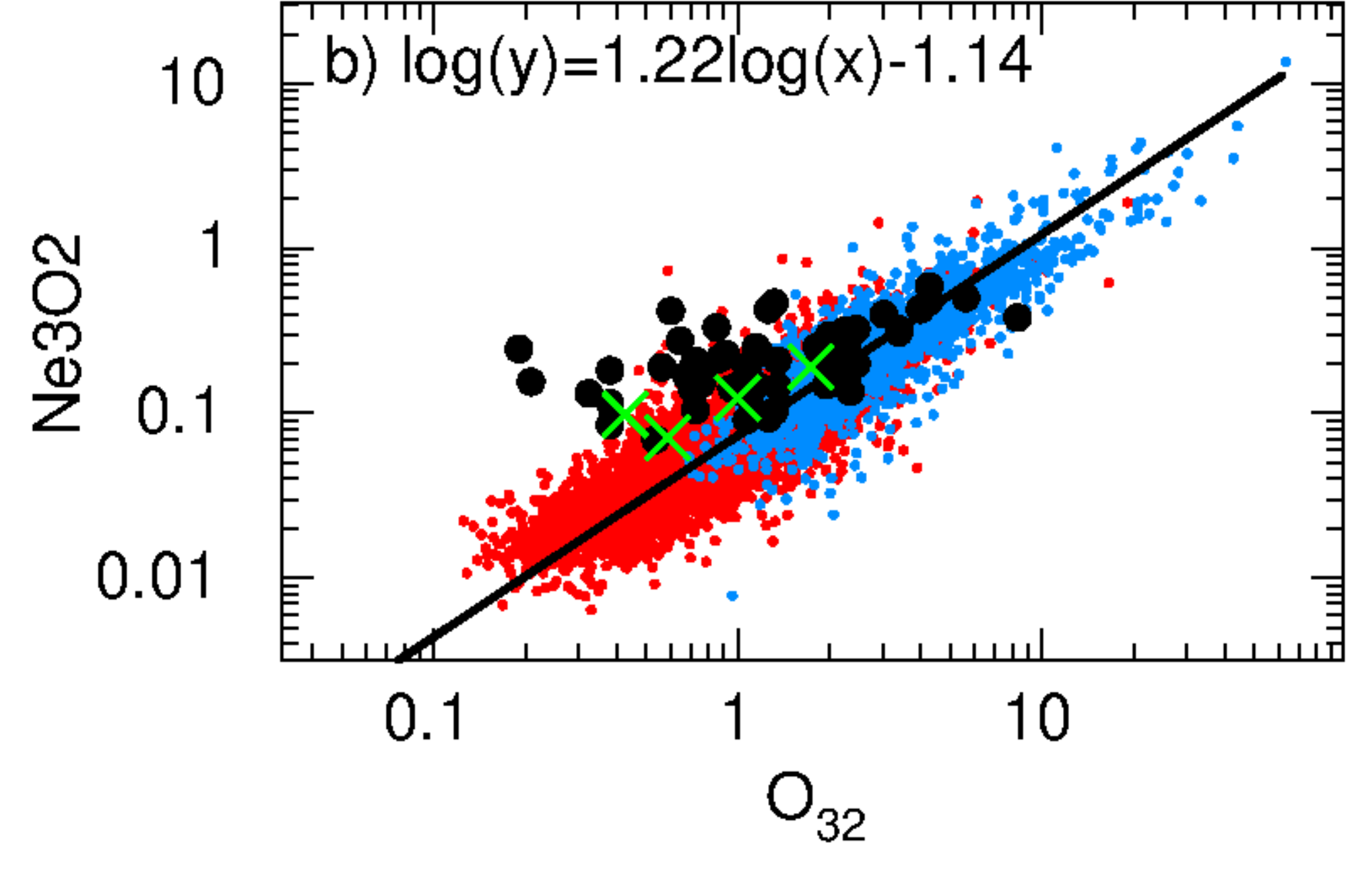}
\caption{Dependences of Ne3O2 = [Ne~{\sc iii}]~$\lambda$3868/[O~{\sc ii}]~$\lambda$3727 on {\bf (a)} stellar masses $M_\star$ and {\bf (b)} the O$_{32}$ = 
[O~{\sc iii}]~$\lambda$5007/[O~{\sc ii}]~$\lambda$3727. The solid line in 
{\bf (b)} is the maximum likelihood relation. The individual 
$z$ $\sim$ 2 SFGs from the MOSDEF survey and stacks of MOSDEF spectra 
from \citet{Je20} are shown by black-filled circles and green crosses, 
respectively. The meanings of symbols for CSFGs are the same as in Fig.~\ref{fig1}.}
\label{fig8}
\end{figure*}



\begin{figure*}
\hbox{
\hspace{0.0cm}\includegraphics[angle=-90,width=0.90\linewidth]{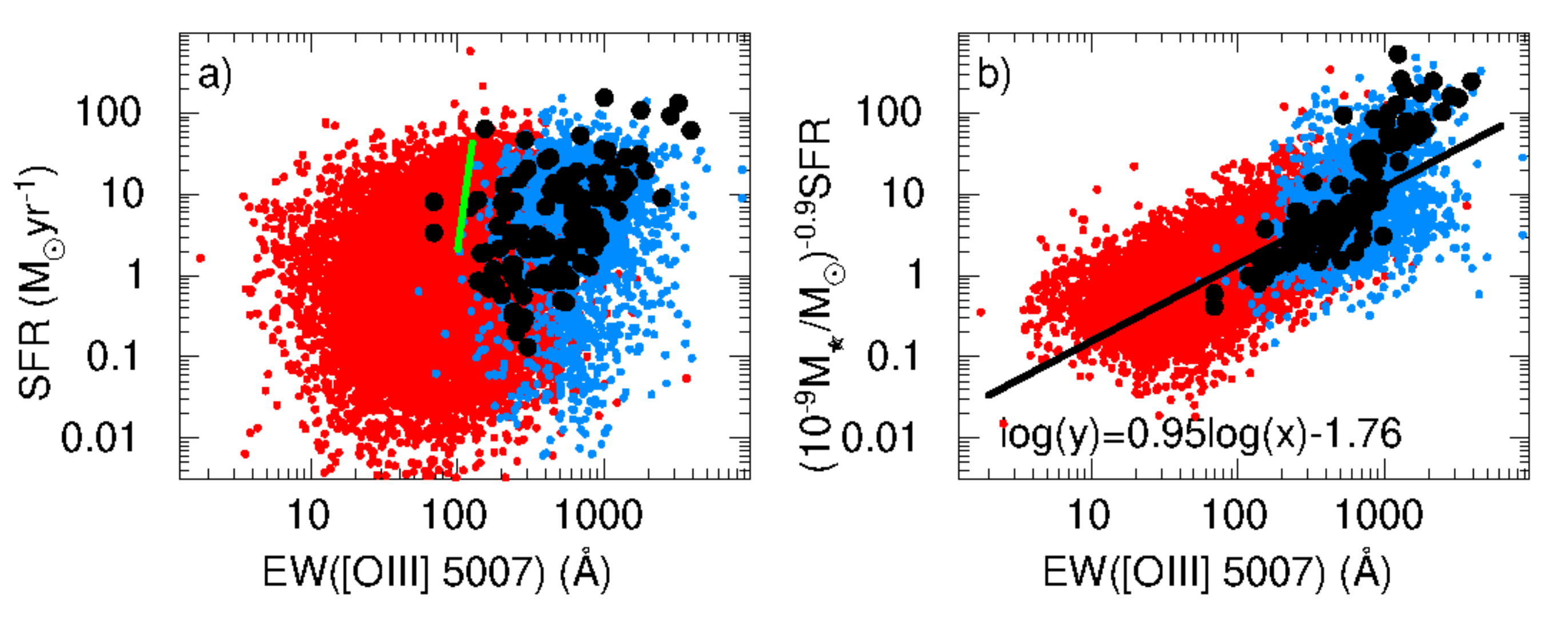}
}
\hbox{
\hspace{0.0cm}\includegraphics[angle=-90,width=0.935\linewidth]{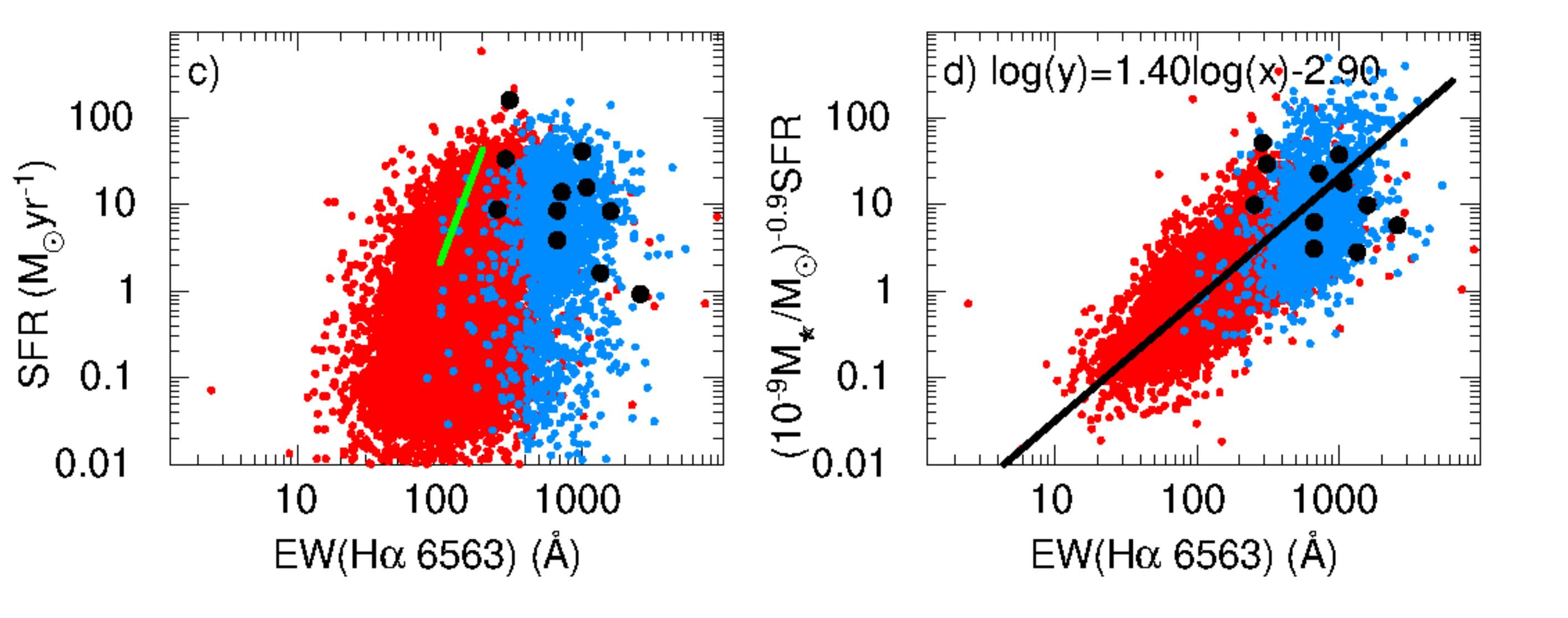}
}
\caption{Dependences of star-formation rates SFR ({\bf (a)}, {\bf (c)})
and of $M_\star^{-0.9}$SFR ({\bf (b)}, {\bf (d)}) on equivalent widths of the
[O~{\sc iii}]~$\lambda$5007 and H$\alpha$~$\lambda$6563 emission lines,
respectively, for samples of CSFGs and high-$z$ 
SFGs. The solid lines in {\bf (b)} and {\bf (d)} are the maximum likelihood relations. 
Galaxies denoted in {\bf (a)} and {\bf (b)} by black-filled circles are 
$z$ $\sim$ 3.5 LBGs by \citet{Ho16}, $z$ $\sim$ 1.3 -- 3.7 
galaxies with high EW([O~{\sc iii}]~+~H$\beta$) by \cite{Ta20},
$z$ $\sim$ 7 galaxies with high EW([O~{\sc iii}]~+~H$\beta$) by \citet{En21} and
$z$ $\sim$ 6.6 galaxies with Ly$\alpha$ emission by \citet{En20},
respectively. Galaxies represented in {\bf (c)} and {\bf (d)} by black-filled
circles are $z$ $\sim$ 5 SFGs by \citet{Ra16}.
Green lines in {\bf (a)} and {\bf (c)} are the relations for $z$ = 1.4 -- 3.8 
SFGs \citep{Re18}. The meanings of symbols for SDSS CSFGs are the same as in 
Fig.~\ref{fig1}.}
\label{fig9}
\end{figure*}


\section{Relations between global parameters for CSFGs and their comparison
with those of high-redshift SFGs} \label{results}

In this section we consider relations between the global parameters of CSFGs,
such as UV luminosities, stellar masses, metallicities, star formation rates, 
specific star-formation rates sSFRs, O$_{32}$ and Ne3O2 ratios, equivalent 
widths EW([O~{\sc ii}]~$\lambda$3727), EW([O~{\sc iii}]~$\lambda$5007), and 
EW(H$\alpha$~$\lambda$6563), ionising photon production 
efficiencies, and the slopes of the UV continua. We also study whether the above-mentioned parameters depend only on a single observationally accessible quantity
or whether a second one is also required, similarly to the
fundamental stellar mass and metallicity relation considered, for example, by \citet[][see also Sect.~\ref{intro}]{Ma10}.
Finally, we compare these relations with the corresponding ones for 
high-redshift SFGs. The expressions of the relations for CSFGs discussed in this
Section are summarised in Table~\ref{tab1} and the corresponding figures
are introduced below.

\subsection{Rest-frame UV absolute magnitudes, stellar masses, and star formation rates}

The relations between the rest-frame UV absolute magnitudes, stellar masses, and 
star formation rates for high-redshift SFGs have been considered in many papers.
\citet{Ka17} found that low-mass galaxies at 3 $<$ $z$ $<$ 6 are forming stars 
at higher rates than seen locally or in more massive galaxies. \citet{A20} 
found that LAEs are typically young low-mass galaxies undergoing one of 
their first bursts of star formation. 

It was found that the sSFR in SFGs with
2.5 $<$ $z$ $<$ 4 is 
4.6 Gyr$^{-1}$ \citep{Co18} and that it increases with redshift from 
$z$ = 2 to 7 \citep[sSFR $\propto$ (1 + $z$)$^{1.1}$, ][]{Da18}. 
\citet{Ta20} concluded that a significant fraction 
of the early galaxy population should be characterised by large sSFRs 
($>$ 200 Gyr$^{-1}$) and low metallicities ($<$~0.1~$Z_\odot$).

\begin{figure*}
\centering
\hspace{0.0cm}\includegraphics[angle=-90,width=1.00\linewidth]{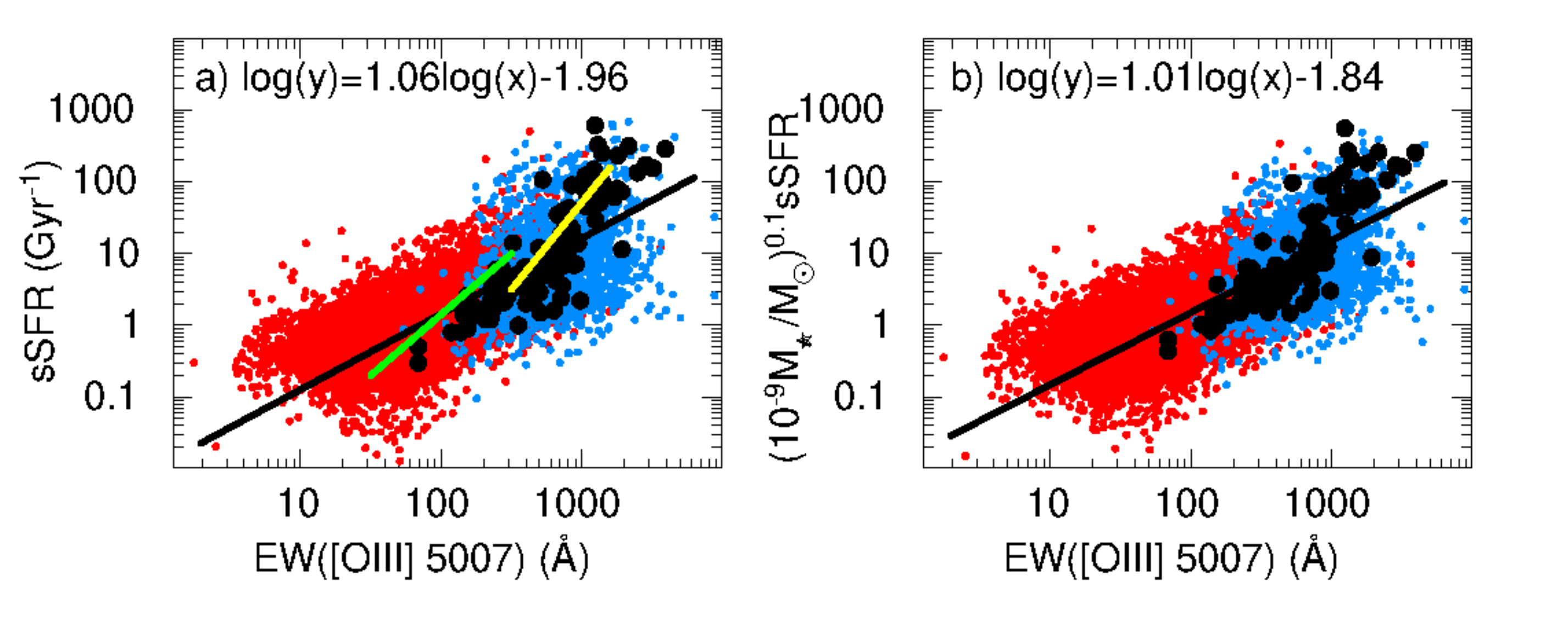}
\hspace{0.0cm}\includegraphics[angle=-90,width=1.00\linewidth]{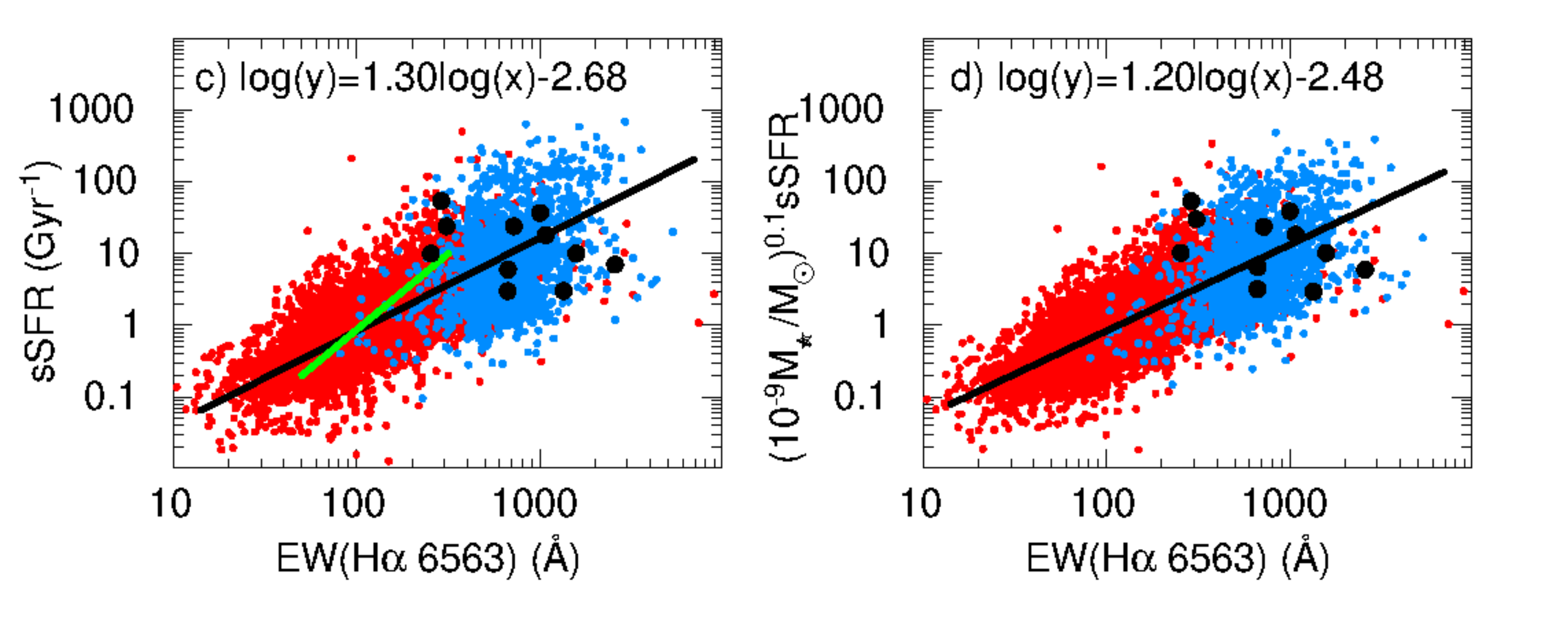}
\caption{Dependences of specific star-formation rates sSFR 
({\bf (a)}, {\bf (c)})
and of $M_\star^{-0.1}$sSFR ({\bf (b)}, {\bf (d)}) on equivalent widths of the
[O~{\sc iii}]~$\lambda$5007 and H$\alpha$~$\lambda$6563 emission lines,
respectively, for samples of CSFGs and high-$z$ 
SFGs. Solid lines are the maximum likelihood relations. 
Galaxies denoted in {\bf (a)} and {\bf (b)} by black-filled circles 
are $z$ $\sim$ 3.5 LBGs by \citet{Ho16},
$z$ $\sim$ 1.3 -- 3.7 galaxies with high EW([O~{\sc iii}]~+~H$\beta$) by 
\cite{Ta20}, $z$ $\sim$ 7 galaxies with high EW([O~{\sc iii}]~+~H$\beta$) 
by \citet{En21} and $z$ $\sim$ 6.6 galaxies with Ly$\alpha$ emission by 
\citet{En20}. 
Galaxies represented in {\bf (c)} and {\bf (d)} by black-filled circles are 
$z$ $\sim$ 5 SFRs by \citet{Ra16}. Yellow and green solid lines in 
{\bf (a)} represent relations of $z$ $\sim$ 2 analogues for $z$ $>$ 6.5 
galaxies \citep{Du20} and $z$ = 1.4 -- 3.8 SFGs \citep{Re18}, respectively,
whereas the relation for $z$ = 1.4 -- 3.8 SFGs \citep{Re18} in {\bf (c)}
is shown by the green line.
The meanings of symbols for SDSS CSFGs are the same as in Fig.~\ref{fig1}.}
\label{fig10}
\end{figure*}

\citet{So16} found that the correlation between the 
rest-frame UV absolute magnitude $M_{\rm FUV}$ at 1500 $\AA$\ and the logarithmic stellar 
mass log $M_\star$ for $z$ = 4 -- 8 SFGs is linear. 
Similarly, \citet{Iy18} found that the relation
log~SFR~--~log~$M_\star$ at $z$ = 6 is linear down to log $M_\star$/M$_\odot$ = 6.
On the other hand, \citet{Sa15} found that star-forming galaxies in the
Cosmic Assembly Near-infrared Deep Extragalactic Legacy Survey (CANDELS), in the redshift range of  
$z$ = 3.5 -- 6.5, follow a nearly unevolving 
correlation between stellar mass and SFR that follows SFR $\sim$ M$_\star$$^a$,
with $a$~=~0.70~$\pm$~0.21 at $z$~$\sim$~4 and 0.54~$\pm$~0.16 at $z$~$\sim$~6.
Similarly, \citet{A20} found that the SFR~--~$M_\star$ relation has negligible 
evolution from $z$~$\sim$~4 to $z$~$\sim$~6.

In Fig.~\ref{fig3}a, we present the relation rest-frame absolute UV magnitude
$M_{\rm FUV}$ -- stellar mass $M_\star$ for our sample of SDSS CSFGs
(red and blue dots), ranging in stellar mass from $\sim$~10$^5$~M$_\odot$ to
$\sim$~10$^{11}$ M$_\odot$, or over six orders of magnitude, and in UV absolute
magnitude from $\sim$ --12 mag to $\sim$ --25 mag, or over five orders of magnitude
in the FUV luminosity. We note that the $M_{\rm FUV}$s in Fig.~\ref{fig3}a are 
attenuated magnitudes, derived from the intrinsic magnitudes and
adopting extinction obtained from the hydrogen Balmer decrement in the SDSS 
spectra, and the \citet{C89} reddening law with $R(V)$ = 3.1. 
All other symbols and lines in Fig.~\ref{fig3}a are related to high-z SFG 
studies, which report `observed' magnitudes, uncorrected for reddening. 
We can see that the locations of high-$z$ SFGs are in good agreement with
the location of our CSFGs with EW(H$\beta$) $\ge$ 100$\AA$\ (blue dots), 
indicating little variation for strongly star-forming galaxies over the redshift
range of $z$ $\sim$ 0 -- 8.

The relations of the SFR to stellar mass $M_\star$ and the  sSFR to stellar mass $M_\star$ are shown in Figs.~\ref{fig3}b 
and \ref{fig3}c, respectively. The SFR for CSFGs ranges over five orders of
magnitude, from $\sim$~0.001~M$_\odot$~yr$^{-1}$ to $\sim$~100~M$_\odot$~yr$^{-1}$,
whereas the sSFR can reach values of up to several hundred Gyr$^{-1}$, indicating
that the major part of the stellar mass has been formed during the 
last $<$~10~Myr.
We see in Figs.~\ref{fig3}b and \ref{fig3}c that CSFGs with 
EW(H$\beta$) $\ge$ 100$\AA$\ and high-$z$ galaxies have very similar properties.

High sSFRs in CSFGs with EW(H$\beta$) $\ge$ 100$\AA$\ are related to the 
presence of high-excitation H~{\sc ii} regions with strong 
[O~{\sc iii}]$\lambda$5007/H$\beta$ ratios (blue dots in Fig.~\ref{fig3}d),
which are similar to those in high-$z$ SFGs (black-filled circles in 
Fig.~\ref{fig3}d). A remarkable feature of these CSFGs with high EW(H$\beta$) and, 
thus, with high-excitation H~{\sc ii} regions is the 
absence of dependence on stellar mass because of very similar metallicities
(12 + logO/H $\sim$ 8.0 with a standard deviation of 0.1, Fig.~\ref{fig2}f). 
To justify this conclusion, 
we show in Fig.~\ref{fig3}d the most metal-poor nearby galaxies with 
12 + logO/H $\sim$ 6.9 -- 7.25 and similar high EW(H$\beta$)s (encircled filled
circles). These galaxies strongly deviate from the sequence of CSFGs with 
EW(H$\beta$) $>$ 100$\AA$, as in Figs.~\ref{fig1}a -- \ref{fig1}b.

\begin{figure*}
\centering
\hspace{0.0cm}\includegraphics[angle=-90,width=0.90\linewidth]{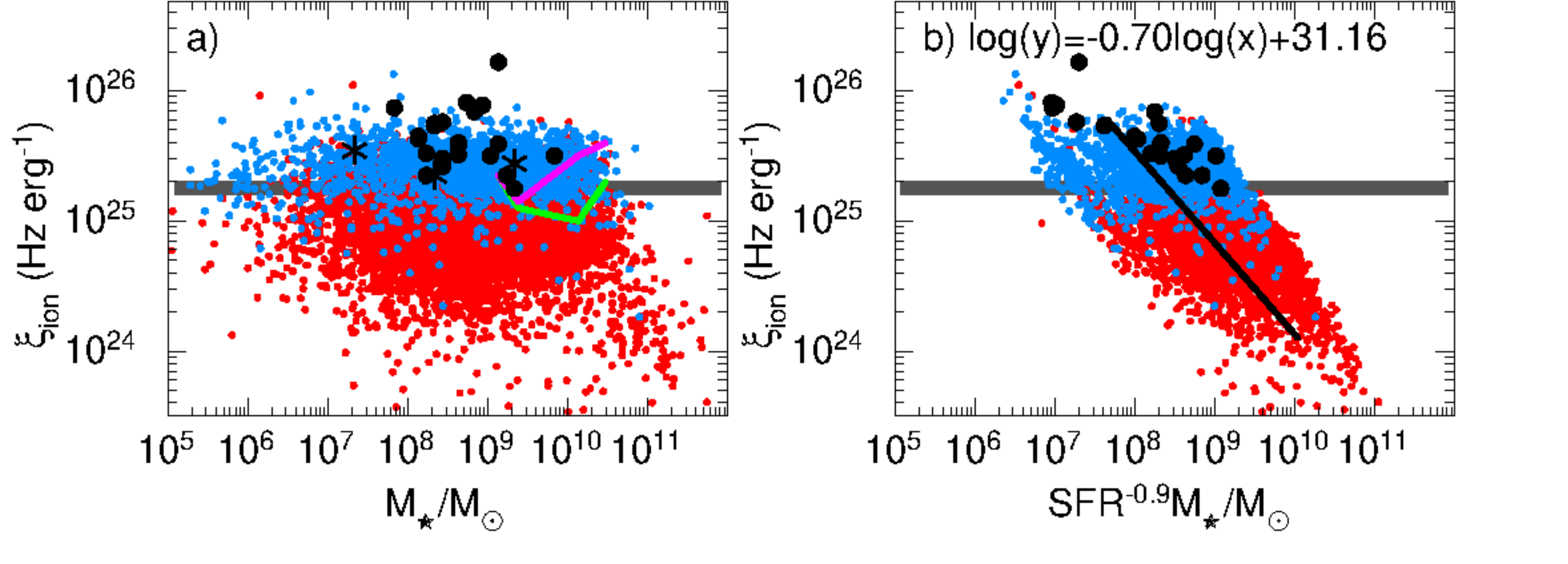}
\caption{Relations between ionising photon production 
efficiences $\xi_{\rm ion}$ and stellar masses $M_\star$ {\bf (a)}
and SFR$^{-0.9}$$M_\star$ {\bf (b)} for samples of CSFGs and high-$z$ 
SFGs. The solid line in {\bf (b)} is the maximum 
likelihood relation whereas stacks for $z$ $\sim$ 1.4 -- 2.6 SFGs \citep{Shi18}
for \citet{Ca94,Ca00} and SMC reddening laws are shown in {\bf (a)} 
by green and magenta 
lines, respectively. Mean $\xi_{\rm ion}$ values of $z$ = 4 -- 5 galaxies 
\citep{La19a} in the three stellar mass bins are shown in {\bf (a)}
by asterisks, whereas $z$ $\sim$ 6.6 SFGs with Ly$\alpha$ emission by
\citet{En20} are represented in both panels by filled circles.
Values of $\xi_{\rm ion}$ assumed in canonical Universe reionisation 
models are shown with a thick horizontal grey line \citep[e.g. ][]{Bo16}.
The meanings of symbols for the SDSS CSFGs are the same as in Fig.~\ref{fig1}.}
\label{fig11}
\end{figure*}

\subsection{Stellar mass and metallicity relation}

A number of studies \citep[e.g. ][]{Ma10,Cu14,Tr14,San15,On16,Cu20,Sa20b} have
revealed an offset of high-$z$ SFGs in the stellar mass-metallicity 
diagram to lower than 12 + logO/H, in the range of 0.15 -- 0.70 dex, when compared to SDSS 
galaxies at $z$~=~0. 
\citet{Tr14} have attributed these differences to 
prominent outflows and massive pristine gas inflows.

It was found in some of these studies that there is a second-order parameter 
in the stellar massmetallicity relation, namely the SFR, which introduces an additional scatter 
to the relation. We note in the introduction to this work that in \citet{Ma10}, 
by replacing the
parameter log~$M_\star$ with log~$M_\star$~--~$\alpha$~$\times$~log~SFR 
in the fundamental mass-metallicity relation, these authors found 
$\alpha$ = 0.32 for a minimum of the data dispersion, whereas \citet{Cu20} 
and \citet{Sa20} derived $\alpha$ = 0.55 and 0.63, respectively.

The relation $M_\star$ -- 12 + logO/H for CSFGs is represented in 
Fig.~\ref{fig4}a. The oxygen abundances of 12 + logO/H were derived either by the 
direct $T_{\rm e}$-method for galaxies, where the [O~{\sc iii}]~$\lambda$4363 emission
line was detected with an accuracy better than 4 $\sigma$, or by the strong-line
method using Eq.~\ref{sel2}, if the [N~{\sc ii}]~$\lambda$6584 emission line
is measured, or by using Eq.~\ref{sel3} otherwise. Furthermore, 
it is only for the mass-metallicity relation, among the galaxies 
with oxygen abundances derived by the strong-line method, that we excluded all 
galaxies with [O~{\sc iii}]~$\lambda$4959/H$\beta$~$<$~1 because 
Eqs.~\ref{sel2} and \ref{sel3} are derived using the method from \citet{Iz15}, namely, the direct 
$T_{\rm e}$-method, only for galaxies with 
[O~{\sc iii}]~$\lambda$4959/H$\beta$~$\ge$~1. 

The black line in Fig.~\ref{fig4}a shows the steep relation for $z$~=~0 SDSS SFGs 
derived by \citet{Ma10}. Our shallower relation for CSFGs is offset to lower
metallicities and extends to lower stellar masses. The difference is not due to 
a redshift effect as our galaxies are also low-redshift systems, with a major
fraction below $z$~=~0.5. It is likely, as suggested by \citet{Cu20}, that at a fixed 
stellar mass, the requirement of an [O~{\sc iii}]~$\lambda$4363 detection selects the most
metal-poor galaxies. However, our CSFG sample includes also
$\sim$~10,000 galaxies without a detected [O~{\sc iii}]~$\lambda$4363 emission
line. These galaxies follow
the same relation as those with a detected [O~{\sc iii}]~$\lambda$4363 line
\citep[fig. 9a in ][]{Iz15}. Possibly, the difference in the \citet{Ma10}
relation is caused by our selection of CSFGs with strong emission lines, whereas
the \citet{Ma10} sample includes 
many galaxies with weak emission lines and likely higher metallicities.
Additionally, different methods were used in different works for oxygen 
abundance determination \citep[e.g. see the discussion in ][]{Iz15} and the 
galaxies from our sample show a wider range of SFR than those in \citet{Ma10}, 
stressing the importance of this additional parameter in this relation. 

For the purposes of comparison, we also show data for $z$~$\sim$~2~--~7 galaxies in 
Fig.~\ref{fig4}. They are in general agreement with the data for
CSFGs but with a higher dispersion, presumably due to different methods used for oxygen
abundance determination and higher uncertainties in emission fluxes 
for high-$z$ galaxies.
To take into account the effect of the secondary parameter in the 
stellar mass -- metallicity relation, the SFR, we show 
in Fig.~\ref{fig4}b the fundamental metallicity 
relation 12 +log O/H -- SFR$^{-a}$$\times$($M_\star$/M$_\odot$), where $a$ = 0.5
was obtained by minimising the dispersion. The linear regression of 
this relation and its expression are shown in the Figure. The derived value of
$a$ for the CSFGs is very similar to $a$ = 0.55 obtained by \citet{Cu20} for SDSS 
galaxies.

\begin{figure*}
\centering
\hspace{0.0cm}\includegraphics[angle=-90,width=0.49\linewidth]{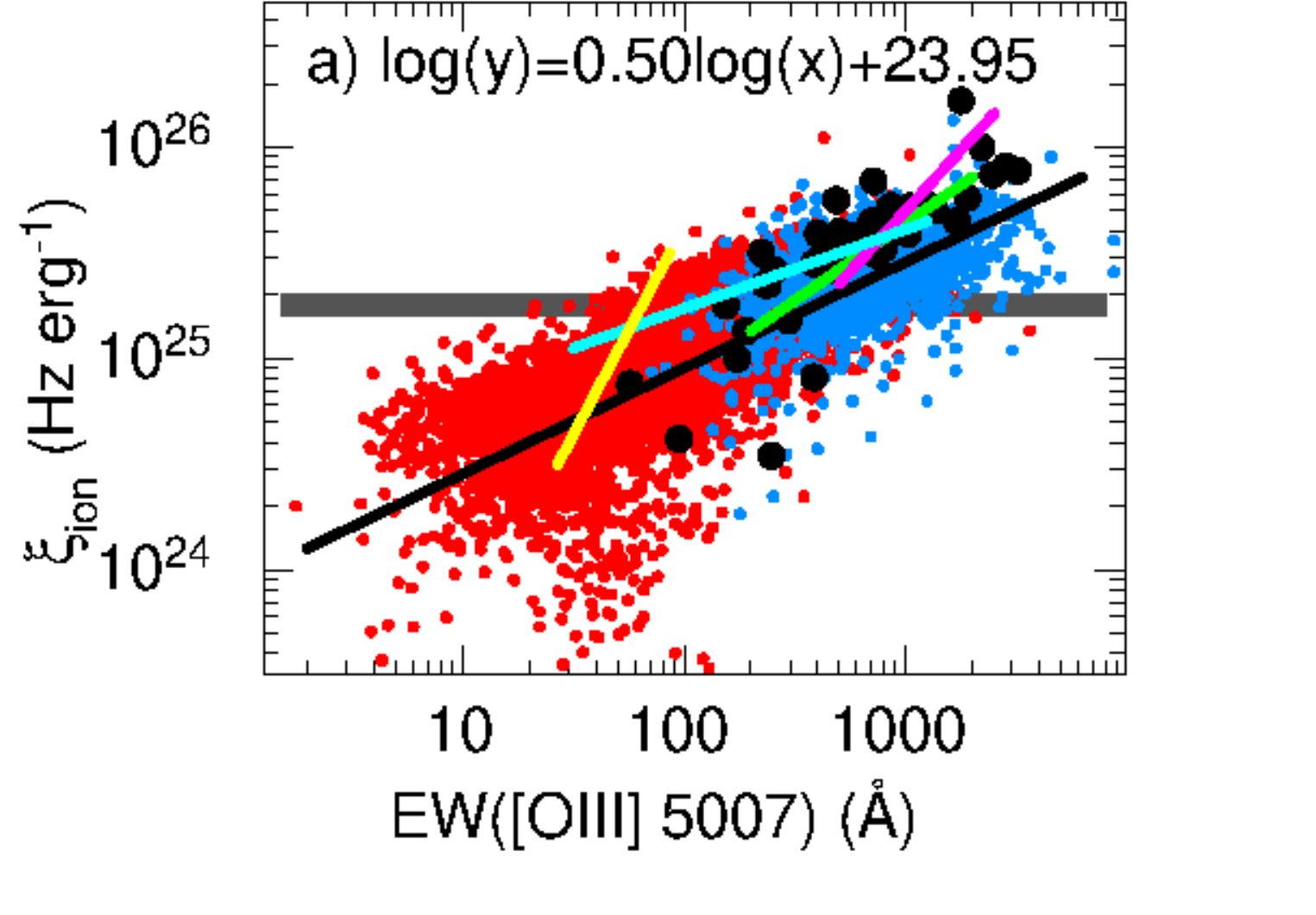}
\hspace{0.0cm}\includegraphics[angle=-90,width=0.49\linewidth]{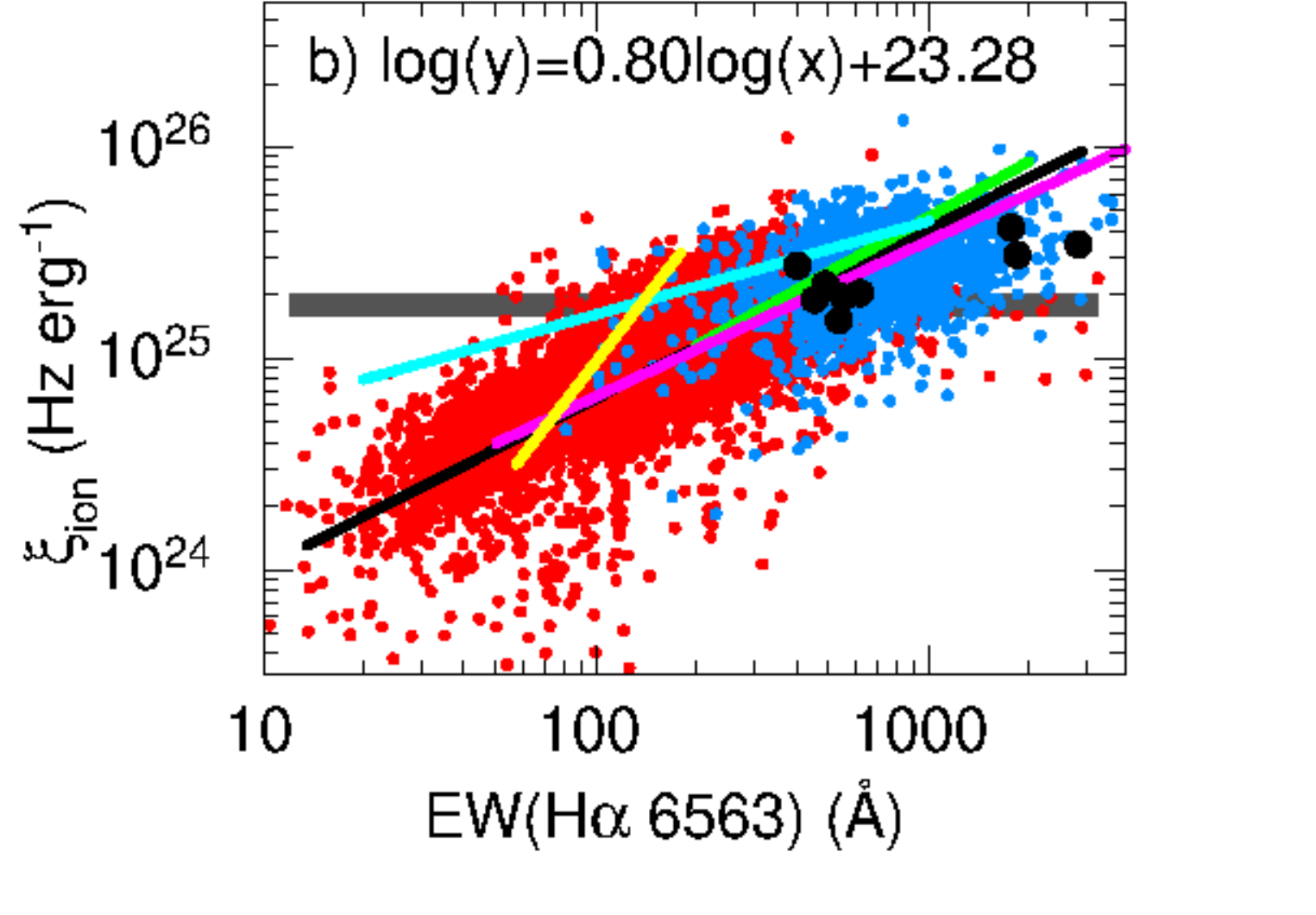}
\caption{{\bf (a)} and {\bf (b)} Relations between ionising photon production 
efficiences, $\xi_{\rm ion}$, and equivalent widths of the
[O~{\sc iii}]~$\lambda$5007 and H$\alpha$~$\lambda$6563 emission lines for 
samples of CSFGs and high-$z$ SFGs. The black solid lines 
are the maximum likelihood relations whereas green, 
magenta, cyan, and yellow solid lines in {\bf (a)} are the relations for
$z$ $\sim$ 1.3 -- 2.7 strong [O~{\sc iii}] emitters \citep{Ta19},
$z$ $\sim$ 0 SFGs with EW([O~{\sc iii}]) $>$ 1000$\AA$ \citep{Ch18},
$z$ $\sim$ 2 lensed galaxies \citep{Em20}, and $z$ $\sim$ 1.4 -- 3.8 galaxies
\citep{Re18}, respectively; LAEs at $z$ $\sim$ 3 \citep{Na20} 
and $z$ $\sim$ 6.6 SFGs with Ly$\alpha$ emission \citep{En20} are shown in 
{\bf (a)} by filled black circles. Mean $\xi_{\rm ion}$ values for 
$z$~=~4~--~5 galaxies \citep{La19a} are shown in {\bf (b)} by filled circles.
The lines in {\bf (b)} are the relations for
$z$ $\sim$ 1.3 -- 2.7 strong [O~{\sc iii}] emitters 
\citep[green line, ][]{Ta19}, $z$ $\sim$ 4 -- 6 SFGs 
\citep[magenta line, ][]{Fa19}, $z$ $\sim$ 2 lensed galaxies 
\citep[cyan line, ][]{Em20}, and $z$~$\sim$~1.4~--~3.8 
\citep[yellow line, ][]{Re18}.
Values for $\xi_{\rm ion}$ assumed in canonical Universe reionisation 
models are shown with a thick horizontal grey line \citep[e.g. ][]{Bo16}.
The meanings of symbols for SDSS CSFGs are the same as in Fig.~\ref{fig1}.}
\label{fig12}
\end{figure*}

\subsection{O$_{32}$, Ne3O2, stellar masses, and equivalent widths of emission 
lines}

We now consider the relations between stellar masses and emission
line properties for CSFGs and compare them with those for high-$z$ SFGs.
Fluxes, equivalent widths of strong emission lines 
(EW([O~{\sc ii}]~$\lambda$3727), and more commonly EW([O~{\sc iii}]~$\lambda$5007)
and EW(H$\alpha$)) and O$_{32}$ have been measured in spectra or are inferred 
from the photometry of many high-redshift galaxies.

In particular, \citet{Fa16} found that the [O~{\sc iii}]~$\lambda$5007/H$\beta$ 
ratio increases progressively out to $z$ $\sim$ 6, whereas \citet{Co18} 
concluded that extreme [O~{\sc iii}]~$\lambda$5007 emission may be a 
common early lifetime phase for star-forming galaxies at $z$~$>$~2.5.
\citet{Du20} also concluded that strong [O~{\sc iii}]~$\lambda$4959,5007
+ H$\beta$ emission appears to be typical in star-forming galaxies at 
$z$~$>$~6.5. They found that extreme Ly$\alpha$ emission starts to emerge at 
high EW([O~{\sc iii}]~$\lambda$5007) $>$ 1000$\AA$.

Using {\sl Spitzer} photometry, \citet{Ra16} derived a rest-frame EW(H$\alpha$+[N~{\sc ii}] + [S~{\sc ii}]) 
for $z$ = 5.1 - 5.4 galaxies of $\sim$~700$\AA$.
On the other hand, \citet{En21} studied the distribution of 
[O~{\sc iii}] + H$\beta$ line strengths at $z$ $\sim$ 7, using a sample of 22 
bright ($M_{\rm FUV}$ $<$ --21 mag) galaxies, and derived a median EW = 692$\AA$
for the sum of these emission lines.
\citet{Fa19} found a tentative anticorrelation between EW(H$\alpha$) and stellar
mass, ranging from 1000$\AA$\ at log($M_\star$/M$_\odot$) $<$ 10 to below 
100$\AA$\ at log($M_\star$/M$_\odot$) $>$ 11. \citet{Tr20} similarly concluded
that H$\beta$+[O {\sc iii}] rest-frame equivalent widths in 
$z$ $\sim$ 3 -- 4 galaxies
tend to be higher in lower-mass systems. They also suggested that strong 
[O~{\sc iii}]$\lambda$5007 emission signals an early episode of intense stellar 
growth in low-mass galaxies and many, if not most galaxies at $z$ $>$ 3 go 
through this starburst phase.

\begin{figure*}
\centering
\hspace{0.0cm}\includegraphics[angle=-90,width=0.4\linewidth]{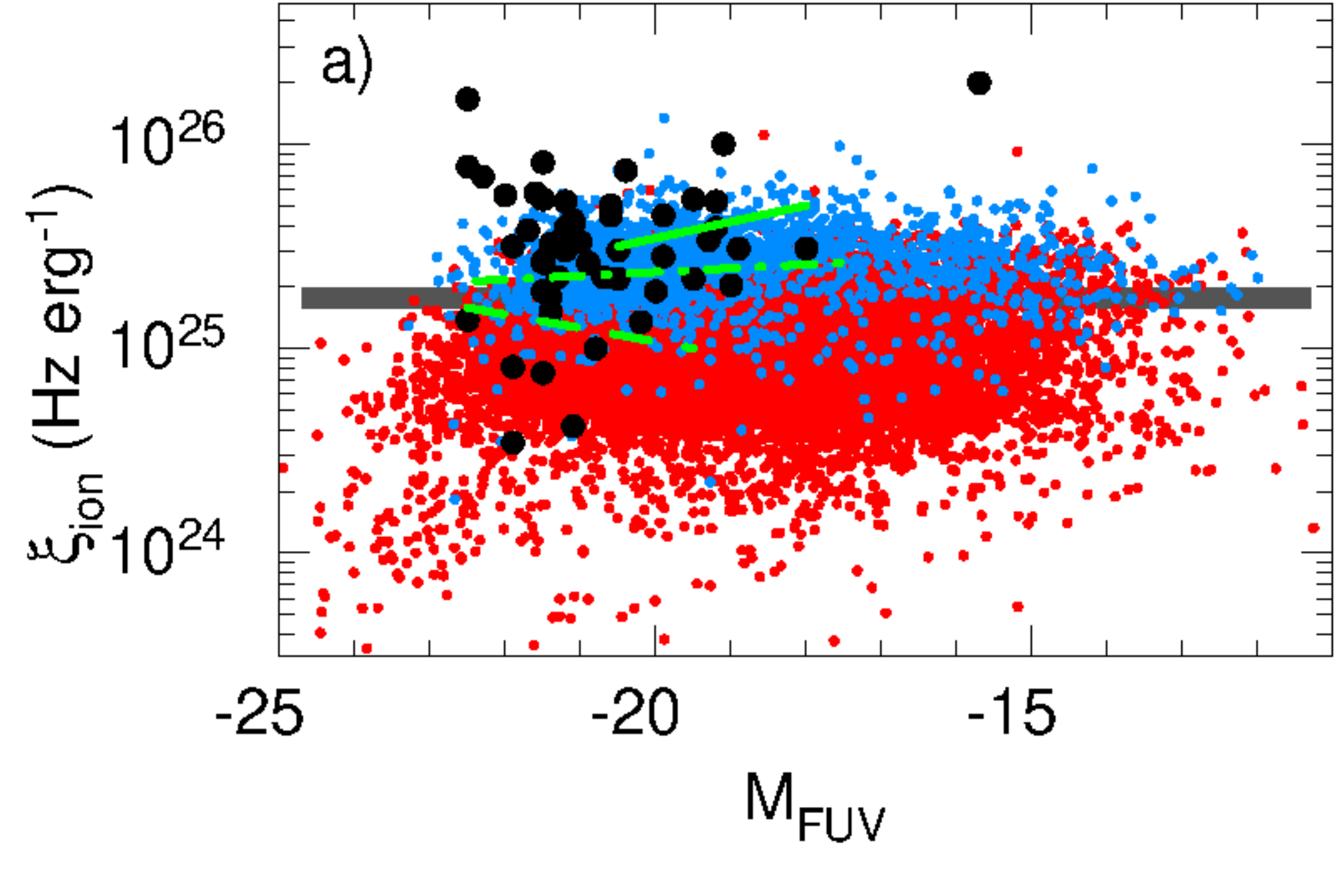}
\hspace{0.0cm}\includegraphics[angle=-90,width=0.4\linewidth]{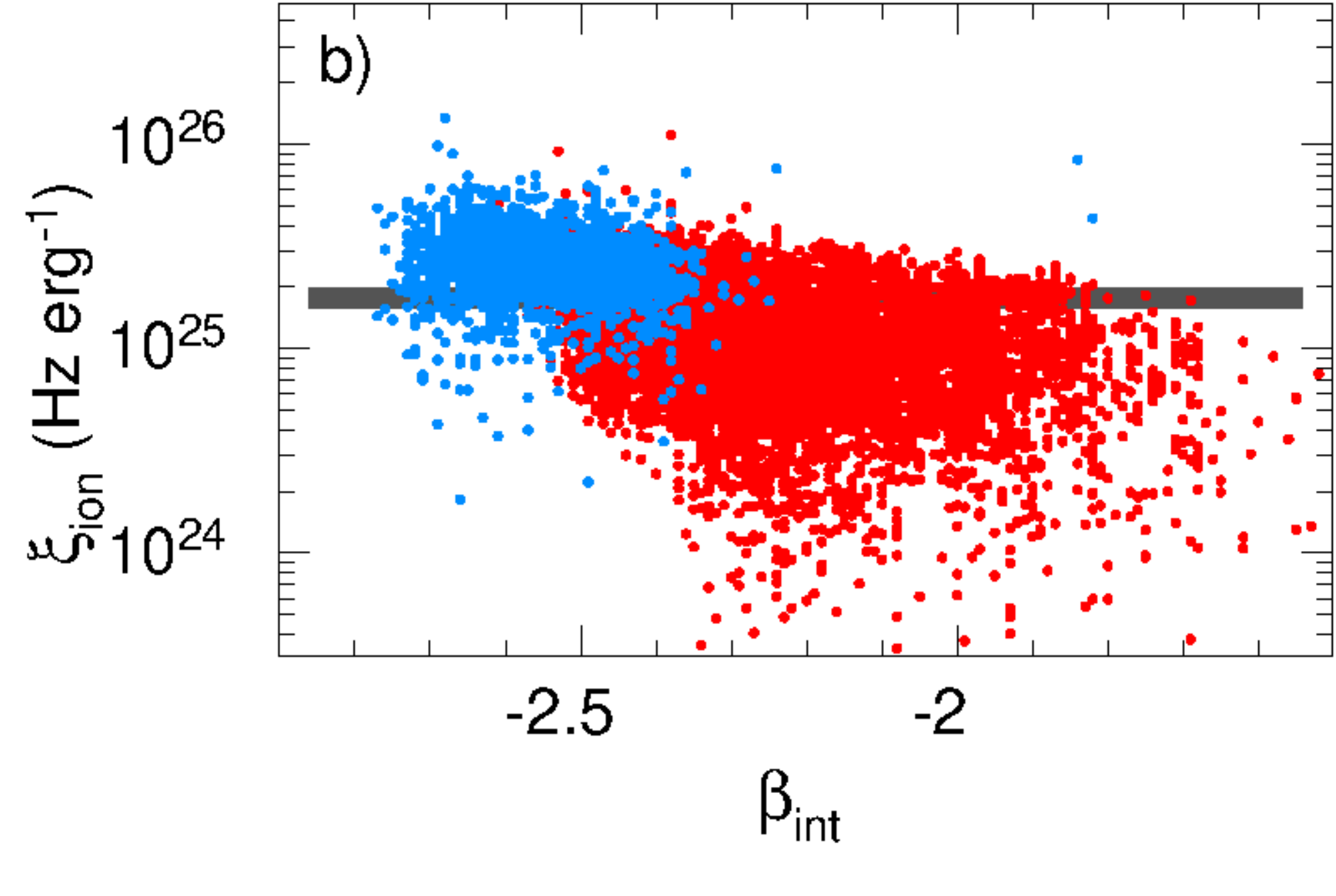}
\hspace{0.0cm}\includegraphics[angle=-90,width=0.4\linewidth]{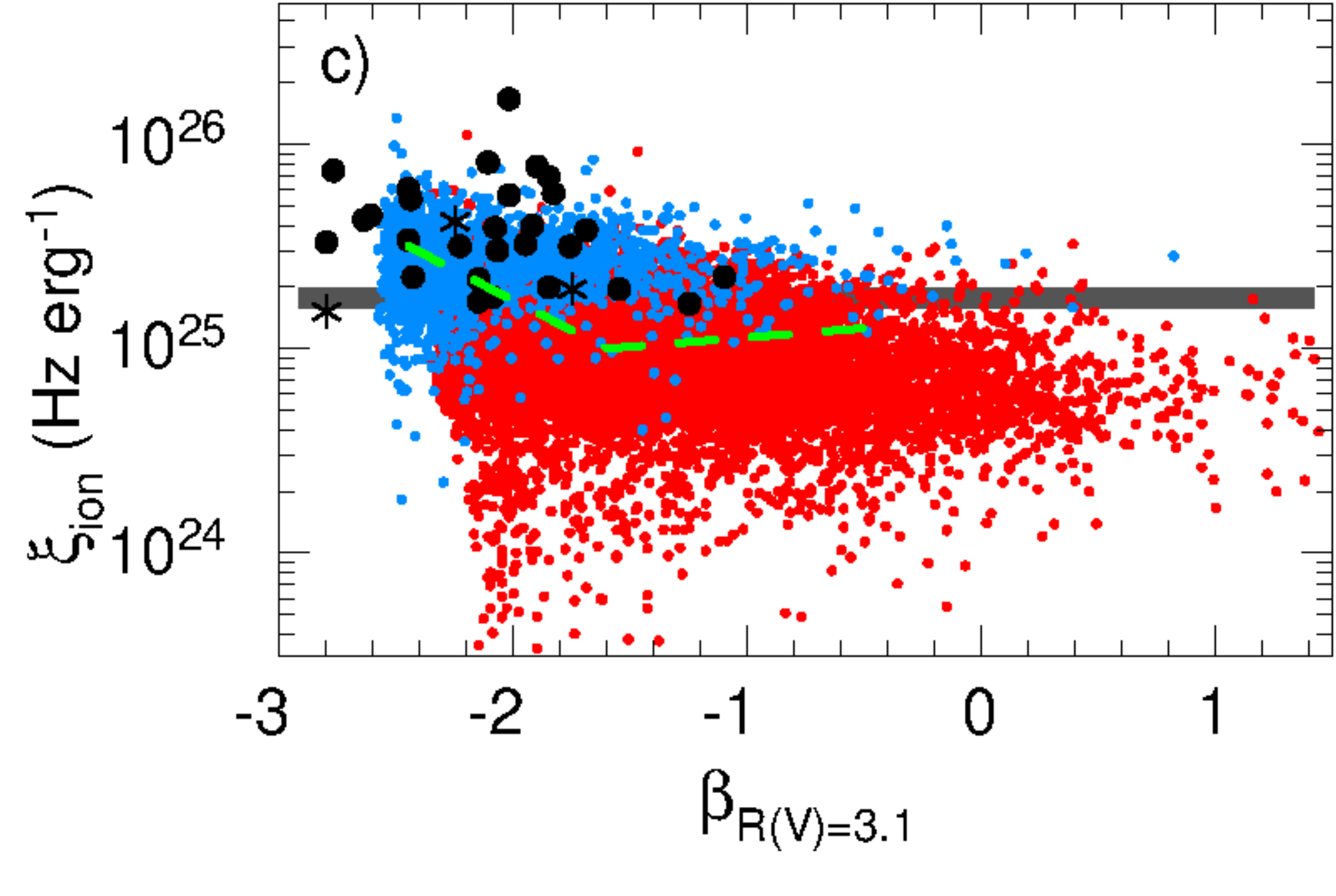}
\caption{{\bf (a)} Dependence of the ionising photon production 
efficiency $\xi_{\rm ion}$ on the UV absolute magnitude $M_{\rm FUV}$. 
The LAEs at $z$ $\sim$ 3 \citep{Na20}, $z$ $\sim$ 6.6 SFGs with Ly$\alpha$ 
emission \citep{En20}, mean $\xi_{\rm ion}$ values for 
$z$ = 3.8 -- 5.4 galaxies \citep{Bo16}, mean $\xi_{\rm ion}$ values for 
$z$ = 4 -- 5 galaxies \citep{La19a}, and a mean $\xi_{\rm ion}$ value for 
$z$ = 3.8 -- 5.4 faint galaxies with high EW(Ly$\alpha$) \citep{Ma20} are shown 
by filled circles.
The relation for $\sim$ 2 galaxies \citep{Shi18}, for $z$ $\sim$ 9 -- 10 galaxies
\citep{Bo19} and for $z$ $\sim$ 3 faint Ly$\alpha$ emitters \citep{Na18} are
shown by a green dashed, dash-dotted and solid lines, respectively.
{\bf (b)} Dependence of the ionising photon production efficiency $\xi_{\rm ion}$ 
on the intrinsic UV slope $\beta$.
{\bf (c)} Dependence of the ionising photon production efficiency 
$\xi_{\rm ion}$ on the intrinsic UV slope $\beta$ reddened with $R(V)$ = 3.1. 
Filled circles are $z$ $\sim$ 3.8 -- 5.4 galaxies \citep{Bo16} and 
$z$ $\sim$ 6.6 SFGs with Ly$\alpha$ emission \citep{En20}, asterisks are 
mean $\xi_{\rm ion}$ values of $z$ = 4 -- 5 galaxies \citep{La19a}.
The relation for $z$ $\sim$ 2 galaxies \citep{Shi18} is shown by a green dashed line.
Values of $\xi_{\rm ion}$ assumed in canonical Universe reionisation 
models are shown (as in Figs.~\ref{fig11} and \ref{fig12} 
in grey \citep[e.g. ][]{Bo16}.
The meanings of symbols for the SDSS CSFGs are the same as in Fig.~\ref{fig1}.}
\label{fig13}
\end{figure*}

\citet{PA18} studied a large sample of LAEs in the redshift range of 
$\sim$ 2 to 6. They found that LAEs with the highest rest-frame 
equivalent widths are the smallest and most compact galaxies. Finally,
it was suggested that SFGs with the strongest emission lines are characterised 
by high O$_{32}$ ratios and high sSFRs, and that they may, in fact, be the main
contributors to the reionisation process of the Universe \citep[e.g. ][]{Na20}.

In Fig.~\ref{fig5}a, we show the relation between the stellar mass and the O$_{32}$
ratio. O$_{32}$ at fixed $M_\star$ is higher for CSFGs with high EW(H$\beta$),
as expected, because the ionisation parameter is higher in younger starbursts.
High-$z$ galaxies in the figure (black symbols and green line) have mainly 
O$_{32}$ $>$ 1 and are located in the region of massive CSFGs 
($M_\star$ $>$ 10$^8$ M$_\odot$), with high EW(H$\beta$)~$\ge$~100$\AA$.
However, no low stellar mass galaxies are present in the current high-$z$ 
samples.

The O$_{32}$ ratio for CSFGs slowly increases with decreasing $M_\star$,
attaining O$_{32}$ $>$ 10 in a significant fraction of galaxies with $M_\star$~$<$~10$^7$~M$_\odot$,
although such high O$_{32}$ are present in CSFGs with stellar masses of up
to 10$^{9}$~M$_{\odot}$.
This behaviour implies a dependence of the relation on a second 
parameter, namely, on the SFR. 
Introducing the parameter SFR$^{-a}$$\times$($M_\star$/M$_\odot$) instead of the 
stellar mass alone, we show in Fig.~\ref{fig5}b the fundamental relation with 
$a$ = 0.9, where $a$ is obtained by minimising the scatter in the data.
The linear regression of the fundamental relation is shown by a solid line.
We note that adopting $a$ = 1 would correspond to the relation 
O$_{32}$ -- sSFR$^{-1}$.

The relations between stellar mass and equivalent widths 
EW([O~{\sc ii}]~$\lambda$3727), EW([O~{\sc iii}]~$\lambda$5007), and
EW(H$\alpha$) are shown in Fig.~\ref{fig6}. As for high-$z$ galaxies, the data 
are available 
mainly for [O~{\sc iii}]~$\lambda$5007 (black symbols in Fig.~\ref{fig6}c,d)
and, to a lesser extent for H$\alpha$, but not for 
[O~{\sc ii}]~$\lambda$3727. 

It is notable that there is weak correlation between EWs and the stellar mass  
for the entire CSFG sample (blue and red dots in Figs.~\ref{fig6}a,c,e), 
whereas it is tight for high-$z$ galaxies in \ref{fig6}c.
This can be explained by the lack of galaxies with low SFRs in 
the sample of high-$z$ SFGs.
Figs.~\ref{fig6}b,d,f represent the fundamental relations of EW -- 
SFR$^{-a}$$\times$($M_\star$/M$_\odot$), with $a$ = 0.9,
indicating that EWs increase with sSFRs while bearing in mind that the relation 
transforms to O$_{32}$ -- sSFR$^{-1}$ if $a$ is set to 1.0.

A comparison of O$_{32}$ -- $M_\star$ and EW -- $M_\star$ relations
(Figs.~\ref{fig5} and \ref{fig6})
indicates that they behave in a similar manner, namely, that both O$_{32}$ and EWs 
decrease with increasing $M_\star$. Moreover, both quantities 
increase with decreasing starburst age. Therefore, it is expected that 
they would tightly correlate. 
We show in 
Fig.~\ref{fig7}a the O$_{32}$ - EW([O~{\sc iii}]~$\lambda$5007) relation for
CSFGs, which indeed reveals  a tight correlation. We note
that the distributions of high-$z$ galaxies are in agreement with this relation.
The relation in Fig.~\ref{fig7}a can be used, for example, to estimate the
O$_{32}$ ratio and the [O~{\sc ii}]~$\lambda$3727 line flux, if the 
characteristics of the [O~{\sc iii}]$~\lambda$5007 line (flux and 
equivalent width) are known.

In Fig.~\ref{fig7}b, we present the relation between the ratio 
Ne3O2 = [Ne~{\sc iii}]~$\lambda$3868 / [O~{\sc ii}]~$\lambda$3727 and
EW([O~{\sc iii}]~$\lambda$5007) for $\sim$~9500 CSFGs with detected 
[O~{\sc ii}]~$\lambda$3727 and [Ne~{\sc iii}]~$\lambda$3868 emission lines.
This relation can be considered as an alternative 
to the relation in Fig.~\ref{fig7}a. Combining the data from Fig.~\ref{fig7}a and 
Fig.~\ref{fig7}b, we find Ne3O2 = 0.089~$\times$~O$_{32}$. Although
the [Ne~{\sc iii}]~$\lambda$3868 emission line is around ten times weaker than the 
[O~{\sc iii}]~$\lambda$5007 emission line, the relation in 
Fig.~\ref{fig7}b can be useful because the Ne3O2 ratio is almost independent on 
uncertainties in extinction at variance to the O$_{32}$ ratio.

\citet{Je20} considered the properties of ionised neon emission in z $\sim$ 2 
SFGs drawn from the MOSFIRE Deep Evolution Field (MOSDEF) survey. They found a  
Ne3O2 anticorrelation with 
stellar mass and a considerable offset of Ne3O2 to higher values compared
to $z$ $\sim$ 0 main-sequence galaxies. 
In Fig.~\ref{fig8}a, we compare the location of CSFGs and $z$ $\sim$ 2 MOSDEF
galaxies on the Ne3O2 -- $M_\star$ diagram. We find that CSFGs with 
EW(H$\beta$)~$\ge$~100$\AA$\ and $M_\star$ $>$ 10$^8$ M$_\odot$ occupy the same
region in the diagram as the MOSDEF galaxies (black filled circles). However, 
they extend
to much lower stellar masses, down to 10$^6$~M$_\odot$, and show an 
anticorrelation with $M_\star$ down to 10$^7$~M$_\odot$. 
These facts imply that the ionising radiation
in CSFGs and in MOSDEF SFGs, with stellar masses $M_\star$~$>$~10$^8$~M$_\odot$, 
shares the same properties.

\citet{Je20} also found that Ne3O2 correlates positively with O$_{32}$ in 
$z$ $\sim$ 2 galaxies. Their data at a fixed O$_{32}$ is offset towards 
higher Ne3O2 when compared with local SFGs. They concluded that the ionising 
spectrum in $z$ $\sim$ 2 SFGs is harder compared to $z$ $\sim$ 0 galaxies
resulting in stronger [Ne~{\sc iii}]~$\lambda$3868 emission of Ne$^{2+}$ with 
a high ionisation potential.
In Fig.~\ref{fig8}b, we show the  relation Ne3O2 -- O$_{32}$ for CSFGs (blue and
red dots) and $z$ $\sim$ 2 MOSDEF SFGs (black-filled circles). 
Both Ne3O2 and O$_{32}$ were proposed by \citet{LR14} as estimators for the ionisation
parameter $U$ in local SFGs. Using their relations, we find that 
log $U$ in CSFGs (Fig.~\ref{fig8}b) to be in the range of --3.5 to --1.5. We compare the distributions in log $U$ of the MOSDEF and CSFGs samples
for O$_{32}$ $>$ 1,  and find they are similar, again indicating close properties of ionising radiation.
However, there is a considerable difference between galaxies with 
low-excitation H~{\sc ii} regions, characterised by O$_{32}$ $<$ 1. CSFGs
follow the relation Ne3O2~$\sim$~0.1~O$_{32}$, which is expected because the Ne
abundance is around five to six times lower than that of oxygen \citep[e.g. ][]{I06}.
On the other hand, the 
Ne3O2/O$_{32}$ ratios are $\sim$~1 in the MOSDEF galaxies with the lowest 
O$_{32}$ $\la$ 0.3.
Such high ratios are difficult to explain through stellar ionising radiation alone, even if 
adopting a very low stellar metallicity \citep{Je20}. However, 
the Ne3O2/O$_{32}$ ratios in MOSDEF stacked spectra are very similar to those 
of CSFGs.

The diagrams depicting the SFR versus 
EW([O~{\sc iii}]~$\lambda$5007) and EW(H$\alpha$) for
the entire sample are shown in Figs.~\ref{fig9}a and \ref{fig9}c, 
respectively. From these Figures, it is clear that correlations between these 
quantities for CSFGs, as well as for the high-$z$ SFGs that mainly populate the 
region of CSFGs with high EW(H$\beta$)s, are weak. On the other hand, tight fundamental relations are at play between $M_\star^{-a}$$\times$SFR and 
EW([O~{\sc iii}]~$\lambda$5007) and EW(H$\alpha$)
(Figs.~\ref{fig9}b and \ref{fig9}d), with a high $a$~=~0.9. 
This indicates a strong
dependence on the stellar mass $M_\star$. We note that a fraction of CSFGs 
with EW(H$\beta$)~$\ge$~100$\AA$\ (in Figs.~\ref{fig9}c,d) have 
EW(H$\alpha$)~$\la$~300$\AA$\ because the H$\alpha$ lines in the spectra of 
these galaxies are clipped.

The relations in Figs.~\ref{fig9}b and \ref{fig9}d with a $a$ = 1.0 would
correspond to sSFR -- EW([O~{\sc iii}]~$\lambda$5007) and 
sSFR~--~EW(H$\alpha$) relations, which we consider in 
Figs.~\ref{fig10}a and \ref{fig10}c. 
The fundamental relations in Figs.~\ref{fig10}b and 
\ref{fig10}d only weakly depend on the secondary parameter $M_\star$, 
because of the small $a$~=~--0.1.

\begin{figure*}
\hbox{
\hspace{0.0cm}\includegraphics[angle=-90,width=0.41\linewidth]{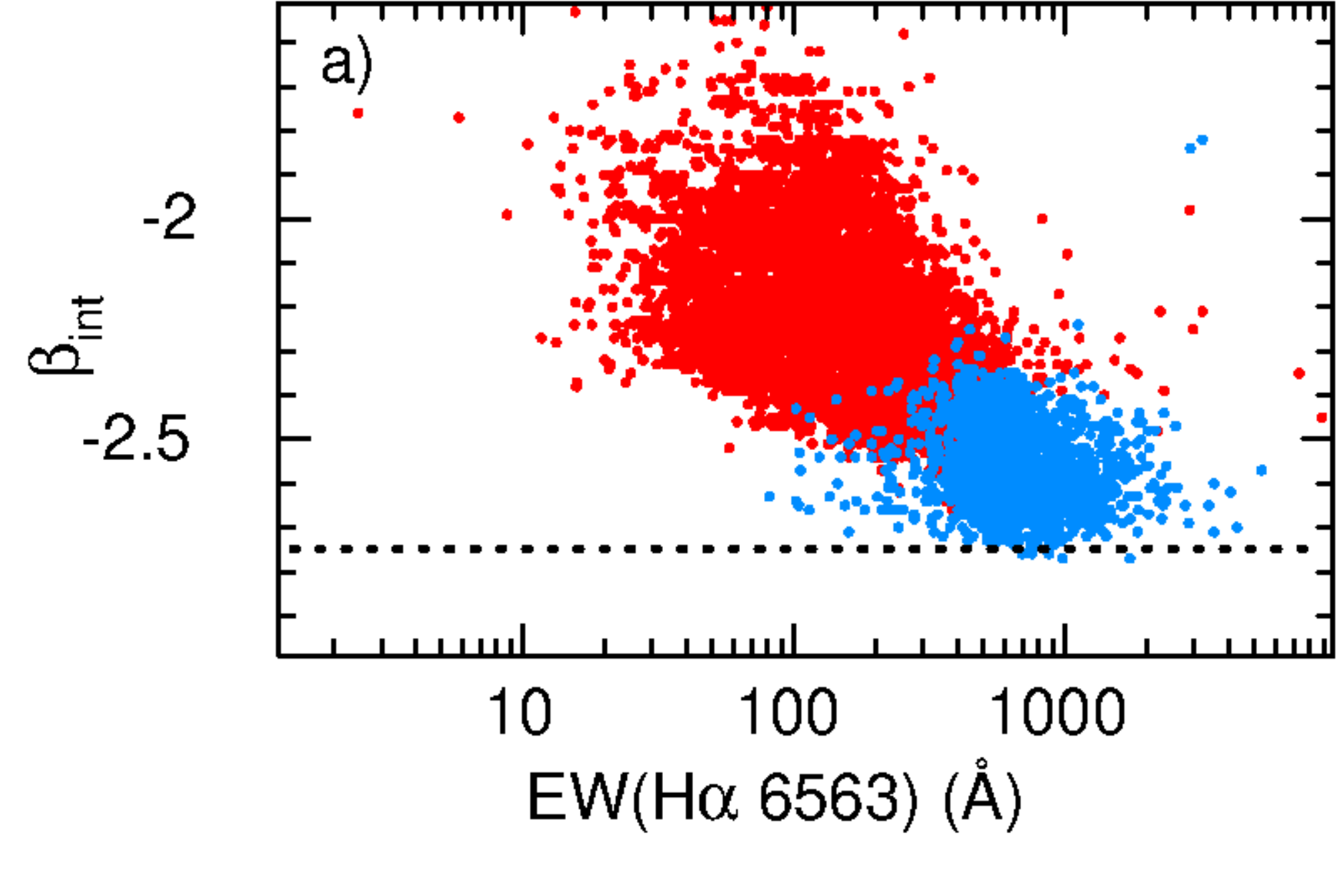}
\hspace{0.0cm}\includegraphics[angle=-90,width=0.41\linewidth]{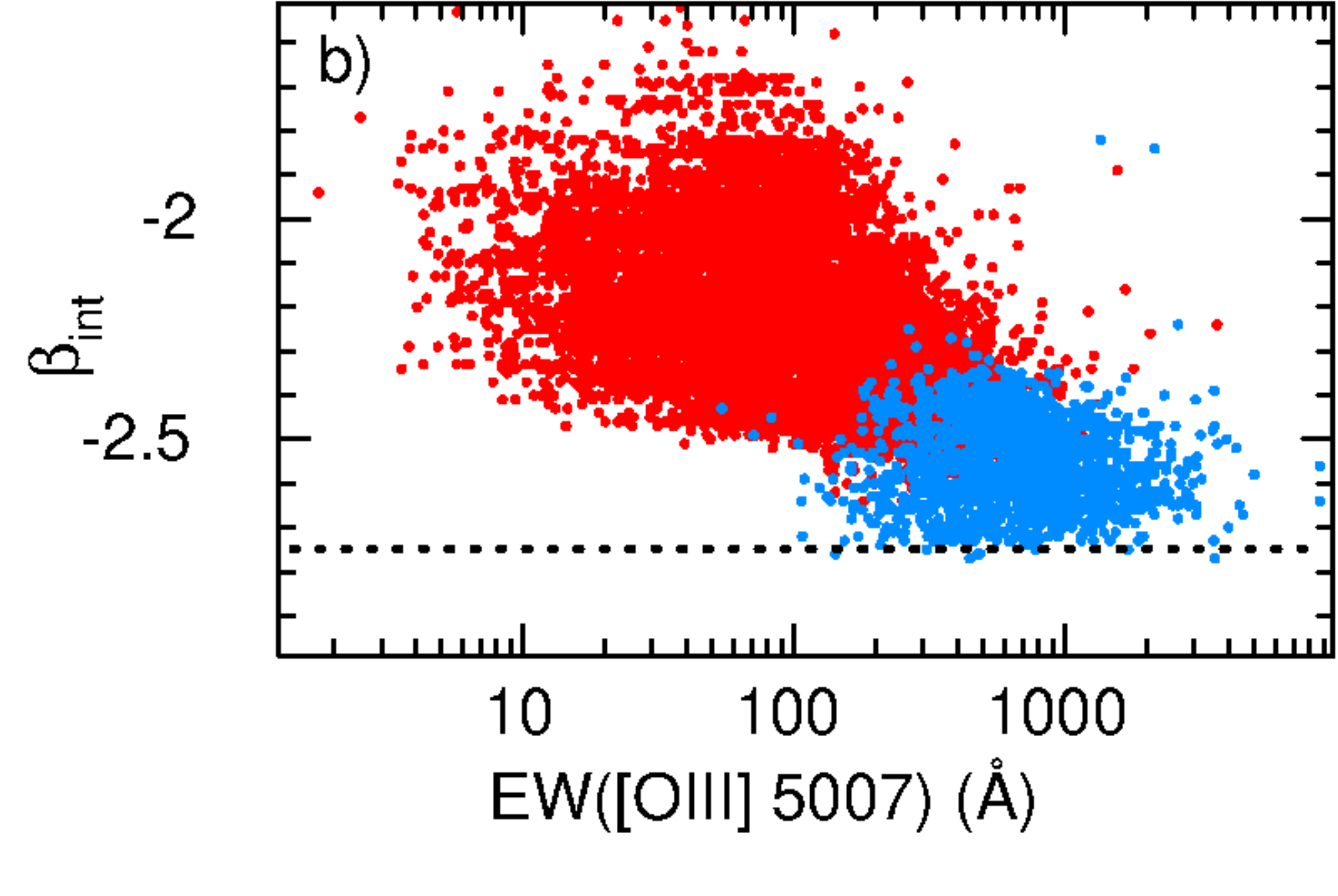}
}
\hbox{
\hspace{0.0cm}\includegraphics[angle=-90,width=0.41\linewidth]{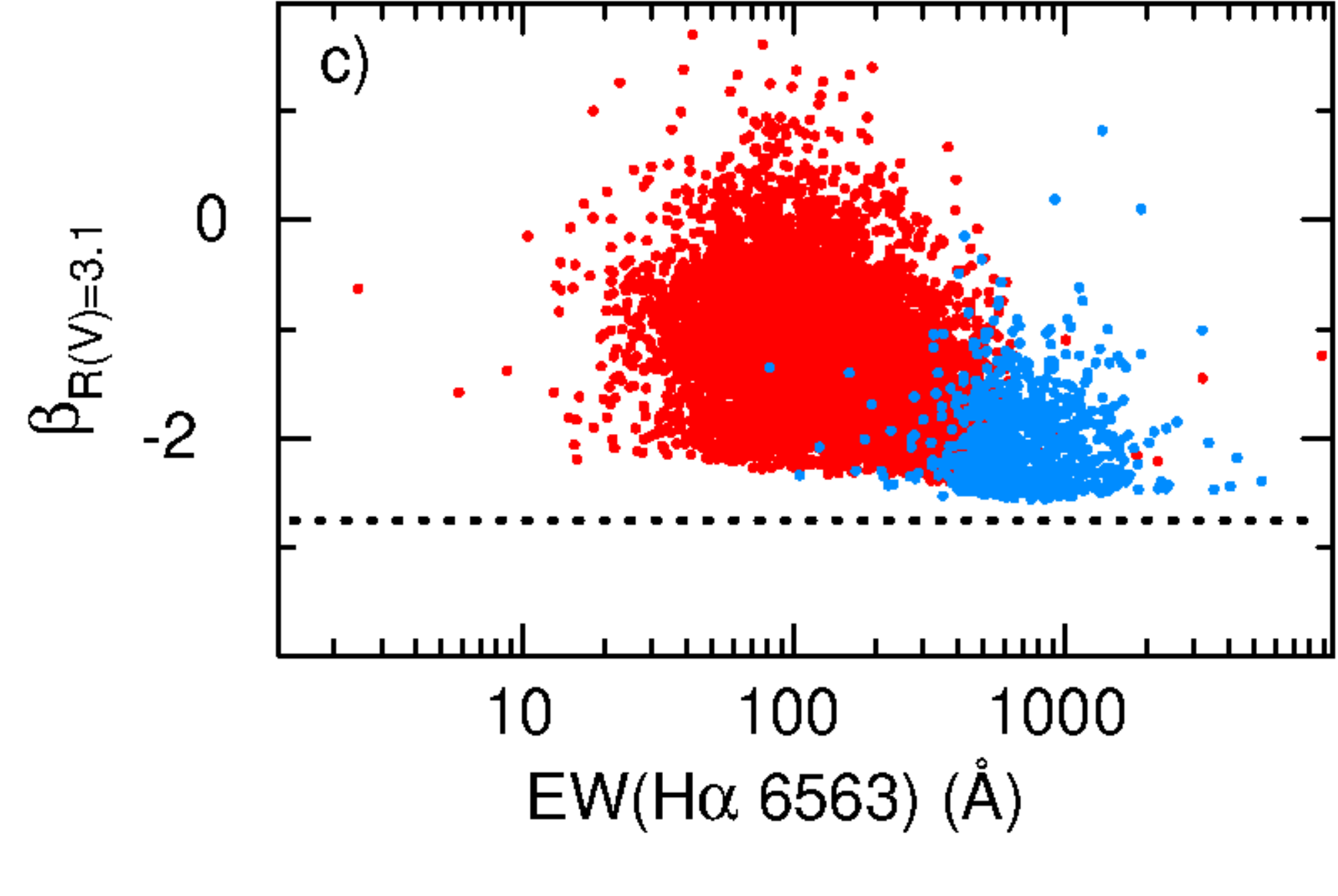}
\hspace{0.0cm}\includegraphics[angle=-90,width=0.41\linewidth]{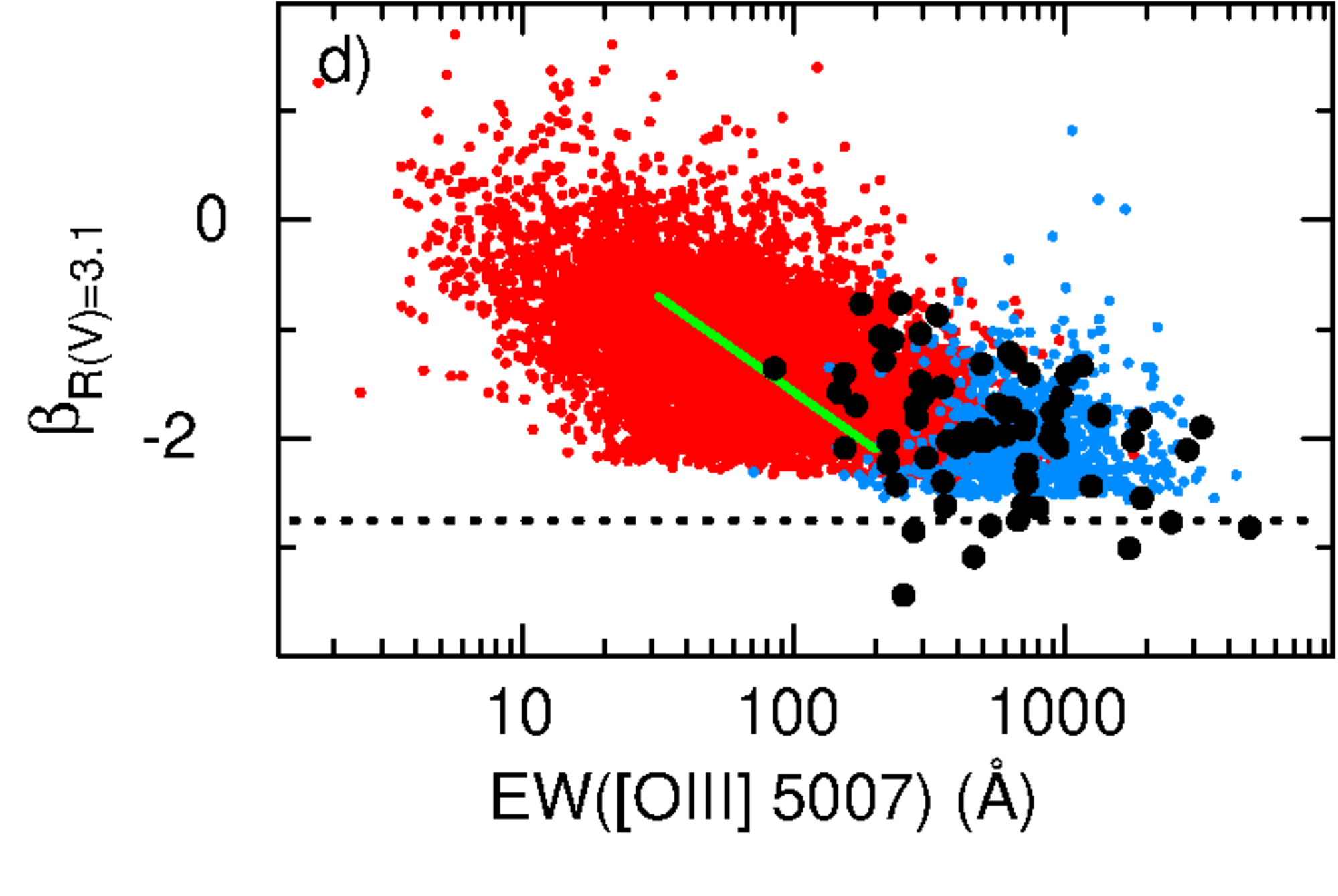}
}
\hbox{
\hspace{0.0cm}\includegraphics[angle=-90,width=0.41\linewidth]{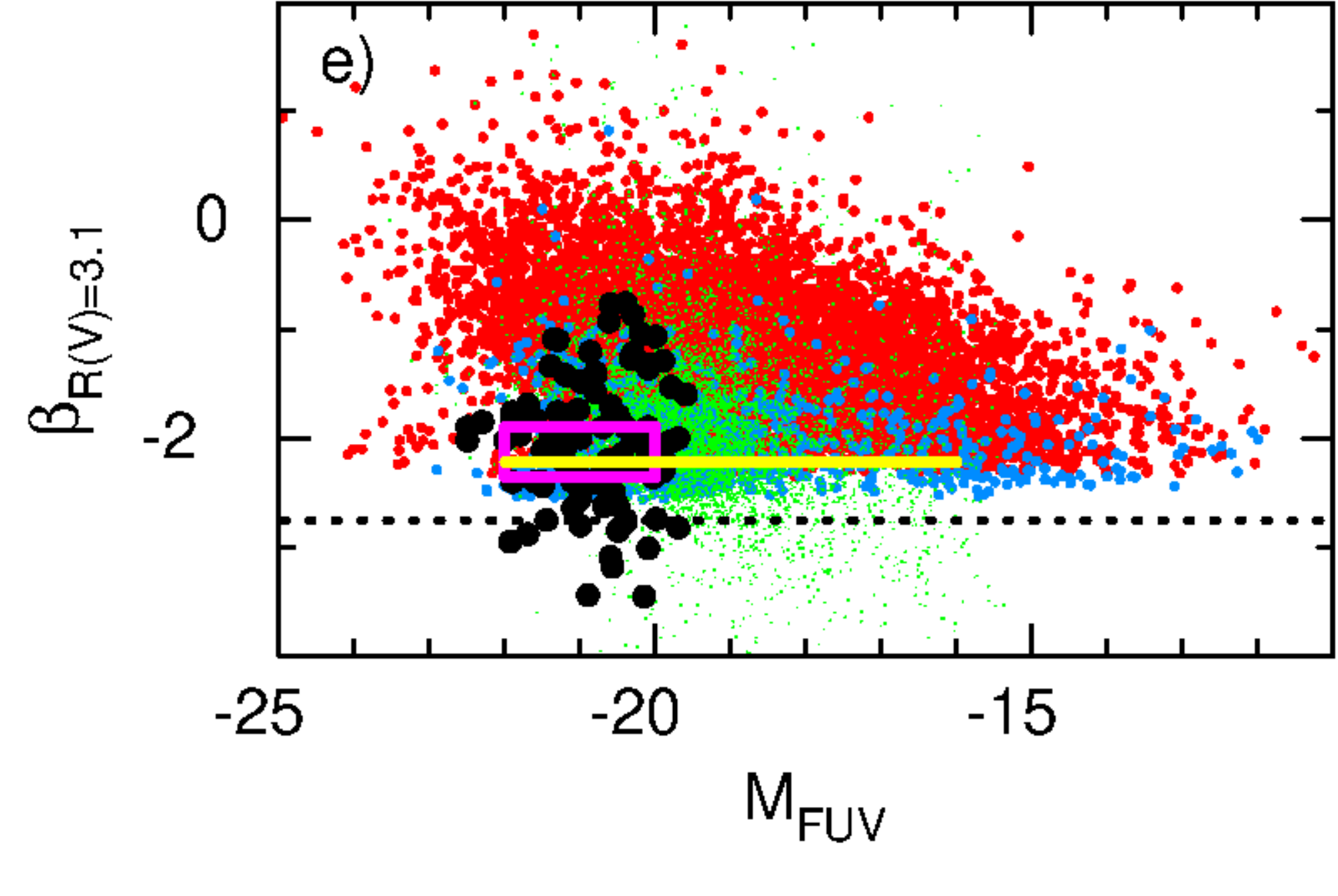}
\hspace{0.0cm}\includegraphics[angle=-90,width=0.41\linewidth]{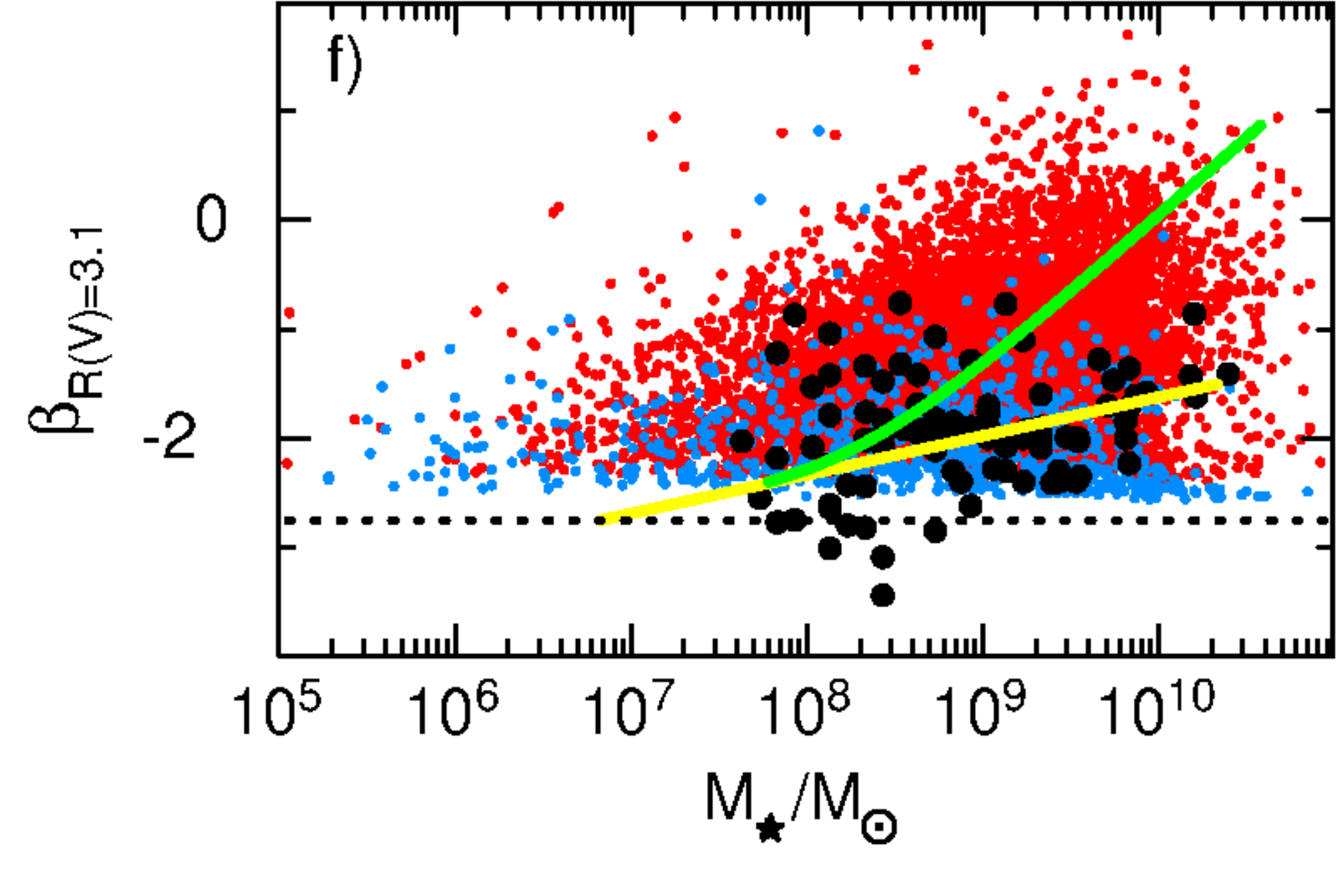}
}
\caption{Dependences of the intrinsic (i.e. with zero extinction) UV SED slopes 
$\beta_{\rm int}$ on {\bf (a)} the H$\alpha$ equivalent widths and 
{\bf (b)} the [O~{\sc iii}]~$\lambda$5007 equivalent widths; {\bf (c)} and {\bf (d)}
are the same as {\bf (a)} and {\bf (b)}, but for the SEDs reddened with the 
extinction coefficients derived from the hydrogen Balmer decrements and
assuming the \citet{C89} reddening law with $R(V)$ = 3.1. The green line in 
{\bf (d)} is the relation for $z$~=~1.4~--~3.8 SFGs \citep{Re18} 
whereas $z$~=~3.5 LBGs \citep{Ho16}, $z$~$\sim$~7 galaxies with high 
EW([O~{\sc iii}]~+~H$\beta$) \citep{En21} and $z$~$\sim$~6.6 SFGs with 
Ly$\alpha$ emission \citep{En20} are shown by filled circles. 
{\bf (e)} Dependence of 
$\beta_{R(V)=3.1}$ on the absolute FUV magnitudes $M_{\rm FUV}$ calculated using the
intrinsic SEDs. Green dots are for $z$~$\sim$~4~--~8 SFGs \citep{Bo12,Bo14},
black-filled circles are for 
$z$~$\sim$~6 SFGs \citep{Pe18}, for $z$~$\sim$~6.6 SFGs with Ly$\alpha$ 
emission \citep{En20}, for median values of $z$~$\sim$~2~--~5 LAEs 
\citep{San20}, for $z$~$\sim$~7 SFGs with high EW([O~{\sc iii}]~+~H$\beta$) 
\citep{En21} and for $z$~=~5.7~--~6.6 LAEs with extremely steep UV continua
\citep{Ji20}. The 
magenta rectangle delineates the location of stacks for $z$~$\sim$~10 SFGs 
\citep{Wi16}. {\bf (f)} Dependence of $\beta_{R(V)=3.1}$ on stellar masses 
$M_\star$. Black-filled circles are 
for $z$~$\sim$~3.5 LBGs \citep{Ho16}, for median values of $z$~$\sim$~2~--~5 
LAEs \citep{San20}, for $z$ $\sim$ 7 SFGs with high EW([O~{\sc iii}]+H$\beta$) 
\citep{En21}, for $z$~$\sim$~6.6 SFGs with Ly$\alpha$ emission \citep{En20}
and for stacks of SFGs at $z$~$\sim$~4~--~6 \citep{Fu20}, 
respectively. Yellow solid lines in {\bf (e)} and {\bf (f)} represent 
relations for $z$~=~6~--~9 galaxies \citep{BC20}, whereas the green solid line
in {\bf (f)} is the relation for $z$~=~1.5~-~3.5 LBGs by \citet{Bo20}. 
Dotted horizontal 
lines in all panels indicate the lowest modelled intrinsic $\beta_{\rm int}$ 
of --2.75. The meanings of symbols for SDSS CSFGss in all panels are the same as 
in Fig.~\ref{fig1}.}
\label{fig14}
\end{figure*}

\subsection{Relations including the ionising photon production efficiency 
$\xi_{\rm ion}$}

Together with the escape fraction of ionising photons from galaxies, 
the ionising photon production efficiency $\xi_{\rm ion}$ is one of the 
important parameters characterising the ability of high-$z$ SFGs to ionise the
intergalactic medium during the epoch of reionisation at $z$ $\ga$ 6. 

The ionising photon production efficiency $\xi_{\rm ion}$ has been derived
in several studies for high-$z$ SFGs, with 
log($\xi_{\rm ion}$/[Hz erg$^{-1}$])
values in the range of $\sim$ 25.1 -- 25.8 \citep{Bo16,Ha18,DB19,Shi18,Em20}.
On the other hand, \citet{Ma20} used a sample
of 35 $z$ $\sim$ 4 -- 5 continuum-faint LAEs and measured a very high
log($\xi_{\rm ion}$/[Hz erg$^{-1}$]) = 26.28, implying a more efficient production
of ionising photons in lower-luminosity Ly$\alpha$ selected galaxies, possibly
produced by extremely low-metallicity stellar populations in very young
starbursts. Thus, in most of these studies, the derived $\xi_{\rm ion}$s are above
the canonical value required for reionising the early Universe and marked in Figs.~\ref{fig11} - \ref{fig13} by a broad grey 
horizontal bar \citep{Bo16}.

To better understand the contribution of dwarf galaxies to the ionising 
background and reionisation, \citet{Em20} measured the $\xi_{\rm ion}$ of low-mass 
galaxies (10$^{7.8}$ -- 10$^{9.8}$M$_\odot$) in the redshift range of $1.4<z<2.7$. They do not find any 
strong dependence of log($\xi_{\rm ion}$) on stellar mass, far-UV 
magnitude, or UV spectral slope, whereas \citet{Na18} show that $\xi_{\rm ion}$
increases for fainter objects in $z$ $\sim$ 3 LAEs and \citet{Shi18} derive
that $\xi_{\rm ion}$ is large in galaxies with high O$_{32}$ ratios.

On the other hand, \citet{Em20}, \citet{En21}, \citet{Fa19}, \citet{Ta19}, and
\citet{Na20} find a correlation between log($\xi_{\rm ion}$) 
and the equivalent widths of H$\alpha$ and [O~{\sc iii}] $\lambda$5007,
confirming that these quantities can be used to estimate $\xi_{\rm ion}$.

We now consider relations between $\xi_{\rm ion}$ and some global
parameters of our CSFGs to check whether these relations are consistent with 
similar relations for high-redshift SFGs and whether the relations for low-redshift
CSFGs can be used to predict $\xi_{\rm ion}$ in galaxies at any redshift.

In Fig.~\ref{fig11}a, we present the diagram $\xi_{\rm ion}$ -- $M_\star$ for the
entire CSFG sample. Some data for high-$z$ galaxies are present as well and they
are mostly located in the region of CSFGs with high EW(H$\beta$) $\ge$ 100$\AA$.
A considerable fraction of our galaxies, including almost all galaxies with 
EW(H$\beta$) $\ge$ 100$\AA$, are characterised by high values of $\xi_{\rm ion}$, 
above the canonical value \citep{Bo16} shown by the grey horizontal strip. This 
value is suggested
by models of the reionisation of the Universe. 
We find that no correlation exists between
$\xi_{\rm ion}$ and $M_\star$ for the entire sample, excluding the use of stellar mass for predicting
$\xi_{\rm ion}$.

However, the dispersions of the galaxy distributions in SFR bins are considerably
smaller (not shown), implying a strong dependence of the $\xi_{\rm ion}$ -- $M_\star$ 
relation on the secondary parameter SFR. This strong dependence is indeed seen in the 
fundamental relation $\xi_{\rm ion}$ -- SFR$^{-a}$$M_\star$, with $a$ = 0.9, 
in Fig.~\ref{fig11}b. Adopting $a$ = 1 would reduce the fundamental relation to 
$\xi_{\rm ion}$ -- sSFR$^{-1}$. Thus, the relation in Fig.~\ref{fig11}b can be used for
predicting $\xi_{\rm ion}$. However, its application requires two quantities, namely, 
$M_\star$ and the SFR.

Figure~\ref{fig12} shows the $\xi_{\rm ion}$ -- EW([O~{\sc iii}]~$\lambda$5007) and 
$\xi_{\rm ion}$ -- EW(H$\alpha$~$\lambda$6563) relations for the entire sample 
of CSFGs. For comparison,  some data for high-$z$ galaxies
are also included. They are in agreement with those for CSFGs.
Both panels display tight correlations between
$\xi_{\rm ion}$ and EW([O~{\sc iii}]~$\lambda$5007), and EW(H$\alpha$~$\lambda$6563). The linear regressions are shown by solid black lines. 
Thus, both EW([O~{\sc iii}]~$\lambda$5007) and EW(H$\alpha$) are good indicators of $\xi_{\rm ion}$ in SFGs at any redshift. In particular, $\xi_{\rm ion}$ is above the canonical value in galaxies with
EW([O~{\sc iii}]~$\lambda$5007) and EW(H$\alpha$)~$\ga$~300$\AA$\ 
(Figs.~\ref{fig12}a -- \ref{fig12}b). We note, however, that we may see a
considerable number of both CSFGs and high-$z$ SFGs with low EWs, but with
$\xi_{\rm ion}$s above the canonical value. We attribute this result to
uncertainties in EW measurements and SED modelling.

On the other hand, $\xi_{\rm ion}$ does not show any correlation with the
absolute magnitude $M_{\rm FUV}$ (Fig.~\ref{fig13}a). For comparison, we also
show high-$z$ galaxy data, which are in general agreement with the data for CSFGs  
at the bright end of $M_{\rm FUV}$. 
Only two high-$z$ galaxies are outliers, located above the distribution of CSFG galaxies.

We next consider relations between $\xi_{\rm ion}$ and the 
intrinsic ($\beta_{\rm int}$) and obscured ($\beta_{R(V)=3.1}$) slopes of the UV
continuum (Figs.~\ref{fig13}b and \ref{fig13}c). As expected, almost all
CSFGs with EW(H$\beta$) $\ge$ 100$\AA$\ (blue dots) are located above the
canonical value and the data for 
high-$z$ SFGs are in agreement with those for CSFGs (black symbols and the
dashed green line in 
Fig.~\ref{fig13}c), if the \citet{C89} reddening law, with $R(V)$ = 3.1 and
extinctions derived from the Balmer decrement, are adopted for the latter 
galaxies.

\subsection{Relations involving the UV continuum slope $\beta$}\label{beta1}

The slope of the UV continuum (see Sect.~\ref{beta}) is a common characteristic derived from the
photometric and spectroscopic observations of high-redshift SFGs. Thus,
\citet{Bo12,Bo14} derived a slope $\beta$ for a large sample of
$z$ $\sim$ 4 -- 8 galaxies and reported a steepening of the slope at lower 
luminosities. These results are in agreement with what was found by \citet{Ka17}, who obtained
$\beta$ $<$ --2 for all galaxies with $M_\star$~$<$~10$^8$~M$_\odot$.
\citet{BC20} obtained $\beta$ for galaxies at 
$z$ = 6 -- 9 in the Frontier Field cluster MACSJ0416.1-2403. No 
correlation was seen between $\beta$ and the rest-frame UV magnitude 
at $\lambda$ = 1500$\AA$. Instead, they found a strong 
correlation between $\beta$ and stellar mass, with lower mass galaxies 
exhibiting bluer UV slopes, but no trend was seen between 
$\beta$ and sSFR.
Finally, \citet{Wi16} derived $\beta$ $\sim$ --1.9 - --2.3 for 
$z$ $\sim$ 10 galaxies.

In this section, we consider whether the UV slopes of CSFGs are consistent with those of high-redshift SFGs. Fig.~\ref{fig14}a and \ref{fig14}b present
the dependences of the intrinsic slope $\beta_{\rm int}$ on the rest-frame
equivalent widths of the H$\alpha$ and [O~{\sc iii}] $\lambda$5007 emission lines. The dotted horizontal lines indicate the lowest value of $\beta$~$\sim$~--2.75,
derived from the stellar continuum of the youngest bursts. 
We note that $\beta_{\rm int}$ attains its minimum value at 
EW(H$\alpha$)~$\sim$~700$\AA$\ (corresponding to EW(H$\beta$)~$\sim$~200$\AA$)
and then increases again at higher EW(H$\alpha$).
This is caused by the contribution of nebular continuum in the UV 
range, characterised by a shallower slope than the stellar slope. This continuum
contribution is relatively high at large EW(H$\alpha$)~$\ga$~700$\AA$\ and 
should be taken into account \citep*[see also ][]{Ra10}. 

Figs.~\ref{fig14}c and \ref{fig14}d show the obscured $\beta$s vs.
the H$\alpha$ and [O~{\sc iii}] $\lambda$5007 equivalent
widths, adopting the \citet{C89} reddening law, with $R(V)$ = 3.1. The 
distribution of high-redshift SFGs in Fig.~\ref{fig14}d is in agreement with 
that for CSFGs, excluding high-$z$ objects with the steepest slopes.

Similarly, there is agreement between the CSFGs and high-redshift SFGs
in the diagrams of $\beta_{R(V)=3.1}$ -- $M_{\rm FUV}$ and 
$\beta_{R(V)=3.1}$ -- $M_\star$, respectively 
(Figs.~\ref{fig14}e and \ref{fig14}f), excluding some high-$z$ galaxies with 
lowest $\beta$s. Again, this indicates an overall agreement between the 
properties of CSFGs and high-redshift galaxies.

\section{Conclusions} \label{summary}

In this study, we discuss the properties of a sample of $\sim$ 25,000 
compact star-forming galaxies (CSFGs) at redshift $<$ 1
from Data Release 16 (DR16) of the Sloan Digital Sky survey (SDSS).
The properties of the CSFGs are compared to those of star-forming galaxies 
at high redshift ($z$ $\ga$ 1.5), including the relations for equivalent widths 
of the strongest emission lines EW([O~{\sc ii}] $\lambda$3727), EW(H$\beta$), 
EW([O~{\sc iii}] $\lambda$5007), and EW(H$\alpha$). Our results are as follows.

1. The sample of CSFGs includes galaxies in a wide range of stellar masses from
$\sim$ 10$^6$M$_\odot$ to $\sim$ 10$^{11}$M$_\odot$ and is characterised by high 
star formation activity with SFR up to 
$\sim$ 100 M$_\odot$ yr$^{-1}$ and  sSFR up to
several hundred Gyr$^{-1}$. The equivalent widths of the strongest emission
lines [O~{\sc iii}] $\lambda$5007 and H$\alpha$ are
in excess of $>$ 1000 $\AA$\ in $\sim$ 1200 galaxies.

2. Our comparison shows that the properties of our $z < 1$ CSFGs are very similar 
to all known properties of high-redshift galaxies at $z$ $\sim$ 1.5 -- 10. We find
no differences between these two types of objects, indicating similar 
physical characteristics of massive stellar populations in these galaxies and showing CFSGs to 
be good local analogues of the high-redshift objects.

In conclusion, we find that CSFGs can be studied in much more detail compared to high-$z$ SFGs thanks to 
their proximity and the ability to characterise them based on a much wider range of physical 
characteristics, such as stellar masses, star formation rates, and UV
luminosities. Therefore, they can be used to predict the physical properties of 
high-redshift galaxies to be studied by future ground-based and space telescopes, including the James Webb 
Space Telescope.

\begin{acknowledgements}
 Y.I.I. and N.G.G. thank the hospitality of the Max-Planck 
Institute for Radioastronomy, Bonn, Germany.    
They acknowledge support from the 
National Academy of Sciences of Ukraine (Project ``Dynamics of particles and
collective excitations in high energy physics, astrophysics and quantum
microsystems'').
Funding for the Sloan Digital Sky Survey (SDSS) has been provided by the 
Alfred P. Sloan Foundation, the Participating Institutions, the National 
Aeronautics and Space Administration, the National Science Foundation, 
the U.S. Department of Energy, the Japanese Monbukagakusho, and the Max Planck 
Society. The SDSS Web site is http://www.sdss.org/.
The SDSS is managed by the Astrophysical Research Consortium (ARC) for the 
Participating Institutions. The Participating Institutions are The University 
of Chicago, Fermilab, the Institute for Advanced Study, the Japan Participation 
Group, The Johns Hopkins University, Los Alamos National Laboratory, the 
Max-Planck-Institute for Astronomy (MPIA), the Max-Planck-Institute for 
Astrophysics (MPA), New Mexico State University, University of Pittsburgh, 
Princeton University, the United States Naval Observatory, and the University 
of Washington.
\end{acknowledgements}












\end{document}